\documentclass[a4paper,11pt]{article}
\usepackage{jheppub}
\usepackage{graphicx}
\usepackage{amsmath}
\usepackage{amsfonts}
\usepackage{amssymb}
\usepackage{paralist}
\usepackage{slashed}
\usepackage[colorlinks=true,linkcolor=blue]{hyperref}
\usepackage{bm}
\usepackage{empheq}
\def\sss{\scriptscriptstyle}
\def\half{\frac 12}

\newcommand{\Tr}[1]{\textrm{Tr}\left[#1\right]}
\def\Gammac{\Gamma^\mathrm{conv}}

\def \bx {\bm{x}}

\def \bq {{\bm{q}}}

\def \cf {C_F}
\def \nc {N_c}
\def \ca {C_A}
\def \nf {N_f}

\def \df {d_F}
\def \da {d_A}

\def \bk {\mathbf{k}}

\def \bp {{\bm{p}}}

\newcommand{\Tint}[1]{{\hbox{$\sum$}\!\!\!\!\!\!\!\int\,}_{\!\!\!\!\raise-0.9ex\hbox{$\scriptstyle{#1}$}}}
\def\onetwo{{1\lra2}}
\def\twotwo{{2\lra2}}

\def \cc {\mathcal{C}}
\def\siml{{\ \lower-1.2pt\vbox{\hbox{\rlap{$<$}\lower6pt\vbox{\hbox{$\sim$}}}}\ }}
\def\simg{{\ \lower-1.2pt\vbox{\hbox{\rlap{$>$}\lower6pt\vbox{\hbox{$\sim$}}}}\ }}

\def \bx {{\bm{x}}}

\def \bq {\mathbf{q}}

\def \crr {C_R}

\def \bfnabla {\boldsymbol{\nabla}}
\def \bk {\mathbf{k}}

\def \bp {{\bm{p}}}

\def \cc {\mathcal{C}}

\def \als {\alpha_{\mathrm{s}}}

\def \m2   {\mu^{2 \epsilon}}

\newcommand{\order}[1]{\mathcal{O}\left(#1\right)}

\def\siml{{\ \lower-1.2pt\vbox{\hbox{\rlap{$<$}\lower6pt\vbox{\hbox{$\sim$}}}}\ }}
\def\simg{{\ \lower-1.2pt\vbox{\hbox{\rlap{$>$}\lower6pt\vbox{\hbox{$\sim$}}}}\ }}
\def\lqcd{\Lambda_\mathrm{QCD}}

\def\nn {\nonumber}

\def\bv {\mathbf{v}}

\def\pp {p_\perp}

\def\qp {q_\perp}

\def\bqp {\mathbf{q}_\perp}
\def\kp {k_\perp}
\def\bkp {\mathbf{k}_\perp}
\def\ql {\hat{q}_{\sss L}}
\def\etad{\eta_{\sss D}}
\def\mmf {m_\infty^2}
\def\mmg {M_\infty^2}
\def\dep {\delta E_\bq}
\def\depq {\delta E_{\bq+\bk}}
\def\depqm {\delta E_{\bq-\bk}}
\def\calr {\mathcal{C}_R}
\def\cala {\mathcal{C}_A}
\def\lra{\leftrightarrow}
\def\ul{\underline}
\def\Eq#1{Eq.~\eqref{#1}}

\def \bh {\mathbf{h}}
\def \bff {\mathbf{F}}
\def \b {\mathbf{b}}

\def \OO {\mathcal{O}}

\def \pl {p^+}
\def \pmm {p^-}
\def \qm {q^-}
\def \qll {q^+}
\def \km {k^-}
\def \kl {k^+}

\def \cc{{\mathcal C}}

\def\x{{\bm x}}
\def\p{{\bm p}}
\def\q{{\bm q}}
\def\k{{\bm k}}

\def\u{{\bm u}}

\def\D{{\bm D}}

\def\nfd {n_{\sss F}}
\def\nbe {n_{\sss B}}
\def\PP {\delta\!f}

\def\pll {p_{\sss L}}
\def\qhat {\hat{q}}
\def\LO{{\mathrm{LO}}}
\def\NLO{{\mathrm{NLO}}}
\def\coll{{\mathrm{coll}}}
\def\qpt {\tilde{q}_\perp}
\def\md {m_{\sss D}}
\def\ra{{$r,a$}\,}

\def\Isc{\frac{\hat{q}(\delta E)}{g^2 \crr}}
\def\mupp{\mu_{\perp}}
\def \mull{\mu_\parallel}
\def\ie{\textit{i.e.}}
\newcommand{ \beqe}{\begin{empheq}[box=\fbox]{equation}}
\newcommand{ \beqa}{\begin{empheq}[box=\fbox]{align}}

\def\Ref#1{ref.~\cite{#1}}

\title{Jet-Medium Interactions at NLO in a Weakly-Coupled Quark-Gluon
Plasma}
\author[1]{Jacopo Ghiglieri,}
\author[2]{Guy D. Moore,}
\author[3]{and Derek Teaney}

\affiliation[1]{Institute for Theoretical Physics, Albert Einstein
Center,\\ University of Bern, Sidlerstrasse 5, 3012 Bern, Switzerland}
\affiliation[2]{Institut f\"ur Kernphysik, Technische Universit\"at Darmstadt\\
Schlossgartenstra\ss e 2, D-64289 Darmstadt, Germany}
\affiliation[3]{Department of Physics and Astronomy, Stony Brook University,\\
Stony Brook, New York 11794-3800, United States}
\emailAdd{jacopo.ghiglieri@itp.unibe.ch}
\emailAdd{guy.moore@physik.tu-darmstadt.de}
\emailAdd{derek.teaney@stonybrook.edu}

\abstract{
  We present an extension to next-to-leading order in the strong
  coupling constant $g$ of the AMY effective kinetic approach to
  the energy loss of high momentum particles in the quark-gluon plasma.
  At leading order, the transport of jet-like particles is determined 
  by elastic scattering with the thermal constituents, and by
  inelastic collinear splittings induced by the medium.
  We reorganize this description into collinear splittings,
  high-momentum-transfer scatterings,  drag and diffusion, and
  particle conversions (momentum-preserving identity-changing
  processes).
  We show that this reorganized description remains valid to 
  NLO in $g$, and compute the appropriate modifications of the drag, diffusion,
  particle conversion, and inelastic splitting coefficients.
  In addition,   a new kinematic regime opens at NLO
  for wider-angle collinear bremsstrahlung.
  These semi-collinear emissions smoothly interpolate between the
  leading order high-momentum-transfer scatterings
  and collinear splittings.  To organize the calculation, we introduce a set of
  Wilson line operators on the light-cone which determine the diffusion and
  identity changing coefficients, and we show how to evaluate these
  operators  at NLO.
}

\begin{document}
\maketitle
\section{Introduction}

Jets are a key observable in the relativistic heavy-ion program~\cite{Roland:2014jsa,d'Enterria:2009am,Majumder:2010qh,Mehtar-Tani:2013pia}.
Advances in reconstructing jets at the LHC \cite{Aad:2010bu,Chatrchyan:2011sx}
challenge our ability to understand the difference
between jet development in the hot medium created in a heavy ion
collision, compared to development in the vacuum or near-vacuum
environment of a proton-proton collision.  While early theoretical
studies concentrated on understanding the leading
hadron in a jet (see \Ref{Burke:2013yra} for an overview), the more inclusive jet reconstructions which are now
possible experimentally demand a theoretical description of the full jet
evolution, including the evolution of all radiated daughters.

Several groups have put forward modeling frameworks for doing this
\cite{Majumder:2013re,Zapp:2011ya,Schenke:2009gb,Renk:2010zx,Armesto:2009fj}.  It is fair to say that these approaches have
some commonalities.  Generally they separate the excitations into
high-energy partons associated with the jet, and low-energy partons or
a scattering medium with a characteristic energy scale $T$ (the local medium
temperature).  Then, one attempts to follow the evolution of the high energy
partons, which will eventually create the hadrons reconstructed as a
jet.  The jet partons are considered to interact with the medium in two
important ways.  They scatter elastically, and they are induced to
radiate or split.  Different frameworks differ in whether both
possibilities are considered, and in exactly how the splitting processes
are computed (how is long-distance coherence handled?  Is the radiated
daughter assumed to have a small fraction of the energy?  What model for
the medium interactions, and what other approximations are made?).

Typically the division of processes into distinct types -- here elastic
scattering and inelastic radiation -- is justified at leading order, but
at subleading orders they often cannot be clearly distinguished.  What happens
to the treatment of jet-medium interaction at subleading order?  Is it
possible to pursue a next-to-leading order (NLO) calculation, in the sense
that the elastic and splitting interactions between the jet partons and
the medium are treated beyond leading perturbative order?%
\footnote%
    {Some authors have used the term NLO in a different sense: for instance
      that the initial parton producing processes or the final
      fragmentation processes are treated at NLO, though the medium
      interactions are still leading order~\cite{Vitev:2009rd,He:2011pd}, or
      within the Higher Twist formalism \cite{Xing:2014kpa,Kang:2014ela}, or
      that higher-order, double-logarithmic corrections to the
      jet-quenching parameter are
      considered and resummed~\cite{Liou:2013qya,Blaizot:2013vha,Blaizot:2015lma}.
      Here by NLO we mean a
      beyond-leading-order treatment of the way the jet interacts with
      the medium.}
In this paper we explore this question by extending a framework where it
is clearly posed -- the AMY/McGill/MARTINI approach
\cite{Arnold:2002ja,Arnold:2002zm,Jeon:2003gi,Qin:2007rn,
Schenke:2009ik,Schenke:2009gb},
where there is a clear power-counting
prescription for determining what is leading and subleading order for
the jet-medium interaction.  The
approach starts with the assumptions that the medium
is thermal and weakly coupled, so the interactions between jet partons
and the medium can be computed in thermal perturbation theory.  One also
assumes that the medium is thick, such that the formation times of
processes under consideration are shorter than the scale of variation of
the medium.  This approximation has sometimes been criticized, and it
can be improved upon without overturning the rest of the approach
\cite{CaronHuot:2010bp}.  But for the processes which will be most
interesting here -- processes involving small momentum transfer or
intermediate opening angles -- the scattering or formation times are
relatively short, so this should be considered a separate issue.

The philosophy of the framework is as follows.  We follow one or more
``hard'' approximately on shell partons traversing the medium.
We assume that the medium has a local temperature $T \gg
\lqcd$, and distinguish a parton as hard if its energy $E$ satisfies
$\exp(-E/T) \ll 1$.  Particles failing this criterion are assumed to
join the thermal medium; but no attempt is made to track the
back-reaction on the medium properties~\cite{Iancu:2015uja}.

The jet parton evolution and jet-medium interactions are dictated by
finite temperature perturbation theory.  Whereas vacuum perturbation
theory is an expansion in the strong fine-structure constant
$\als = g^2/4\pi$, this expansion is spoiled by soft-particle
statistical functions $\nbe(\omega\sim gT) \sim 1/g$ entering
in thermal Feynman graphs. These soft contributions
must be resummed  to obtain a finite  leading order answer, and give rise
to subleading corrections suppressed by a single power of $g$.
We have recently shown how to compute these
subleading corrections in the context of hard real \cite{Ghiglieri:2013gia} and
virtual \cite{Ghiglieri:2014kma} photon production.  Here we extend that
treatment to the case of jet-medium interactions.

As a scattering environment, the essential attribute of QCD (or any
gauge theory) is that there is a large cross section for
small-momentum-transfer scattering processes.  These are responsible for
the high rate of particle splitting.  They also cause complications when
including elastic scattering, since they give a large rate of small
momentum exchanges, both in the transverse and longitudinal components
of the momentum.  At next-to-leading (NLO) order, new processes arise,
which can be understood physically as overlap and interference between
sequential scattering processes and as scatterings with the emission or
absorption of soft ($E\sim gT$) excitations \cite{simonguy}.  These
contribute both
to transverse momentum broadening and to longitudinal momentum loss and
broadening.  They are most easily computed in a way which does not
cleanly separate them into elastic and inelastic processes, and indeed
it is not clear that the distinction is important or well posed.  And
they overlap with the infrared limits of both the elastic scattering and
the splitting processes.  However, frequent and small momentum exchanges
need not be separately identified and tracked.  In traversing enough
medium to significantly modify a jet, the jet partons will undergo
several such soft processes, in which case a statistical description
should be sufficient.  This motivates an approach in which we give a
Fokker-Planck (Langevin) description of soft scatterings, as drag and diffusion processes.

The philosophy of our approach will therefore be the following.  We will
introduce infrared scales $\mupp$, $\mull$.  All scattering
and emission processes which change a jet parton's momentum by more than
$\mu_{\perp,\parallel}$ will be handled explicitly.  All processes which
change momentum by less than this scale, including the NLO effects
alluded to above, will be incorporated as momentum diffusion and drag
coefficients, which can be neatly defined in field theory as
correlators of field strength operators on light-like
Wilson lines.  We will compute these drag and diffusion coefficients at
the NLO level, as well as providing an NLO accurate procedure for
computing the larger-transfer elastic and splitting processes.
We will show explicitly how to perform a matching so that
the choice of the scale $\mu$ drops out in the final results.

$\!$The drag and diffusion coefficients account for momentum
exchange with the medium through soft gauge-boson exchange.
Soft fermion exchange with the medium
can change the identity of a quark to a gluon and \textsl{vice versa}.  We
call such identity changing processes \textsl{conversion processes},
and introduce a medium coefficient (analogous to  the transverse momentum
broadening coefficient $\hat q$ or the energy loss $\hat e$)
which  parameterizes this conversion rate. As with the drag and diffusion
coefficients, all identity changing scattering processes with
momentum exchange greater than $\mu$ will be treated explicitly, while
identity changing scatterings with small momentum
transfer are incorporated into the conversion rate.
The conversion rate will be defined as a correlator of soft
fermionic operators on light-like Wilson lines.

The calculation of the processes involving soft momentum
transfers (drag, diffusion, and conversions) and of the corresponding
light-cone Wilson line correlators
requires 
a resummation scheme known as the Hard Thermal Loop (HTL) effective
theory \cite{Braaten:1989mz,Braaten:1991gm},  which is the QCD analog
of the Vlasov equations \cite{Blaizot:2001nr}.
These formalisms are well known to be computationally complex, and at first sight, any 
calculation beyond leading order in the coupling
 would seem extremely challenging (see \cite{simonguy} for an
 example). However, Caron-Huot has shown \cite{CaronHuot:2008ni} that
HTL correlators (and statistical correlators more generally)
simplify greatly when computed at light-like
separations,
which are exactly those that must be evaluated to determine the
energy loss, diffusion, conversion, and collinear radiation rates of highly
energetic particles propagating in a plasma.

Intuitively, these simplifications
can be seen to arise because
the energetic partons are propagating
almost exactly along the light cone. Hence they are probing an essentially
undisturbed plasma, at least as far as the soft, classical background
is concerned.  Informally, we can say that this background
``can't keep up'' with the hard  particles traversing the plasma.
Thus, the soft correlations that the latter probe
are statistical in nature rather
than  dynamical. Those simplifications, as we shall show,
are at the base of the NLO extension being presented.

A pedagogical review of these recent developments in the understanding
of HTLs has been presented by
two of us in \cite{Ghiglieri:2015zma}. There the main
results of this paper, \ie\ the reorganization
of the kinetic theory we have mentioned, as well as the results of
the computations to NLO of the needed rates and coefficients, have
been partly anticipated. Due
to the review nature of \cite{Ghiglieri:2015zma}, the presentation
there has been more pedagogical and most details and technical aspects
have been omitted for the sake of brevity and clarity. Here we
will present the detailed derivation of the reorganization
of the kinetic approach, as well as the explicit calculations of
the coefficients and rates. Furthermore, \cite{Ghiglieri:2015zma}
was limited to a plasma of gluons only, again for ease of illustration.
We advise readers unfamiliar with Hard Thermal Loops, and especially
with the recent developments discussed before, to explore
\cite{Ghiglieri:2015zma} first.

The paper is organized as follows: in Sec.~\ref{sec_lo_amy} we
present the LO framework in the standard
formulation, which divides into elastic ($\twotwo$) and inelastic
($\onetwo$) processes. Readers familiar with that approach can skip directly
to Sec.~\ref{sec_reorganize}, where we introduce our reorganization in
terms of large-angle scatterings, diffusion, conversion and collinear processes.
In Sec.~\ref{sec_nlo_intro} we give an overview of the NLO corrections,
which are dealt with in detail in Sec.~\ref{sec_nlo_coll} for collinear
processes, Sec.~\ref{sec_long_diff} for diffusion, Sec.~\ref{sec_nlo_conv} for
conversion and finally Sec.~\ref{sec_semi} for the semi-collinear processes,
which first contribute at NLO and smoothly interpolate between the other three.
A summary is presented in Sec.~\ref{sec_concl}, together with our conclusions.
Extensive technical details are to be found in the appendices, such as the NLO
calculation of longitudinal momentum diffusion. 

\smallskip

\noindent
\fbox{\begin{minipage}{0.98\textwidth}
\hspace{1.2em}%
The paper is rather long and detailed.  Some readers will be more
interested in learning how to apply its results to their own
(numerical) treatment of jet modification, without necessarily
following all the details.  In the online version of the paper, we
have put boxes around the key equations which must be included in
an implementation of our results.  These readers can skip most of the
text and focus on these boxed equations.
\end{minipage}}

\section{The leading-order kinetic approach}
\label{sec_lo_amy}

Our aim is to track the time evolution of a small number of highly-energetic
jet-like particles as they propagate through a medium.
We will refer to energies and momenta of order $E$ as \emph{hard},
of order $T$ as \emph{thermal} and
of order $gT$ as \emph{soft}\footnote{
The notation and terminology used here is summarized in App.~\ref{appnotate}, 
and closely follows our previous work~\cite{Ghiglieri:2013gia}.}. 
The hard particles are very close to the mass shell, with energy
$p^0\sim p\sim E$ and virtuality
$|p_0-p| \siml g^2 T$.  We will assume that this
energy is large enough that $\exp(-E/T)\ll1$ and can be neglected, but we will not treat
$T/E$ as an explicit expansion parameter. Thus,  for instance, we do not distinguish
between a rate that is of order $g^4 T$ and one that is $g^4\sqrt{TE}$. 
Moreover, we will often find convenience in using light-cone
coordinates, specifically those defined by the hard four-vector $P$. If, without loss of generality
in an isotropic medium,
$\p$ points in the $\hat{z}$ direction, then for a generic vector $K$ we can define
$k^{-}\equiv k^0- k^z$ and $k^+ \equiv \frac{k^0 + k^z}{2}$.
This normalization, already adopted in \cite{Ghiglieri:2013gia}, is nonstandard,
but we find it convenient because
$dk^0 dk^z = dk^+ dk^-$, and because we will frequently encounter cases
in which $k^-\approx0$, in which case $k^z \approx k^0 \approx k^+$ with our conventions.
The transverse coordinates are written as $\bkp$, with modulus $\kp$.

\begin{figure}[ht]
\begin{center}
\includegraphics[width=10cm]{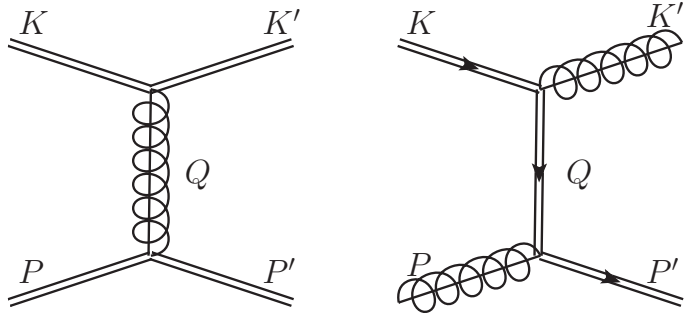}
\end{center}
\caption{Typical diagrams contributing to $\twotwo$ processes at LO. Double lines represent
 hard or thermal particles, which have at least one momentum component of the order of the temperature
or larger. Parallel double lines without arrows can be either gluons or quarks. When
particle identities need to be specified, quarks are identified by the fermion flow arrow and gluons
by the curly line. In all diagrams in the paper, time is understood to flow
from left to right.}
\label{fig_22}
\end{figure}

\begin{figure}[ht]
\begin{center}
\includegraphics[width=7cm]{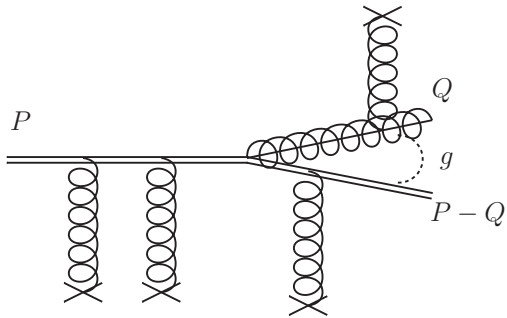}
\end{center}
\caption{A typical diagram contributing to $\onetwo$ processes at LO. The single curly line is a soft
gluon. The crosses represent the soft thermal scattering centers -- see, for instance, \Ref{Arnold:2002ja}. }
\label{fig_coll}
\end{figure}

Let us start from the effective kinetic theory
developed in \cite{Arnold:2002zm}. The Boltzmann equation  reads
\begin{equation}
\label{boltzmann}
\left(\frac{\partial }{\partial t}+\bv\cdot\bfnabla_\bx\right)f^a(\bp,\bx,t)=-C_a^{2\lra2}[f]-C_a^{1\lra2}[f],
\end{equation}
where $f^a(\bp,\bx,t)$ is the phase space distribution for a single color and
helicity state quasiparticle of type $a$ ($f^a=dN^a/(d^3xd^3p)$). In the collision operator, at leading
order in the coupling $g$, one needs to account for $\twotwo$ and  effective
$\onetwo$ processes. The $\twotwo$ rates are given by the simple $\twotwo$ diagrams
of QCD, such as those shown in Fig.~\ref{fig_22}, which also establishes our
graphical conventions.
The $\onetwo$ rates describe the
collinear radiation from the jet-like particles, which is induced by multiple  soft
scatterings with the background plasma, see Fig.~\ref{fig_coll}.
Although apparently suppressed by powers of $g$, multiple scatterings contribute
at leading order under the provision that:
\begin{inparaenum}[(a)]
\item the momenta of the hard lines are nearly on shell
  and collinear to each other (\ie \, $\theta\siml g$, where $\theta$ is the emission angle\footnote{\label{foot_angle} In the case where
$P$ and $Q$ are both thermal, such as
when dealing with the thermal photon rate, then
the angle is of order $g$. In the case of interest, \ie\ $P$ hard, there are two different possibilities.
If either $Q$ or $P-Q$ are thermal, \ie\ there is a hierarchical separation between the emitted particles,
then the angle is again of order $g$. If instead the splitting is more democratic, with no hierarchical
separation, then the angle can become as small as $g \sqrt{T/E}$.}),
and
\item the momenta $K$ of the soft gluons are space-like  $\kl,\kp\sim g T$ and $k^-\siml g^2T$.
\end{inparaenum}
A complete leading order treatment of collinear radiation must consistently
resum these soft scatterings to account for the \emph{Landau-Pomeranchuk-Migdal}
(LPM) effect
\cite{Baier:1994bd,Baier:1996kr,Zakharov:1996fv,Zakharov:1997uu,Arnold:2002ja}.

In detail, the collision operator reads (dropping for brevity the spacetime dependence, which is local)
\begin{eqnarray}
C_a^\twotwo[f](\p)
&=&
\frac {1}{4|\p|\nu_a} \sum_{bcd}
\int_{\k\p'\k'}
\left| {\cal M}^{ab}_{cd}(\p,\k;\p',\k') \right|^2 \>
(2\pi)^4 \, \delta^{(4)}(P + K - P' - K')
\nonumber\\ && \hspace {0.9in} {} \times
\Bigl\{
f^a(\p) \, f^b(\k) \, [1{\pm}f^c(\p')] \, [1{\pm}f^d(\k')]
\nonumber\\ && \hspace {1.0in} {}
-
f^c(\p') \, f^d(\k') \, [1{\pm}f^a(\p)] \, [1{\pm}f^b(\k)]
\Bigr\} \,,
\label{eq:collision22}
\\
\noalign{\hbox{and}}
C_a^\onetwo[f](\p)
&=& \frac{(2\pi)^3}{2|\p|^2\nu_a} \sum_{bc}
\int_0^\infty dp'\> dq' \;
\delta (|\p| - p' - q' ) \;
\gamma^{a}_{bc}(\p;p'\hat\p,q'\hat\p)
\nonumber\\ && \hspace{3em} {} \times
\Bigl\{
f^a(\p) \, [1{\pm}f^b(p' \hat\p)] \, [1{\pm}f^c(q' \hat\p)]
-
f^b(p' \hat\p) f^c(q' \hat\p) \, [1{\pm}f^a(\p)]
\Bigr\}
\nonumber \\ &+&
\frac {(2\pi)^3}{|\p|^2\nu_a} \sum_{bc}
\int_0^\infty dq \> dp' \;
\delta (|\p| + q - p' ) \;
\gamma_{ab}^{c}(p'\hat \p;\p,q \, \hat\p)
\nonumber\\ && \hspace{2em} {} \times
\Bigl\{
f^a(\p) \, f^b(q \hat\p) [1{\pm}f^c(p' \hat\p)]
-
f^c(p' \hat\p) \, [1{\pm}f^a(\p)][1{\pm}f^b(q \hat\p)]
\Bigr\},\quad
\label{eq:collision12}
\end{eqnarray}
where the sum runs over the species $bc(d)$ in the
scattering/splitting event, and the splitting kernel $\gamma^{a}_{bc}$
is defined in Eq.~(5.1--5.4) of Ref.~\cite{Arnold:2002zm}, see also
\Eq{onetwocollisionq}. We are also using the shorthand notation
\begin{equation}
\label{lorentzint}
\int_\k\ldots\equiv\int\frac{d^3k}{2k(2\pi)^3}\ldots
\end{equation}
for the Lorentz-invariant integration.\ The matrix elements $\mathcal{M}$ and
transverse-momentum integrated matrix elements
$\gamma$ will be discussed in the following. $\nu^a=2d_R$ is the degeneracy
of the particle $a$: two spin degrees of freedom and $d_R$ color degrees of freedom,
where $d_R$ is the dimension of the representation of $a$. For quarks it is $d_F=N_c$,
for gluons $d_A=N_c^2-1$.

The hard particles are very dilute, and therefore we only need to
track the interactions of these modes with the thermal and soft constituents.
This can be done by defining $\PP$
\begin{equation}
\label{defdeltaf}
f^a(\bp,\bx,t)=n^a(\p,T(\bx,t),\u(\bx,t))+\PP^a(\p,\x,t),
\end{equation}
and linearizing the Boltzmann equation in this quantity.
Here $n$ is the (local) equilibrium distribution, written generally as a function of the local
temperature $T$ and flow velocity $\u$. In the following we will work
in the local rest frame
where $n$ becomes the Fermi--Dirac distribution $\nfd(p)$
or the Bose--Einstein distribution $\nbe(p)$.
Substituting \Eq{defdeltaf} in the collision operator and
dropping terms which are of order $e^{-p/T}$ yields
\begin{eqnarray}
C_a^\twotwo[\PP](\p)
&=&
\frac {1}{4|\p|\nu_a} \sum_{bcd}
\int_{\k\p'\k'}
\left| {\cal M}^{ab}_{cd}(\p,\k;\p',\k') \right|^2 \>
(2\pi)^4 \, \delta^{(4)}(P + K - P' - K')
\nonumber\\ && \hspace {0.0in} {} \times
\Bigl\{
\PP^a(\p) \, n^b(k) \, [1{\pm}n^c(p'){\pm}n^d(k')]
-
\PP^c(\p') \, n^d(k')
\, [1{\pm}n^b(k)]
\nonumber\\ && \phantom{\times\Bigl\{}  {}
-
n^c(p') \, \PP^d(\k')
\, [1{\pm}n^b(k)]
\Bigr\} \,,
\label{eq:collision22exp}
\\
\noalign{\hbox{and}}
C_a^\onetwo[\PP](\p)
&=& \frac{(2\pi)^3}{2|\p|^2\nu_a} \sum_{bc}
\int_0^\infty dp'\> dq' \;
\delta (|\p| - p' - q' ) \;
\gamma^{a}_{bc}(\p;p'\hat\p,q'\hat\p)
\nonumber\\ && \hspace{-0.4cm} {} \times
\Bigl\{
\PP^a(\p) \, [1{\pm}n^b(p'){\pm}n^c(q' )]
-
[\PP^b(p'\hat\p ) n^c(q' )+n^b(p' ) \PP^c(k'\hat\p )]
\Bigr\}
\nonumber \\ &+&
\frac {(2\pi)^3}{|\p|^2\nu_a} \sum_{bc}
\int_0^\infty dq \> dp' \;
\delta (|\p| + q - p' ) \;
\gamma_{ab}^{c}(p'\hat \p;\p,q \, \hat\p)
\nonumber\\ && \hspace{0em} {} \times
\Bigl\{
\PP^a(\p) \, n^b(q )-
\PP^c(p'\hat\p)[1{\pm}n^b(q)]
\Bigr\}.\quad
\label{eq:collision12exp}
\end{eqnarray}
These are to be used in a linearized Boltzmann equation for the hard
components $\PP^a(\p)$,
\begin{equation}
\label{boltzmann2}
\left(\frac{\partial }{\partial t}+v_\x\cdot
\bfnabla_x\right)\PP^a(\p,\x,t)=-C_a^\LO[\PP]=-C_a^{2\lra2}[\PP]-C_a^{1\lra2}[\PP] \,.
\end{equation}

In the $\twotwo$ collision integrals, soft gluon and fermion exchanges must be
screened to avoid logarithmic divergences.
At leading order, the bare $t-$ and $u-$channel propagators may be replaced 
with their Hard Thermal Loop counterparts in the
infrared to render the collision integrals finite.
This procedure (which is detailed in Appendix A of \cite{Arnold:2003zc})
provides the leading weak-coupling description for soft
exchanges, and is correct to order $g^2$ for hard exchanges. 
However, while this regularization prescription  provides the correct
leading order answer, it is not easily generalized to NLO. Further,
the approach mixes different physics at different scales. In the next
section we will re-examine the $\twotwo$ collision rates,
incorporating soft $t,u-$channel exchanges into  drag, diffusion, and
conversion coefficients, which cleanly reflect the physics of the
Debye sector.  Then, in Sec.~\ref{sec_nlo_intro}, we will compute
these transport parameters at NLO.

\section{A reorganization of leading order: large-angle scattering, drag and diffusion, and conversions}
\label{sec_reorganize}
\begin{figure}[ht]
\begin{center}
\includegraphics[scale=0.8]{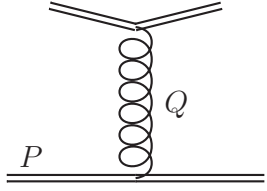}
\end{center}
\caption{The soft limit of a $t-$ or $u-$channel gluon exchange diagram. $P$ is the
hard momentum and $Q$ is the soft gluon momentum.}
\label{fig_diff}
\end{figure}

The leading order picture we have just described, with distinct
$\onetwo$ collinear processes and  $\twotwo$  processes dressed with
HTLs for IR finiteness, starts to be ill-defined  at NLO.  Consider
the soft limit of the  $\twotwo$ processes. In the case of a soft gluon
exchange, as shown in Fig.~\ref{fig_diff},
we obtain a process which changes the hard four-momentum $P$ by a small amount $Q\sim gT$,
\emph{without changing the particle identity}. We call such
a process a \emph{diffusion process}, since, as described in Sec.~\ref{sub_diff}, they can be treated in a diffusion approximation.

\begin{figure}[ht]
\begin{center}
\includegraphics[scale=0.8]{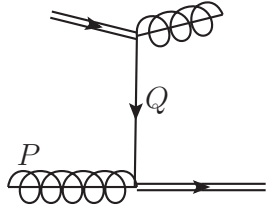}
\end{center}
\caption{The soft limit of a $t-$ or $u-$channel quark exchange diagram. $P$ is the
hard momentum and $Q$ is the soft quark momentum.}
\label{fig_conversion}
\end{figure}

In the case of a soft quark exchange,
 as shown in
Fig.~\ref{fig_conversion},
one obtains an identity-changing process.
 Here a hard gluon with momentum $P$ is turned into a quark with an almost equivalent momentum, up to
$O(gT)$. We then call these processes \emph{conversion processes}, and we will deal with them
in a different way, inspired by the NLO thermal photon rate \cite{Ghiglieri:2013gia}.

\begin{figure}[ht]
\begin{center}
\includegraphics[width=10cm]{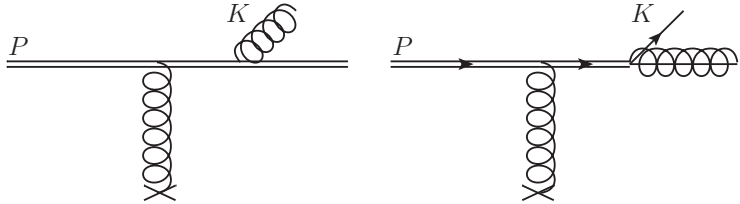}
\end{center}
\caption{The soft-$K$ limits of a $\onetwo$ process. The diagram on the left amounts to
a diffusion process at NLO, whereas the diagram on the right amounts to a conversion process.}
\label{fig_collsoft}
\end{figure}

Now consider a collinear $\onetwo$ process in the limit where one of the hard/thermal legs becomes
soft\footnote{LPM interference is suppressed
in this case~\cite{Ghiglieri:2013gia}.}, 
as shown in Fig.~\ref{fig_collsoft}.
In the first graph, the soft gluon emission contributes
to the (longitudinal) diffusion of the hard particle. Similarly
the soft quark emission contributes to the hard quark conversion rate.
At NLO we
will then need to subtract these limits from the collinear $\onetwo$ region and treat them as part of the diffusion or conversion processes respectively.

To summarize, at leading order we can  rewrite the right-hand side of \Eq{boltzmann2} as
\beqe
\label{boltzmannproc}
-C_a^\LO[\PP]=-C_a^\mathrm{large}[\PP]-C_a^\mathrm{coll}[\PP]
-C_a^\mathrm{diff}[\PP]-C_a^\mathrm{conv}[\PP].
\end{empheq}
Here $C^\mathrm{large}$ is the $\twotwo$ collision operator restricted to large
momentum transfers,  $Q \gg gT$. Defining this scattering rate requires regularization procedure, which we describe in the next section, Sec.~\ref{sub_large}. 
$C^\mathrm{diff}$ notates a diffusion approximation to the collision integral for small momentum transfer, $Q \sim gT$.
 This is discussed in Sec.~\ref{sub_diff}, where the LO longitudinal
 and transverse diffusion coefficients are extracted from the screened $2\leftrightarrow 2$ rates.  
Similarly, $C^\mathrm{conv}$ notates the conversion processes,
and the appropriate LO conversion coefficients are found in  Sec.~\ref{sec_lo_conv}.
The precise value of these diffusion and conversion coefficients depends on the regulator, but the dependence on the regulator cancels to leading order when
$C^{\rm large}, C^{\rm diff}$, and $C^{\rm conv}$ are taken together in
\Eq{boltzmannproc}.
Finally, $C^\mathrm{coll}$ consists of the collinear
$\onetwo$ rates $C^{\onetwo}$ after excluding (or subtracting) the diffusion and conversion-like emissions shown in Fig.~\ref{fig_collsoft}.
These soft emissions (which were originally included in the $C^{\onetwo}$ rates) are limited in phase space to $K\sim gT$, and their exclusion
constitutes an $\OO(g)$ correction. Thus, at leading order $C_a^\mathrm{coll}[\PP]=C^{\onetwo}[\PP]$. We therefore will present the explicit form of $C^\mathrm{coll}$ only when we describe its NLO corrections in Sec.~\ref{sec_nlo_coll}.

\subsection{Large-angle scattering}
\label{sub_large}
In this section, we describe the integration of the $2\leftrightarrow2$ matrix
elements with large momentum transfer, $Q \gg gT$, which enters in
the leading order collision kernel $C_a^{\mathrm{large}}$ in \Eq{boltzmannproc}. This is completely
straightforward, but integrals of the bare matrix elements must be regulated
with some scheme. The cutoff regulator chosen in this section  conveniently
matches with the calculations of the diffusion and conversion coefficients in
Sec.~\ref{sub_diff} and \ref{sec_lo_conv}.

In more detail, to evaluate $C_a^\mathrm{large}[\PP]$, one needs integrate
the matrix elements listed  in Table~\ref{table_mat}, \ie\ the
standard, leading-order QCD matrix elements, summed over all
color and spin indices, with the Mandelstam variables
$s=-(P+K)^2$, $t=-(P-P')^2$ and $u=-(P-K')^2$. 
To regulate  gluon and fermion exchanges in the $t$ and $u$ channels
we use the integration technology of Ref.~\cite{Arnold:2003zc}, which treats each channel differently.
\begin{table}
\begin{center}
\begin{tabular}{|c|@{\quad}l@{\quad}|}
\hline &
\\[-12pt]
$ab \lra cd$ & $\qquad \left|{\cal M}^{ab}_{cd}\right|^2 / g^4$
\\[4pt]
\hline &
\\[-12pt]
$
\begin {array}{c}
q_1 q_2 \lra q_1 q_2 \,,
\\ q_1 \bar q_2 \lra q_1 \bar q_2 \,,
\\ \bar q_1 q_2 \lra \bar q_1 q_2 \,,
\\ \bar q_1 \bar q_2 \lra \bar q_1 \bar q_2
\end {array}
$
&
$
\displaystyle
8\,  \frac{\df^2 \, \cf^2}{\da}
\left( \frac{s^2+u^2}{\ul{t^2}} \right)
$
\\[28pt]
$
\begin {array}{c}
q_1 q_1\lra q_1 q_1 \,, \\
\bar q_1 \bar q_1 \lra \bar q_1 \bar q_1 \>
\end{array}
$
&
$ \displaystyle
8\,  \frac{\df^2 \, \cf^2} {\da}
\left( \frac{s^2+u^2}{\ul{t^2}} + \frac{s^2+t^2}{\ul{u^2}} \right)
+
16\,  \df \, \cf
\left( \cf {-} \frac{\ca}{2} \right) \frac{s^2}{tu}
$
\\[14pt]
$q_1 \bar q_1 \lra q_1 \bar q_1$
&
$ \displaystyle
8\,  \frac{\df^2 \, \cf^2}{\da}
\left( \frac{s^2+u^2}{\ul{t^2}} + \frac{t^2+u^2}{s^2} \right)
+
16\, \df \, \cf
\left( \cf {-} \frac{\ca}{2} \right) \frac{u^2}{st}
$
\\[14pt]
$q_1 \bar q_1 \lra q_2 \bar q_2$
&
$ \displaystyle
8\, \frac{\df^2 \, \cf^2}{\da}
\left( \frac{t^2 + u^2}{s^2} \right)
$
\\[10pt]
$q_1 \bar q_1 \lra g \, g$
&
$ \displaystyle
8\, \df \, \cf^2
\left( \frac{u}{\ul{\ul{t}}} + \frac{t}{\ul{\ul{u}}} \right)
-
8\, \df \, \cf \, \ca
\left( \frac{t^2+u^2}{s^2} \right)
$
\\[14pt]
$
\begin {array}{c}
q_1 \, g \lra q_1 \, g \,,\\ \bar q_1 \, g \lra \bar q_1 \, g
\end {array}
$
&
$ \displaystyle
-8\, \df \, \cf^2
\left( \frac{u}{s}  +  \frac{s}{\ul{\ul{u}}} \right)
+
8\, \df \, \cf \, \ca
\left( \frac{s^2 + u^2}{\ul{t^2}} \right)
$
\\[14pt]
$g \, g \lra g \, g$
&
$ \displaystyle
16\, \da \, \ca^2
\left(
3 - \frac{su}{\ul{t^2}} - \frac{st}{\ul{u^2}} - \frac{tu}{s^2}
\right)
$
\\[8pt]
\hline
\end{tabular}
\end{center}
\vspace*{-5pt}
\caption
{%
\label{table_mat}
Squares of
vacuum matrix elements for $2\lra2$ particle processes
in QCD-like theories, summed over all spins and colors.
$q_1$ and $q_2$ represent fermions of distinct flavors,
$\bar q_1$ and $\bar q_2$ are the associated antifermions,
and $g$ represents a gluon.
}
\end{table}

Singly-underlined matrix elements come from gluon exchange diagrams,
and 
are those that, in the soft limit, give rise to gluonic IR divergences,
corresponding to diffusion processes. Similarly, doubly-underlined matrix elements come from fermion-exchange diagrams
and give rise, in the same limit, to conversion processes.
To illustrate the regularization scheme, let us consider the contribution
from the scattering of different quark species $q_1q_2\lra q_1q_2$, which is given by the square
of a single $t$-channel diagram. The $q_1q_2 \lra q_1q_2$ contribution to $C_{q_1}^\mathrm{large}[\PP]$
reads\footnote{\label{foot_final}
$N.b.$ the sum over $c$ and $d$ in \Eq{eq:collision22} yields a factor of two.}\footnote{
	\label{foot_vector} For ease of illustration,  we are considering for
	large-angle scatterings a simplified case
	where $\PP$ is a function of $p$ rather than of $\p$. Details on the phase space integration
	in the latter case can be found for instance in \cite{Kurkela:2015qoa}.}
\begin{eqnarray}
\nn C_{q_1}^\mathrm{large}[\PP]	&\supset&\frac{g^4}{(2\pi)^3}
\frac{\cf }{16p^2  }\int_{-\infty}^{+\infty} d\omega\int_0^{2p-\omega} dq
\int_{(q-\omega)/2}^\infty dk \theta(q-\vert\omega\vert)\int_0^{2\pi}\frac{d\phi}{2\pi}\frac{s^2+u^2}{t^2}\\
&&\times\Bigl\{
\PP^{q_1}(p) \, \nfd(k) \, [1-\nfd(p-\omega)-\nfd(k+\omega)]\nn\\
&&\hspace{5mm}-\big[
\PP^{q_1}(p-\omega) \, \nfd(k+\omega)
+
\PP^{q_2}(k+\omega) \, \nfd(p-\omega)\big] \, [1-\nfd(k)]
\Bigr\},
\label{el2}
\end{eqnarray}
where the techniques of \cite{Arnold:2003zc} have been followed, by \begin{inparaenum}[(i)] \item eliminating one of the three
integration variables in \Eq{eq:collision22exp} with the momentum-conserving $\delta$-function,
\item shifting one of the remaining ones to $\q\equiv \p-\p'=\k'-\k $, \item introducing
$\omega\equiv p-p'=k'-k$, and \item performing the angular integrations. \end{inparaenum}
The remaining angle $\phi$ represents the azimuthal angle between the
$(\bp,\q)$ plane and the $(\k,\q)$ plane.

The Mandelstam variables become
\begin{equation}
\label{mandelstam}
s=-\frac{t}{2q^2}\left[(p+p')(k+k')+q^2-\cos(\phi)\sqrt{(4pp'+t)(4kk'+t)}\right],\quad
t=\omega^2-q^2,	
\end{equation}
and $p^{'0}=p'$, $k^{'0}=k'$ imply
\begin{equation}
\label{qzcond}
\omega-\hat{\p}\cdot\q=\frac{\omega^2-q^2}{2p},\qquad\omega-\hat{\k}\cdot\q=-\frac{\omega^2-q^2}{2k}.
\end{equation}
It is then easy to see how the 
unscreened
logarithmic divergences 
show up for $\omega,q\sim gT\ll k,p$.
To separate off the divergent region 
(which will match with the diffusion operator 
$C^\mathrm{diff}[\PP]$ described in Sec.~\ref{sub_diff}), 
we change
integration variables from $\omega,q$ to $\omega,\qpt$ with
$\qpt \equiv\sqrt{-t}=\sqrt{q^2-\omega^2}$. 
We can then place an IR cutoff
$T\gg \mu_{\qpt}\gg gT$ on $\qpt$, leaving
\beqa
\nn C_{q_1}^\mathrm{large}[\PP]	\supset&\frac{g^4}{(2\pi)^3}
\frac{\cf}{16p^2}\int_{-\infty}^{p} d\omega\int_{\mu_{\qpt}}^{\sqrt{4p(p-\omega)}} d\qpt \frac{\qpt}{q}
\int_{(q-\omega)/2}^\infty dk \int_0^{2\pi}\frac{d\phi}{2\pi}\frac{s^2+u^2}{t^2}\\
&\times\Bigl\{
\PP^{q_1}(p) \, \nfd(k) \, [1-\nfd(p-\omega)-\nfd(k+\omega)]\nn\\
&\hspace{5mm}-\big[
\PP^{q_1}(p-\omega) \, \nfd(k+\omega)
+
\PP^{q_2}(k+\omega) \, \nfd(p-\omega)\big] \, [1-\nfd(k)]
\Bigr\}.
\label{el2reg}
\end{empheq}
\Eq{el2reg} implicitly depends on the cutoff $\mu_{\qpt}$.
For small $\qpt$ the dominant $\omega$ region is also small, 
and the fermion distribution can be approximated as, $\nfd(p-\omega) \simeq \nfd(p)$.

In the small $\qpt$ regime, the scattering 
rate integrated over
$\omega$ turns out to take a very simple form in terms of this
variable, which is the real motivation for its use.
In addition, the physical interpretation of $\qpt$ in this regime is
 the transverse momentum transferred to the $\p$ particle;
specifically, in terms of the $P$-defined light-cone coordinates, we have for 
soft $Q$
\begin{eqnarray}
\nn	\qpt^2&=&\qp^2\left(1+\frac{\qll}{p}+\order{\frac{g^2T^2}{p^2}}\right),\qquad
\omega=\qll-\frac{\qp^2}{4p}+\order{\frac{g^2T^2}{p^2}},\\
q^z&=&\qll+\frac{\qp^2}{4p}+\order{\frac{g^2T^2}{p^2}}=\omega+\frac{\qpt^2}{2p}+\order{\frac{g^2T^2}{p^2}}.
\label{qptsoft}
\end{eqnarray}

Finally, let us analyze the power counting. Above the cutoff, when angles are large, the contribution
to the collision operator is of order $g^4T$, up to powers of $T/E$.
When $q,\omega\sim gT$, the $\phi$-averaged matrix
element is proportional to $p^2k^2/q^4\sim 1/g^4$, up to corrections, which combined with
$dqd\omega\sim g^2$ and with another $g^2$ coming from the expansion of the curly brackets for small
$Q$ make the
singly-underlined exchanges contribute to LO in the soft region, with a $\ln(g)$ enhancement.
The same happens (without cancellations)
for the doubly-underlined matrix elements. Non-underlined matrix elements with $st$ or $tu$ at the
denominator are suppressed by a further power of $g^2$ in the soft region. Since the integration
is finite, $\mu_{\qpt}$ can be pushed
to zero there for simplicity. Singly  underlined matrix elements with a $u^2$
at the denominator present the same divergences;
they can be dealt with by swapping the $k'$ and $p'$ labels and using the same parameterization.
Matrix elements with $s^2$ at the denominator are not sensitive to
the soft region; hence, at leading order, they can be integrated
without cutoffs as well.

Fermion exchanges, and in particular the log-divergent doubly-underlined
$t$- or $u$-channel exchanges, can be treated with the same techniques and $\mu_{\qpt}$
cutoffs. For illustration, the $t$-channel quark exchange contribution to $q_1\bar q_1\lra gg$ scattering
is
\beqa
\nn C_{q_1}^\mathrm{large}[\PP]	\supset&\frac{g^4}{(2\pi)^3}
\frac{\cf^2}{8p^2}\int_{-\infty}^{p} d\omega\int_{\mu_{\qpt}}^{\sqrt{4p(p-\omega)}} d\qpt \frac{\qpt}{q}
\int_{(q-\omega)/2}^\infty dk \int_0^{2\pi}\frac{d\phi}{2\pi}\frac{u}{t}\\
&\times\Bigl\{
\PP^{q_1}(p) \, \nfd(k) \, [1+\nbe(p-\omega)+\nbe(k+\omega)]\nn\\
&\hspace{5mm}-\big[
\PP^{g}(p-\omega) \, \nbe(k+\omega)
+
\PP^{g}(k+\omega) \, \nbe(p-\omega)\big] \, [1-\nfd(k)]
\Bigr\}.
\label{el2quarkreg}
\end{empheq}
The cancellations of the leading IR behavior in the gluon exchanges,
as well as the matching to the diffusion and conversion processes will be dealt with in the next
sections and in App.~\ref{app_lo_matching}.

\subsection{Diffusion processes}
\label{sub_diff}
In this section we will describe the diffusion collision kernel, $C^\mathrm{diff}$, in greater detail. The cumulative effect 
of a large number of small momentum-transfer collisions that preserve the identity
of the hard particles
can be summarized by a Fokker-Plank equation~\cite{Svetitsky:1987gq,Moore:2004tg}
\beqe
\label{diff}
C_a^\mathrm{diff}[\PP]\equiv-\frac{\partial}{\partial p^i}\bigg[\etad(p)p^i \PP^a(\p)\bigg]
-\frac12\frac{\partial^2}{\partial p^i\partial p^j}
\left[\left(\hat{p}^i\hat{p}^j\ql(p)+\frac12(\delta^{ij}-\hat{p}^i\hat{p}^j)\qhat(p)
\right)\PP^a(\p)\right].
\end{empheq}
App.~\ref{app_lo_matching} directly shows how the diffusion operator arises at leading order from the screened $2\leftrightarrow2$ collisions kernel, \Eq{eq:collision22}.
There are three coefficients that enter in this effective description: $\qhat$ is the
standard \emph{transverse momentum broadening}, $\ql$ is the \emph{longitudinal momentum broadening }
and $\etad$ is the \emph{drag coefficient}. They are defined as\footnote{These
coefficients depend on the species $a$. However, as we shall show,
to leading and next-to-leading orders in $g$ this dependency reduces to a simple Casimir
scaling in the representation of the source $a$, so we drop this label in the text for
simplicity.}
\begin{equation}
\etad(p)=-\frac{1}{\pll}\frac{d\pll}{dt},\qquad	\qhat(p)\equiv\frac{d}{dt}\left\langle (\Delta p_\perp)^2\right\rangle,\qquad
\ql(p)\equiv\frac{d}{dt}\left\langle (\Delta \pll)^2\right\rangle,
\end{equation}
where $\pll$ and $\p_\perp$ are the longitudinal and transverse components relative to the large momentum
$\p$.

These coefficients can be determined  through the interaction rates
\cite{Svetitsky:1987gq,Braaten:1991jj,Moore:2004tg}, \ie
\begin{eqnarray}
\label{dragrate}
\frac{d\pll}{dt}&=&-\int dq^z\, q^z\, \frac{d\Gamma(\p,\p-\q)}{dq^z},\\
\label{longrate}
\ql(p)&=&\int dq^z \,(q^z)^2\, \frac{d\Gamma(\p,\p+\q)}{dq^z},\\
\label{transrate}
\qhat(p)&=&\int d^2q_\perp\, \qp^2\, \frac{d\Gamma(\p,\p+\q)}{d^2\qp},
\end{eqnarray}
where $\Gamma(\p,\p\pm\q)$ is the transition rate from  initial hard momentum $\p$ to final
hard momentum\footnote{Since the exchanged momentum is soft by construction, there is no ambiguity
in the identification of the hard outgoing line} $\p\pm\q$, with $\q$ soft.
A regulator which cuts off
the  $Q$ integrations is implicit, and the  values of these coefficients will in general depend
on the chosen scheme. 
Rather than determining $\Gamma(\p,\p\pm \q)$ 
and evaluating
the integrals in these equations directly, it is convenient (especially at NLO)
to use field-theoretical
definitions for the coefficients in \Eq{diff}

The transverse scattering rate $d\Gamma/d^2\qp$ at
large momentum is traditionally parameterized by $\cc(\qp)$
\begin{equation}
\label{defcq}
\lim_{p\to\infty}\frac{d\Gamma(\p,\p+\q_\perp)}{d^2\qp}= \frac{\cc(\qp)}{(2\pi)^2}.
\end{equation}
In $p\to\infty$ limit the hard particle's behavior eikonalizes, 
and $\cc(\qp)$ can be defined in terms of a specific Wilson
loop \cite{CaronHuot:2008ni,Benzke:2012sz} in the $(x^+,x_\perp)$ plane
(for propagation in the  positive $z$ direction).
Using this  Wilson loop definition, $\cc(\qp)$ and $\qhat$
have been evaluated at leading
\cite{Aurenche:2002pd}
and next-to-leading orders \cite{CaronHuot:2008ni}.
In particular, at leading order the
result is
\begin{equation}
\label{C_LO}
\cc_R(\qp) =g^2\crr\int\frac{dq^0dq_z}{(2\pi)^2}2\pi\delta(q^0-q_z) G_{rr}^{--}(Q)=
g^2\crr T \frac{\md^2}{\qp^2(\qp^2{+}\md^2)} \,,
\end{equation}
where $R$ labels the representation of the source and
\fbox{$\md^2=g^2T^2(\nc/3+\nf/6)$}
is the leading
order Debye mass.  $\qhat$ then reads at LO
\beqe
\label{qhat_LO}
\qhat=\int\frac{d^2\qp}{(2\pi)^2} \qp^2 \cc_R(\qp)
=g^2\crr T\int\frac{d^2\qp}{(2\pi)^2}
\frac{\md^2}{(\qp^2{+}\md^2)}=\frac{g^2\crr T\md^2}{2\pi}\ln\frac{\mu_{\qpt}}{\md} \,,
\end{empheq}
where, since we are in the $p\to \infty$ limit, $\qpt=\qp$ and  we
have used $\mu_{\qpt}$ as UV regulator.

What makes the Wilson loop
definition particularly attractive is that it can
be evaluated~\cite{CaronHuot:2008ni} using the (much simpler)
Euclidean, dimensionally-reduced Electrostatic QCD (EQCD)
\cite{Braaten:1994na,Braaten:1995cm,Braaten:1995jr,Kajantie:1995dw,Kajantie:1997tt}. This made the NLO computation possible~\cite{CaronHuot:2008ni}, and
opened the door to recent
non-perturbative lattice measurements~\cite{Laine:2013lia,Panero:2013pla}. These formal definitions,
as well as those for related light-front operators,
are summarized in App.~B of \cite{Ghiglieri:2013gia} and 
reviewed in \cite{Ghiglieri:2015zma}.

These techniques and results,
like most eikonal expansions, are based on a large momentum
expansion, $p \gg T$ or $gT$.  In App.~\ref{app_lo_matching} we
study the finite-$p$ corrections, showing that $T/p$ suppressed
corrections are really corrections in $gT/p$; and the first correction
involves vanishing odd integrands, so the first nonzero corrections from this
expansion are $\OO(g^2)$ even for $p\sim T$, and are
therefore irrelevant at the level of precision we are seeking here.
Therefore we can use the leading (and later, subleading) order
calculations in the strict Wilson-line limit which we have just
discussed.

To fully specify the diffusion operator $C^{\mathrm{diff}}$ in \Eq{diff}  we also need to
evaluate the longitudinal diffusion and drag coefficients, $\ql$ and $\etad$. To this end, we will
first compute the diffusion coefficient $\ql$ and then use
fluctuation-dissipation relations to determine the drag (see below).
At the practical level, we  introduce a Wilson-line based
definition for $\ql$ in the $p\to\infty$ limit, or equivalently at leading
order
in $T/p$.\footnote{%
    In the following, $\qhat$ and $\ql$ are understood to be in the infinite-momentum
    limit unless otherwise specified.}
In App.~\ref{app_wline} we will give a more formal
justification for our definition, whereas in
App.~\ref{app_lo_matching} we show that, as in the previous paragraph, finite-momentum corrections
start at $\OO(g^2)$, and are thus irrelevant to  current accuracy.

Intuitively, longitudinal momentum diffusion occurs because the longitudinal force
along the particle's trajectory has a nonzero correlator.
Experience with $\qhat$ and heavy quark diffusion~\cite{CasalderreySolana:2006rq,Gubser:2006nz} suggests that
$\ql$ should be given by a lightlike longitudinal force-force
correlator.  The force is determined by the electric field in the
direction of propagation, which 
motivates the following operator definition for $\ql$ in the large
momentum limit:
\begin{eqnarray}
\nn\ql&=&\frac{g^2}{d_R}\int_{-\infty}^{+\infty}dx^+\, \mathrm{Tr}\left\langle
U_R(-\infty,0,0_\perp;x^+,0,0_\perp)\bar v_\mu v_\nu F^{\mu\nu}(x^+,0,0_\perp)
\right.\\
\label{defqlong}
&&\hspace{3cm}\left.U_R(x^+,0,0_\perp;0,0,0_\perp)F^{\rho\sigma}(0)U_R(0;-\infty,0,0_\perp)
\bar v_\rho v_\sigma\right\rangle.
\end{eqnarray}
Here $v\equiv(1,0,0,1)$ and $\bar v\equiv(1/2,0,0,-1/2)$ are null vectors that are chosen to maintain our  light-cone conventions (\ie\ $\pl\equiv -\bar v\cdot P$,
$\pmm\equiv-v\cdot P$), and
$F^{\mu\nu} \bar v_\mu v_\nu=F^{+-}=E^z$ is the electric field along the
propagation direction.  We are
using a matrix notation, so that $F^{\mu\nu}=F^{\mu\nu\,a}T^a_R$, and
$U_R$ is a straight Wilson line in the
representation $R$ of the source
\begin{equation}
\label{defwilson}
U_R(Y;X)=P\exp\left(-ig \int_0^1 ds\, (Y-X)\cdot A_R(s(Y-X)+X) \right).
\end{equation}
Gauge fields and matrices
in the Wilson lines are both to be understood as path ordered.
\Eq{defqlong}  comes from the eikonal approximation, \ie\
the replacement of the highly energetic particle with momentum $\p$ with a Wilson line in the appropriate
representation along its classical trajectory.
We also note that this definition of $\ql$
has the correct
``amplitude times conjugate amplitude'' structure required to enter in a rate. 

\begin{figure}[ht]
\begin{center}
\includegraphics[width=5cm]{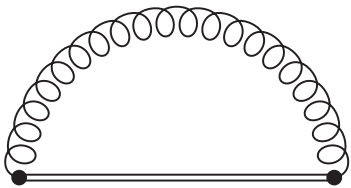}
\end{center}
\caption{The leading-order soft contribution to $\ql$. The Wilson lines before and after the two
black dots, which represent the $F^{+-}$ vertices, cancel at leading order, whereas the one between
the two dots always turns into an adjoint line, which we have represented as a double line.
The curly line is a soft HTL gluon.}
\label{fig_lo_soft}
\end{figure}

\begin{figure}
\begin{center}
\includegraphics[width=9cm]{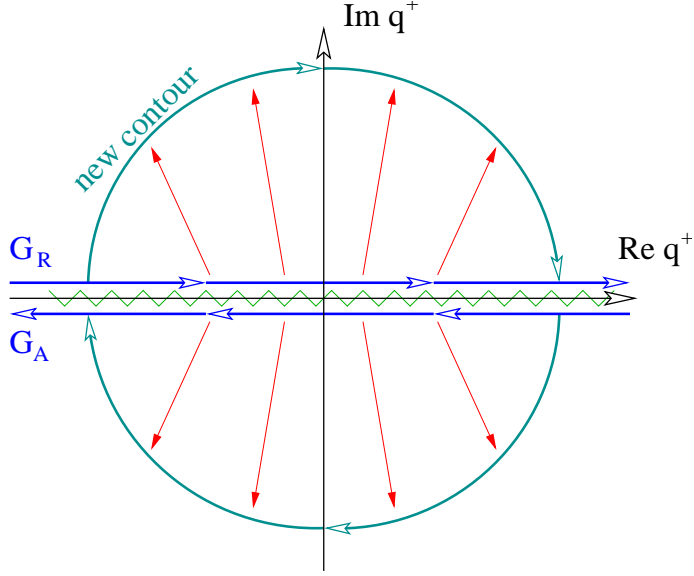}
\end{center}
\caption{\label{fig_contour}  Integration contour in the complex $\qll$
integration, and the deformation we use to render $\qll \gg gT$. $G_R$
runs above the real axis and $G_A$ below. }
\end{figure}

We now evaluate \Eq{defqlong} at LO: we simply contract the two $F$ fields, obtaining a
forward Wightman correlator, \ie\ the diagram shown in Fig.~\ref{fig_lo_soft},
which reads
\begin{equation}
\label{lo}	\ql
=g^2\crr \int_{-\infty}^{+\infty}dx^+
\int\frac{d^4Q}{(2\pi)^4}e^{-iq^-x^+}(\qll)^2G^{--\,>}(Q),
\end{equation}
where $G^>(Q)$ is the HTL-resummed forward propagator and the integral is understood to run
over  soft momenta only.
The $x^+$ integration sets $\qm$ to zero and, as we show in App.~\ref{app_lo_matching},
brings this expression into agreement with the one obtained from the rate-based definition
in \Eq{longrate}. 
Note that only the
even-in-$\qll$ part of $G^>(\qll,\qm=0,\qp)$ contributes to the integral.  
Then, using the fluctuation-dissipation theorem,
$G^>(Q) = (1 + n_B(q^0)) \rho(Q)$ with $\rho(Q)=G_R(Q) - G_A(Q)$, we expand 
for small $q^0=q^+ \sim gT$ to find
\begin{equation}
\label{lo2}	\ql
=g^2\crr
\int\frac{d\qll d^2\qp}{(2\pi)^3}T\qll (G^{--}_R(\qll,\qp)-G^{--}_A(\qll,\qp)),
\end{equation}
up to an $\OO(g^2)$ correction.
Numerical integration is straightforward, using the HTL
propagators given in App.~\ref{app_props}. Beyond leading order, however,
one would be plagued with intricate multi-dimensional numerical integrals.
However, as we anticipated in the introduction, we can perform
the $\qll$ integration (and similar ones elsewhere)
by resorting to the analyticity sum rule
techniques developed in \cite{CaronHuot:2008ni,Ghiglieri:2013gia}.\footnote{
	The following derivation has been anticipated in \cite{Ghiglieri:2015zma}.} Since retarded (advanced)
two-point functions are analytic in the upper (lower) half-plane in any time-like or light-like
variable, we can deform the integration contours away from the real axis onto $\calr$
($\vert\qll\vert
\gg gT$, $\mathrm{Im}\,\qll>0$)
and $\cala$ ($\vert\qll\vert
\gg gT$, $\mathrm{Im}\,\qll<0$),
as depicted in Fig.~\ref{fig_contour}.
Along the arcs the longitudinal and transverse propagators simplify greatly, \ie
\begin{equation}
\label{arcexpand}
G^{--}_R(P)\to
\frac{i}{(\qll)^2}\left(1+\frac{q^-}{\qll}\right)\frac{2\qll q^--\mmg}
{2\qll q^--\qp^2-\mmg}\Big\vert_{\calr},	
\end{equation}
where
\fbox{$\mmg\equiv\md^2/2$}
is the gluon asymptotic thermal mass. The end result is then
\beqe
\label{lofinal}	\ql
=g^2\crr T
\int\frac{d^2\qp}{(2\pi)^2}\frac{\mmg}{\qp^2+\mmg}=
\frac{g^2\crr T }{2\pi}\mmg\ln\frac{\mu_{\qpt}}{M_\infty},
\end{empheq}
where contributions smaller than $1/\qll$ in \Eq{arcexpand} are not needed,
as they would only give rise to power-law terms in the cutoff on $\qll$ which would then cancel
against contributions from larger scales. As in the $\qhat$ case, we have used $\mu_{\qpt}$ as a transverse
regulator,
since the $(\qll,\qp)$ and $(\omega,\qpt)$ coordinates differ by
$\OO(g^2)$ (all $\OO(g)$ corrections vanish under integration).

The sum rule we have just obtained is the bosonic equivalent of the one presented
in \cite{Ghiglieri:2013gia}. Let us remark that the longitudinal and transverse
contributions to $G_R^{--}(Q)$ contain poles at $\qll=\qm/2\pm i\qp$ ($q^2=0$), which, being on
both sides of the complex plane, appear to violate analyticity. However their residue cancels
in the sum of longitudinal and transverse components. As observed in \cite{CaronHuot:2008ni},
they are artifacts of the decomposition into Lorentz-variant longitudinal and transverse modes
and their contribution has to vanish in all gauge-invariant quantities.

We also remark that the same result~\eqref{lofinal} has been obtained in a different
way in \cite{Peigne:2007sd} for energy loss, which is related by an Einstein relation.
As shown there, once the difference in regularization
between  $\qp<\mu_{\qpt}$ and $q<\mu_q$ is taken into account, \Eq{lofinal} agrees with the
numerical results of Braaten and Thoma \cite{Braaten:1991jj} for $v\to1$.

Having determined $\ql$, the drag coefficient $\etad(p)$ 
is constrained by the requirements that the Fokker-Planck description be
equivalent to the Boltzmann one and that interactions with the medium tend
to drive the hard excitations towards equilibrium~\cite{Arnold:1999uza,Arnold:1999va,Moore:2004tg}. Since
we have taken a classical particle approximation for the hard
particles, the equilibrium form is $\PP(\p)\propto
\exp(-p/T)$.  
The drag (in a given regularization scheme) is determined from
$\ql(p)$ and $\qhat(p)$ 
by adjusting the value of $\etad(p)$ so that \Eq{diff} approaches  equilibrium, \ie\ its
right-hand side vanishes for $\PP(\p)\propto \exp(-p/T)$. 
Since $\qhat$ and $\ql$ are $p$-independent up to $\OO(g^2)$, the  equilibration condition yields to following relation:
\beqe
\label{equilibrate}
\etad(p)=\frac{\ql}{2Tp}+\frac{1}{2p^2}(\qhat-2\ql).
\end{empheq}
The consistency of this condition is verified by direct computation of $\etad(p)$
and $\ql$ at leading order in App.~\ref{app_lo_matching}.
Inserting this relation between the coefficients into the diffusion
equation, \Eq{diff}, we find
\begin{equation}
C_a^\mathrm{diff}[\PP] = -\left[\frac{\PP(\p)}{Tp}
  + \frac{2T+p}{2pT} \frac{d\PP(\p)}{dp^z}+\half\frac{d^2\PP(\p)}{d(p^z)^2}\right]
\ql
-\left[-\frac{1}{2p}\frac{d\PP(\p)}{dp^z}
+\frac{1}{4}\nabla^2_{\pp}\PP(\p)\right] \qhat,
\label{diffexplo}
\end{equation}
which is our final form for the diffusion operator.

We end by making a few remarks about the equilibrium condition and $\etad$.
The first term in \Eq{equilibrate} comes from the simple Einstein relation that arises in the infinite
momentum limit, \ie\ $\ql=-2T d\pll/dt+\OO(1/p)$. The relative $\OO(1/p)$ terms
can then be obtained by imposing equilibration on \Eq{diff}. In App.~\ref{app_lo_matching}
we will show how, at leading order, those $1/p$ terms can be determined explicitly, how the diffusion
picture matches exactly with $C^\mathrm{large}[\PP]$ at large $Q$ and how different cutoff schemes
can be implemented.  It is also worth
stressing that the $1/p$ terms to \Eq{equilibrate} and equivalently to $\etad$ do not come
from a $T/p$ expansion, but only from the $g\ll1$ expansion. Up to relative $\OO(g^2)$, there are no
$1/p^2$ terms. Finally we remark that, in the simpler case where $\PP$ is a function
of $p$ rather than $\p$, as in footnote~\ref{foot_vector}, the contribution proportional
to $\qhat$ vanishes in Eq.~\eqref{diffexplo}.

\subsection{Conversion processes}
\label{sec_lo_conv}

The conversion-process part of the collision operator can be simplified as
\beqa
\label{defconvq}
C_{q_i}^\mathrm{conv}[\PP]=&\PP^{q_i}(\p)\Gammac_{q\to g}(p)
-\PP^g(\p)\frac{d_A}{d_F}\Gammac_{g\to q}(p),\\
\label{defconvqbar}
C_{\bar q_i}^\mathrm{conv}[\PP]=&\PP^{\bar q_i}(\p)\Gammac_{\bar q\to g}(p)
-\PP^g(\p)\frac{d_A}{d_F}\Gammac_{g\to \bar q}(p),\\
C_{g}^\mathrm{conv}[\PP]=&\sum_{i=1}^{\nf}\bigg\{\PP^{g}(\p)\bigg[
\Gammac_{g\to q_i}(p)+\Gammac_{g\to \bar q_i}(p)\bigg]\nn\\
\label{defconvgluon}
&\hspace{2cm}-\frac{d_F}{d_A}\bigg[\PP^{q_i}(\p)\Gammac_{q\to g}(p)+
\PP^{\bar q_i}(\p)\Gammac_{\bar q\to g}(p)\bigg]\bigg\},
\end{empheq}
representing a rate for each species to disappear due to conversion to
another type, and a rate for that species to appear due to the
conversion of another type to the type in question.
The \emph{conversion rates} $\Gammac$ describing these processes can
depend on momentum $p$, and they also implicitly depend on the
regularization scheme. They do not, however, depend  on the exchanged momentum
at leading or next-to-leading order. To see this, consider
\Eq{el2quarkreg} for
$\omega,\qpt\sim g$. One has that the statistical factors, once expanded
for $g\ll1$, yield
\begin{equation}
\label{statexpand}
\bigg\{\PP^{q_1}(p) \, \nfd(k)  [1+\nbe(k)]-
\PP^{g}(p) \, \nbe(k)  [1-\nfd(k)]\bigg\}\left(1+\order{\frac{\omega}{T},\frac{\omega}{p}}\right).
\end{equation}
Similarly, as we shall show in more detail in App.~\ref{app_lo_conv}, the HTL-resummed
and $\phi$-averaged matrix elements, once expanded for small $Q$, are to leading order
even in $\omega$, up to $\OO(\omega/T,\omega/p)$ corrections.\footnote{Non-underlined fermion exchange
matrix elements, such as $u/s$ in $q_1g\lra q_1g$ scattering, are  suppressed by two powers of $g$.}
Furthermore, the $(\omega,\qpt)$ and $(\qll,\qp)$ coordinates are equivalent up to
another odd-in-$\omega$ correction, as shown in \Eq{qptsoft}.
All these odd, subleading corrections vanish upon $d\omega$ integration,
so corrections first arise from $\omega^2/T^2$ type corrections or the
product of two $\omega/T$ corrections, which are both safely NNLO.  

At leading order the rates can simply be obtained from the aforementioned
even-in-$\omega$ term in the HTL-resummed, doubly underlined matrix elements,
whereas at next-to-leading order soft-gluon loop corrections need to be considered.
To this end, we  find it convenient to define the conversion rates in terms
of gauge-invariant Wilson line operators, following the
work on the soft contribution to the photon rate in \cite{Ghiglieri:2013gia}, where we showed
that the leading and next-to-leading order soft contributions were obtained from similar operators.
Physically, the amplitude for a quark to convert to a gluon involves
a quark propagating in from an early initial time, and being converted
to a gluon by the insertion of a quark destruction operator $\psi$.
The rate is the product of this amplitude with its conjugate, and we must
integrate over the time difference between the quark annihilation
event in the amplitude and in its conjugate.  Eikonalizing, the
propagation of the quark turns into a fundamental Wilson line, while
the gluon which propagates between the earlier and later
$\psi,\bar\psi$ insertion is represented by an adjoint line.  This
leads to the following Wilson-line representations for the conversion
processes:
\begin{eqnarray}
\Gammac_{q\to g}(p)&=&-\frac{ g^2}{8d_Fp}\int_{-\infty}^{+\infty}dx^+\left\langle\mathrm{Tr}
\big[
U_F(-\infty,0,0_\perp;x^+,0,0_\perp)T^a \bar\psi(x^+,0,0_\perp)\slashed{v} \right.\nn\\
\label{convratefermion}
&&\times\left.
U_A(x^+,0,0;0,0,0_\perp)\psi(0)T^b U_F(0;-\infty,0,0_\perp)\big]\right\rangle,\\
\Gammac_{g\to q}(p)&=&-\frac{ g^2}{8d_Ap}\int_{-\infty}^{+\infty}dx^+\left\langle\mathrm{Tr}
\big[
U_A(-\infty,0,0_\perp;x^+,0,0_\perp)T^a \bar\psi(x^+,0,0_\perp)\slashed{v} \right.\nn\\
\label{convrategluon}
&&\times\left.
U_F(x^+,0,0;0,0,0_\perp)\psi(0)T^b U_A(0;-\infty,0,0_\perp)\big]\right\rangle,\\
\Gammac_{\bar q\to g}(p)&=&\Gammac_{q\to  g}(p),\qquad \Gammac_{g\to \bar q}(p)=\Gammac_{g\to  q}(p).
\label{convratesqbar}
\end{eqnarray}
The traces appearing in the rates are over the Dirac and color indices.
At leading order the rates read
\beqa
\label{convratefermionlo}
\Gammac_{q\to g}(p)\Big\vert_\LO=&
-\frac{ g^2\cf}{8p}\int\frac{d^4Q}{(2\pi)^4}\Tr{\slashed{v} S^>(Q)}
2\pi\delta(\qm)\nn\\
=&\frac{ g^2\cf}{4p}\int\frac{d^2\qp}{(2\pi)^2}\frac{\mmf}{\qp^2+\mmf}=
\frac{ g^2\cf\mmf}{8\pi p}\ln\frac{\mu_{\qpt}}{\mmf},\\
\Gammac_{g\to q}(p)\Big\vert_\LO=&
\frac{d_F}{d_A}\Gammac_{q\to g}(p)\Big\vert_\LO,
\label{convrategluonqbarlo}
\end{empheq}
where we have used the light-cone sum rule obtained
in \cite{Besak:2012qm,Ghiglieri:2013gia}. \fbox{$\mmf\equiv g^2\cf T^2/4$} is
the asymptotic mass of quarks. The
$\mu_{\qpt}$ regulator is the same used in the large angle and diffusion regions. In
App.~\ref{app_lo_conv} we show how the evaluation of the appropriate part of the
HTL-resummed $2\lra2$ collision operator in this momentum region leads to the same result.

\section{Next-to-leading order corrections:  Overview}
\label{sec_nlo_intro}
The reorganization we have presented in the previous section allows us to introduce
$\OO(g)$ corrections to the collision operator. For convenience we identify two different
sources, \ie\ \emph{loop corrections} and \emph{mistreated regions}. The former arise by adding
a soft gluon loop to a diagram, which, in the finite-temperature power counting, gives rise to
an $\OO(g)$ contribution.
The latter instead originate from integrating over $\OO(g)$ regions
of the leading-order phase space where one particle becomes soft, without being treated correctly as
an HTL quasiparticle. One such example is mentioned at the end of Sec.~\ref{sec_reorganize} and in
Fig.~\ref{fig_collsoft}, where a soft, final-state gluon in a $\onetwo$ process gives rise to a finite
contribution to the LO $\onetwo$ collision
operator. 
As we shall show, this indeed represents
an $\OO(g)$ region of the $\onetwo$ phase space; its  evaluation, as well as the evaluation
of all such mistreated regions, requires the identification of the limiting behavior of the LO calculation
in that region. Such behavior will then have to be subtracted from the proper, HTL-resummed, evaluation
of that region, which, in the example of Fig.~\ref{fig_collsoft}, will be done when dealing with $\ql$ at NLO.

In the large-angle region, loop corrections are suppressed by a factor of $g^2$, as long as
the momentum transfer stays large. But the LO evaluation, in the form of Eqs.~\eqref{el2reg} and
\eqref{el2quarkreg}, mistreats the region where an incoming gluon is soft. This region
will be properly addressed in the \emph{semi-collinear region}, which we shall introduce later on.
We defer other considerations on the necessary subtraction to that point and to
Sec.~\ref{sec_semi}.%
\footnote{\label{foot_qhat_subtr}
  At the NLO level there is also a linear in $\mupp$ divergence in
  evaluating $\qhat$, which is canceled by a linear in $\mupp$
  soft-gluon effect in the hard scattering regime,
  see \cite{CaronHuot:2008ni,Arnold:2008vd}.  This divergence and
  mistreatment simply cancel; so will not discuss it further,
  directing the interesting reader to those papers for details.}

In the collinear region, we will encounter both loop corrections and subtraction
regions. The former arise from adding extra soft gluons to the scatterings that broaden the
hard particles, inducing their splitting. They correspond to the NLO corrections to $\cc(\qp)$
\cite{CaronHuot:2008ni},
which have been already mentioned after \Eq{defcq}. The asymptotic masses of the hard
particles also receive $\OO(g)$ corrections that contribute at NLO.
In Sec.~\ref{sec_nlo_coll}
we will discuss in detail those corrections, as well as three mistreated regions: the aforementioned
overlap with the diffusion region, an altogether equivalent one with the conversion sector and finally
one with the semi-collinear region.

In the diffusion sector, \Eq{diff} remains valid to NLO. Its coefficients $\etad$, $\ql$ and
$\qhat$ all receive $\OO(g)$ loop corrections. Those to $\qhat$ are known \cite{CaronHuot:2008ni}.
In Sec.~\ref{sec_long_diff} we will set up the calculation of the $\OO(g)$ corrections to $\ql$,
through the field-theoretical definition~\eqref{defqlong} and the causality-based sum rules.
The details of the  evaluation will be presented in App.~\ref{app_nlo}. It requires the subtraction of a
mistreated $\OO(g)$ region in its LO evaluation, as well as  of the  aforementioned diffusion
limit of the collinear sector. Finally, $\etad$ can be determined through
the equilibration condition~\eqref{equilibrate}.

In the conversion sector, the operators defined in Eqs.~\eqref{convratefermion}-\eqref{convratesqbar}
receive $\OO(g)$ loop corrections from the addition of one extra soft gluon. In Sec.~\ref{sec_nlo_conv}
we will show how these operators are equivalent up to NLO to their abelian counterparts. Hence, the
$\OO(g)$ corrections can be extracted from the soft-sector
contribution to the NLO photon rate in \cite{Ghiglieri:2013gia}. In this case too there are subtractions
from mistreated regions in the LO conversion and collinear rates.

Finally, a new kinematical region enters at NLO, the aforementioned \emph{semi-collinear} region. It
corresponds to medium-induced splittings with  larger virtuality, transverse momenta
and opening angle,
respectively of order $gT^2$, $gT^2$ and $\sqrt{g}$.%
\footnote{%
  Up to respective factors
  of $T/E$ and $\sqrt{T/E}$ in a democratic splitting case, similarly
  to footnote~\ref{foot_angle}.}
Other differences with respect to the
collinear region are that the kinematics now allow  the soft gluons
to be either space-like or time-like (hence the overlap with the soft limit of the large-angle region),
that LPM interference is suppressed and that the soft gluons can change the small minus component
of the hard/thermal particles' momentum. We will deal with this sector in detail in Sec.~\ref{sec_semi}. As
mentioned, we will have to subtract the mistreated overlap regions of the large-angle and collinear
regions.

We conclude this overview by sketching the form of the NLO corrections:
\beqa
\label{cnloall}
\delta C_a[\PP]=&\delta C^\mathrm{coll}_a[\PP]
+\delta C^\mathrm{diff}_a[\PP]
+ \delta C^\mathrm{conv}_a[\PP]
+ \delta C^\mathrm{semi-coll}_a[\PP]
\end{empheq}
Here and in what follows $\delta$ refers to an NLO contribution.
The first term consists of the loop correction to the collinear
sector. In the second term
the form of the diffusion equation~\eqref{diffexplo} remains unchanged, but the 
parameters, $\hat q$ and $\hat q_{L}$, receive NLO corrections from soft loops.
In particular, the corrections to the longitudinal diffusion coefficient,
$\delta \ql(\mupp^\NLO)$, depends logarithmically 
on an ultraviolet cutoff, $\mupp^\NLO$.
Similarly, the momentum dependence of the conversion rates remains unchanged, $\propto 1/p$, but 
the overall magnitude of the rate depends logarithmically on $\mupp^\NLO$. 
This dependence on the ultraviolet cutoff in the diffusion and conversions collision kernel
cancels in the complete kernel when the semi-collinear emission rates are
included. In the semi-collinear case,  $\mupp^\NLO$ serves as an infrared
cutoff limiting the semi-collinear emission of soft quarks and gluons.

To compute each of the collision operators in $\delta C$, the  phase space
regions which were mistreated at LO must be subtracted as counterterms. This
replaces the mistreated LO terms with the full NLO result, and generally
removes power divergences in soft loop integrals:
\begin{align}
\label{diffcorr}
\delta C^\mathrm{diff}_{a}[\PP] =& \Delta C^\mathrm{diff}_{a}[\PP]-
\delta C_{a\;\mathrm{coll\,subtr.}}^\mathrm{diff}[\PP]
-\delta C_{a\;\mathrm{diff\,subtr.}}^\mathrm{diff}[\PP]
,\\
\label{convcorr}
\delta C^\mathrm{conv}_a[\PP] = & \Delta C^\mathrm{conv}_{a}[\PP]-
\delta C_{a\;\mathrm{coll\,subtr.}}^\mathrm{conv}[\PP]
-\delta C_{a\;\mathrm{conv\,subtr.}}^\mathrm{conv}[\PP],\\
\delta C^\mathrm{semi-coll}_a[\PP] =&\Delta C^\mathrm{semi-coll}_{a}[\PP]-
\delta C_{a\;\mathrm{coll\,subtr.}}^\mathrm{semi-coll}[\PP]-
\delta C_{a\;\mathrm{large\,subtr.}}^\mathrm{semi-coll}[\PP].
\label{semicorr}
\end{align}
In each case, the subtraction terms arise from a mistreatment 
in a specific region of phase space
from one of the four LO collision kernels in \Eq{boltzmannproc}.
For example, in the first line 
$\Delta C^\mathrm{diff}$ treats
the diffusion process with NLO accuracy by including the appropriate soft
loops, while the two counterterms arise because the LO collinear 
and LO diffusion collision kernels give incomplete contributions to 
the diffusion process at NLO. 

We will devote the next four sections to evaluating in turn the four
contributions to the NLO collision operator given in \Eq{cnloall}.

\section{The collinear region}
\label{sec_nlo_coll}

Here we discuss the NLO corrections, and subtractions, needed to
establish splitting processes to this order.  But for completeness and
context, and to set notation, we begin by presenting the leading-order
result.

\subsection{Leading-order recapitulation}

At LO $C_a^\mathrm{coll}[\PP]=C_a^{\onetwo}[\PP]$,
which is \cite{Jeon:2003gi}
\beqa
\nn-C_{q,\bar q}^\coll[\PP]=&\int_{-\infty}^{+\infty}d\omega \PP_{q,\bar q}((p+\omega)\hat\p)
\frac{d\Gamma^{q}_{qg}(p+\omega,\omega)}{d\omega}\Big\vert_\mathrm{coll}-\PP_{q,\bar q}(\p)
\frac{d\Gamma^{q}_{qg}(p,\omega)}{d\omega}\Big\vert_\mathrm{coll}\\
\label{onetwocollisionq}
&+\PP_{g}((p+\omega)\hat\p)
\frac{\da}{\df}\frac{d\Gamma^{g}_{q\bar q}(p+\omega,\omega)}{d\omega}\Big\vert_\mathrm{coll},\\
\nn-C_{g}^\coll[\PP]=&\int_{-\infty}^{+\infty}d\omega 
\PP_{g}((p+\omega)\hat\p)\frac{d\Gamma^{g}_{gg}(p+\omega,\omega)}{d\omega}\Big\vert_\mathrm{coll}\\
\nn&
+\left[\sum_{i=1}^{\nf}\bigg(
\PP_{q_i}((p+\omega)\hat\p)+\PP_{\bar q_i}((p+\omega)\hat\p)\bigg)\right]\frac{\df}{\da}
\frac{d\Gamma^{q}_{qg}(p+\omega,\omega)}{d\omega}\Big\vert_\mathrm{coll}\\
&-\PP_{g}(\p)\left(\nf
\frac{d\Gamma^{g}_{q\bar q}(p,\omega)}{d\omega}\Big\vert_\mathrm{coll}+\theta(p-2\omega)
\frac{d\Gamma^{g}_{gg}(p,\omega)}{d\omega}\Big\vert_\mathrm{coll}\right),
\label{onetwocollision}
\end{empheq}
where the $\theta$-function multiplying the last term prevents a double counting
of the $gg$ final states (equivalently one may use a $\half$ symmetry factor).
$\Gamma^a_{bc}(p,\omega)=\Gamma(\p,\hat{p}\omega)$ is the rate for a particle $a$ with hard
momentum $\p$ to emit ($\omega>0$) or absorb ($\omega<0$) a gluon (quark in the case
$\Gamma^g_{q\bar q}$) with energy (longitudinal momentum) $\omega$.\footnote{In keeping with
the notation in the other sections, we label $\omega$ the longitudinal component of one
of the outgoing momenta.}
$\PP_{q,\bar q}$ is either $\PP_{q_i}$ or $\PP_{\bar q_i}$:
\Eq{onetwocollisionq} applies both for quarks
and antiquarks, provided a consistent labeling of $\omega$ in $\Gamma^g_{q\bar q}$ is chosen.
At leading order these rates read \cite{Arnold:2002ja,Jeon:2003gi}%
\footnote{%
  The distribution functions in \cite{Jeon:2003gi} are summed over spin, color and flavor,
  so that the factors of $\da/\df$ and $\df/\da$ vanish in Eqs.~\eqref{onetwocollisionq} and
  \eqref{onetwocollision}. However, the $g\to q\bar q$ rate in \cite{Jeon:2003gi}
  and subsequent references (see \cite{Schenke:2009gb,Schenke:2009ik}) was missing the factor of $d_F/d_A$
  that appears in \Eq{jmcoll}. Indeed, the group-theoretical
  factor for this process should read $\cf \df/\da=T_F=1/2$.}
\beqa
\nn\frac{d\Gamma(p,\omega)}{d\omega}\Big\vert_\coll=&\frac{g^2\crr}{16\pi p^7}(1\pm n(\omega))
(1\pm n(p-\omega))\left\{\begin{array}{cc}
\frac{1+(1-x)^2}{x^3(1-x)^2}& q\to qg\\
\frac{d_F}{d_A}\frac{x^2+(1-x)^2}{x^2(1-x)^2}& g\to q\bar q\\
\frac{1+x^4+(1-x)^4}{x^3(1-x)^3}& g\to gg	
\end{array}\right\}\\
&\times\int\frac{d^2h}{(2\pi)^2}2\bh\cdot\mathrm{Re}\,\bff(\bh,p,\omega),
\label{jmcoll}
\end{empheq}
where  $x\equiv \omega/p$
is the momentum fraction of the outgoing gluon or, in the $q\bar q$ final state,
of one of the two fermions. $\bh\equiv\p\times\q$ is the two-dimensional invariant
describing the transverse separation of the final states.  Note that for QCD, in the cases $q\to qg$
and $g\to q\bar q$ $\crr=\cf$, whereas for $g\to gg$ $\crr =\ca$.
$\bff(\bh,p,\omega)$ determines the transverse evolution of the system;
it is to be determined through an equation which resums multiple soft
interactions. In momentum space it has the form of an integral equation,
whereas in position space it is a differential one. The latter will be described in App.~\ref{sub_nlo_coll}; the former reads \cite{Arnold:2002ja}
\begin{eqnarray}
\nn2\bh&=&i\delta E(\bh,p,\omega)\bff(\bh)+\int\frac{d^2\kp}{(2\pi)^2}
\frac{\mathcal{C}_F(\kp)}{C_F}
\bigg\{(\crr-\ca/2)[\bff(\bh)-\bff(\bh-\omega\bkp)]\\
&&+\frac{\ca}{2}[
\bff(\bh)-\bff(\bh+p\bkp)]+\frac{\ca}{2}[
\bff(\bh)-\bff(\bh-(p-\omega)\bkp)]\bigg\}.
\label{defimplfull}
\end{eqnarray}
For the case of
$g\to q\bar q$, $(\crr-\ca/2)$ multiplies the term with $\bff(\bh-p\bkp)$ rather than
$\bff(\bh-\omega\bkp)$. The equation depends on two inputs,
$\mathcal{C}_F(\kp)$  and $\delta E(\bh,p,\omega)$. The former is $\mathcal{C}(\kp)$ for a fundamental
source, whereas $\delta E$ is the energy difference between the initial and final collinear particles. It reads
\begin{equation}
\label{defdeltaE}
\delta E(\bh,p,\omega)=\frac{h^2}{2p\omega(p-\omega)}+\frac{m^2_{\infty\,\omega}}{2\omega}
+\frac{m^2_{\infty\,p-\omega}}{2(p-\omega)}
-\frac{m^2_{\infty\,p}}{2p},
\end{equation}
where $m_{\infty\,p}^2$ is the asymptotic mass of the particle with momentum $p$, as summarized
in \Eq{loasym}.

\subsection{The collinear sector at next-to-leading order}
\label{sub_intro_nlo_coll}
Subleading corrections to collinear splitting are treated, for the
case of photon production, in \cite{Ghiglieri:2013gia}; the case here
is conceptually similar.  We must identify any NLO corrections to
splitting for generic kinematics; and we must identify any limits of
the kinematics which contribute an $\OO(g)$ faction of the total
splitting rate, but which overlap with the kinematics in another
region we are studying.

The NLO corrections for generic kinematics enter as
two corrections which arise when solving
\Eq{defimplfull}, specifically, $\OO(g)$ corrections to
$\cc(\qp)$ and to the asymptotic masses entering in $\delta E$.
The computation of these masses to NLO has been carried out in
\cite{CaronHuot:2008uw} using Euclidean techniques
and is reviewed in \cite{Ghiglieri:2013gia,Ghiglieri:2015zma}.
It is only due to soft gluons
and depends on the nature of the particle (quark or gluon) through a simple
Casimir scaling:
\beqe
\label{minftyNLO}
\delta \mmf=-g^2\cf \frac{T \md}{2\pi},
\qquad
\delta \mmg=-g^2\ca \frac{T \md}{2\pi}.
\end{empheq}
The NLO collision kernel $\delta \cc(\qp)$ has been computed in
\cite{CaronHuot:2008ni}, as a first application of the mapping
to the Euclidean theory.
All one needs to do to treat generic momenta at NLO is to include
these two corrections into \Eq{defimplfull}.
We review how to do so, using impact-parameter-space methods, in
App.~\ref{sub_nlo_coll}.

\subsection{Subtraction regions}
\label{sub_subtr_coll}

Besides these generic-momentum corrections, there are also corners of
the collinear-splitting kinematics where it starts to overlap with
other processes -- momentum diffusion, identity change, and $\twotwo$
scattering.  Each regime represents an $\OO(g)$ suppressed fraction of
the total contribution from splitting processes, so a correct
leading-order treatment is sufficient.  Unfortunately, in each regime
at least one approximation made in arriving at \Eq{jmcoll},
\Eq{defimplfull}, \Eq{C_LO}, or \Eq{defdeltaE} breaks down.  We handle
this in two steps.  First, we find out what contribution the (naive)
leading-order splitting calculation actually contributes in each
region.  Then, we perform a more complete NLO calculation of the
specific kinematic corner of interest, \textsl{subtracting} the
(naive) leading-order splitting contribution we have found, since it
is already incorporated via the LO splitting treatment.  The remainder
of this section carries out the calculation of the LO splitting
behavior in each kinematical corner.

In each relevant corner, $\delta E\gg \int d^2\kp \mathcal{C}(\kp)$, so that,
physically, the formation time $1/\delta E$ of the collinear particles  
becomes much shorter than the time between collisions, estimated by
$(\int d^2\kp \mathcal{C}(\kp))^{-1}$.  In this case emission
amplitudes associated with different scattering events become
incoherent, and it is sufficient to treat emission as a sum of the
rate arising from each scattering event (LPM suppression is small), up
to $\OO(g)$ corrections which we can neglect.
In the diffusion and conversion cases
this happens because the denominators in \Eq{defdeltaE} become
smaller by a factor of $g$, whereas in the semi-collinear case  $h^2$
becomes larger by $1/g$.  Therefore,
we first obtain the generic solution for $\delta E\gg \int d^2\kp \mathcal{C}(\kp) $: if we solve \Eq{defimplfull}
by substitution as in \cite{Arnold:2001ms,Ghiglieri:2013gia} we have at leading order\footnote{
The $g\lra q\bar q$ case, which has a different color structure in
curly braces, is not dealt with explicitly.}
\begin{equation}
\mathrm{Im}\,\bff(\bh,p,\omega)=\frac{-2\bh}{\delta E(\bh,p,\omega)},
\end{equation}
which in turn yields
\begin{eqnarray}
\nn \mathrm{Re}\,\bff(\bh,p,\omega)&=&\frac{2}
{\delta E(\bh,p,\omega)}\int\frac{d^2\kp}{(2\pi)^2}\frac{\mathcal{C}_F(\kp)}{\cf}
\bigg\{\left(\crr-\frac{\ca}{2}\right)\left[\frac{\bh}{\delta E(\bh)}
-\frac{\bh-\omega\,\bkp}{\delta E(\bh-\omega\,\bkp)}\right]\\
&&\hspace{-2cm}+\frac{\ca}{2}\left[\frac{\bh}{\delta E(\bh)}
-\frac{\bh+p\,\bkp}{\delta E(\bh+p\,\bkp)}\right]+\frac{\ca}{2}
\left[\frac{\bh}{\delta E(\bh)}
-\frac{\bh-(p-\omega)\bkp}{\delta E(\bh-(p-\omega)\bkp)}\right]\bigg\}.
\label{implsubst}
\end{eqnarray}

\subsubsection{The diffusion limit}
Let us now specialize to the soft gluon region, which corresponds to
the diffusion limit. Explicitly, one has $\omega\to gT$ in the $q\lra g q$
and $g\lra gg$ processes. In the case of the latter process, there is also a $(p-\omega)\sim gT$
region which appears in the loss term in \Eq{onetwocollision} but
is  absent from the gain term. Since its contribution is identical (the
loss term and \Eq{jmcoll} are symmetric around $\omega=p/2$) this compensates
for the relative factor of two between $g\lra gg$ gain and loss terms in
\Eq{onetwocollision}, yielding for \Eq{onetwocollision} a limit
of the form of
the $\ql$-proportional part of Eq.~\eqref{diffexplo}\footnote{\label{foot_dlog}
This is a consequence of the form of Eqs.~\eqref{eq:collision12} and
\eqref{onetwocollisionq}-\eqref{onetwocollision}. In their derivation (see Eq.~(2.6) in
\cite{Arnold:2002zm})
one integrates the effective $\onetwo$ matrix elements over the transverse momenta of the final
states, neglecting the small deviations from eikonality in the distribution functions, \ie\
taking $f(\q')\approx f(q'\hat\p)$. This makes the diffusion limit of Eqs.~\eqref{onetwocollisionq}
and \eqref{onetwocollision} insensitive to $\qhat$, as it also happens when $\PP$ is a function
of $p$ only, as we remarked at the end of Sec.~\ref{sub_diff}.
Transverse momentum broadening would enter in the diffusion limit of the collinear sector when
taking the first correction to the eikonal approximation, which would take the form
of $\qp^2\nabla_\perp^2f(q'\hat\p)$, assuming as usual $p\parallel z$, and would thus be suppressed by a
factor of $g^2$. Interestingly, this term would be responsible for the appearance of the \emph{double
logarithm} that has been recently pointed out in \cite{Liou:2013qya,Blaizot:2013vha,Blaizot:2015lma}.}.
We then have
\begin{equation}
\delta E(\bh,p,\omega)=\frac{h^2}{2(p)^2\omega}+\frac{\mmg}{2\omega}+\order{g^2T},
\end{equation}
which indeed is of order $gT$ and larger than the collision operator by a factor of $1/g$.
\Eq{implsubst} can be integrated over $d^2h$, symmetrized and expanded for small
$\omega$ to become
\begin{eqnarray}
\label{implsubstsoft}
&& \int\frac{d^2h}{(2\pi)^2}2\bh\cdot\mathrm{Re}\,\bff(\bh,p,\omega)\Big\vert_\mathrm{soft\,g}
\\
\nn
\hspace{-0.5em} &=& 8p^6\ca x^2(1-2x)\int\frac{d^2\qp}{(2\pi)^2}
\int\frac{d^2\kp}{(2\pi)^2}\frac{\mathcal{C}_F(\kp)}{\cf}
\left[\frac{\bqp}{\qp^2+\mmg}
-\frac{\bqp+\bkp}{(\bkp+\bqp)^2+\mmg}\right]^2,
\end{eqnarray}
where we have relabeled $\bh=p\bqp$ on the r.h.s.\footnote{Due to the properties
of the cross product, $\bh$ and $p\bqp$ have the same modulus but point in different directions, which
is irrelevant in this case.}
and kept the subleading term in $x\sim g$,
which is necessary to match to the diffusion equation. Indeed,
one can check that, upon plugging \Eq{implsubstsoft} in
Eqs.~\eqref{onetwocollision}-\eqref{jmcoll} and expanding consistently for $x\sim g$,
the diffusion structure described in detail in App.~\ref{app_lo_matching} and in particular in
\Eq{diffexpscalar} appears. The subtraction term then reads
\begin{equation}
\label{diffexpsoft}
\delta C_{a\,\mathrm{coll\,subtr}}^\mathrm{diff}[\PP]
=-
\left[\frac{1}{Tp}\PP(\p)+\left(\frac{1}{p}
+\frac{1}{2T}\right)\frac{d\PP(\p)}{dp^z}+\half\frac{d^2\PP(\p)}{d(p^z)^2}\right]
\delta\ql\Big\vert^\mathrm{coll}_\mathrm{subtr.},
\end{equation}
with
\begin{eqnarray}
\nn\delta\ql\Big\vert^\mathrm{coll}_\mathrm{subtr.}&\!\equiv\!&\int_{-\mu_\omega}^{\mu_\omega}
d\omega\,\omega^2
\frac{d\Gamma(p,\omega)}{d\omega}\Big\vert^\mathrm{coll}_\mathrm{soft\,g}=\frac{g^2\crr T}{8\pi p^4}
\int_{-\mu_\omega}^{\mu_\omega}  \frac{d\omega}{\omega^2}
\int\frac{d^2h}{(2\pi)^2}2\bh\cdot\mathrm{Re}\,\bff(\bh,p,\omega)\Big\vert_\mathrm{soft\,g}\\
\nn&&\hspace{-1.3cm}=\frac{g^2\crr \ca T}{\pi}\int_{-\mu_\omega}^{\mu_\omega} d\omega
\int\frac{d^2\qp}{(2\pi)^2}
\int\frac{d^2\kp}{(2\pi)^2}\frac{\mathcal{C}_F(\qp)}{\cf}
\left[\frac{\bqp}{\qp^2{+}\mmg}
-\frac{\bkp{+}\bqp}{(\bkp{+}\bqp)^2+\mmg}\right]^2,\\
&&
\label{collsoftct}
\end{eqnarray}
where we have dropped the statistical factor on $p$. The subleading term in $\omega$ in
\Eq{implsubstsoft}, while vanishing in \Eq{collsoftct},
is critical in obtaining the necessary $\propto p^{-1}$ terms in
\Eq{diffexpsoft}. $\mu_\omega\siml T$ is a UV regulator for this region, as the approximations
we have taken for the derivation of Eq.~\eqref{collsoftct} fail when $\omega\sim T$. Indeed,
there $\delta E$ becomes of the same size of $\int d^2\kp\mathcal{C}(\kp)$ and the LPM effect
intervenes, so that the complete leading-order rate, as given by Eq.~\eqref{jmcoll}, is finite.
\subsubsection{The conversion limit}
We now need to consider the  $q\lra gq$ process with $(p-\omega)\sim gT$ and the $g\lra q \bar q $
one with either $\omega$ or $p-\omega$ soft, which yields again a factor of 2. An altogether
similar treatment then results in
\begin{equation}
\delta C_{q_i\,\mathrm{coll\,subtr}}^\mathrm{conv}[\PP]=
\PP^{q_i}(\p)\delta\Gammac_{q\to g}(p)\Big\vert^\mathrm{coll}_\mathrm{subtr.}
-\PP^g(\p)\frac{d_A}{d_F}\delta\Gammac_{g\to q}(p)\Big\vert^\mathrm{coll}_\mathrm{subtr.},
\label{collsubtrconv}
\end{equation}
and similarly the antiquark and gluon terms have the same structure as their
leading-order counterparts Eqs.~\eqref{defconvqbar}-\eqref{defconvgluon},
with the leading-order conversion rates replaced by subtraction rates $\delta\Gamma$.
These subtraction rates read
\begin{eqnarray}
\delta\Gammac_{a\to b}(p)\Big\vert^\mathrm{coll}_\mathrm{subtr.}&=&\frac{g^2}{4\pi p}
\left\{\begin{array}{cc}
\cf& q\to g\\
\half& g\to q,\bar q
\end{array}\right\}\int_{-\mu_\omega}^{\mu_\omega} d\omega \int\frac{d^2\qp}{(2\pi)^2}
\int\frac{d^2\kp}{(2\pi)^2}
\nn \\ && \hspace{5em} \times
\mathcal{C}_F(\kp)
\left[\frac{\bqp}{\qp^2+\mmf}
-\frac{\bqp+\bkp}{(\bqp+\bkp)^2+\mmf}\right]^2.
\label{jmcollconvfinal}
\end{eqnarray}
Subleading corrections to the expansion of \Eq{implsubst} are not needed in this case.
\subsubsection{The semi-collinear limit}
\label{sub_coll_semi}
As we shall explain in more detail in \ref{sec_semi},
the semi-collinear regime refers to the region where
$\qp^2 \sim gT^2$ and no leg is soft, i.e. $\omega\simg T$, 
$p-\omega\simg T$. This in turn implies that
$h^2/(\omega(p-\omega))\sim gT^2\gg \kp^2,\mmg,\mmf$.  In this case,
we have
\begin{equation}
\delta E\equiv\delta E(\bh,p,\omega) \simeq \frac{h^2}{2p\omega(p-\omega)},
\label{desemicoll}
\end{equation}
which is larger than $\int d^2\kp \cc_F(\kp)$
so that  the integrated and symmetrized version of \Eq{implsubst}
becomes
\begin{eqnarray}
\int\frac{d^2h}{(2\pi)^2}2\bh\cdot\mathrm{Re}\,\bff(\bh)\Big\vert_\mathrm{semi-coll}&=&
\int\frac{d^2h}{(2\pi)^2}
\int\frac{d^2\kp}{(2\pi)^2}\frac{2}{\delta E^2}\frac{\kp^2\mathcal{C}_F(\kp)}{C_F}
\bigg[\left(\crr-\frac{\ca}{2}\right)\omega^2\nn\\
&&\hspace{3.7cm}+\frac{\ca}{2}\left(p^2+(p-\omega)^2\right)\bigg].
\label{colltosemiqh}
\end{eqnarray}
Its equivalent for the $g\to q\bar q$ process can be easily obtained. Since, as we shall show,
the collision operator for the semi-collinear sector is conveniently formulated in the same form
as \Eq{onetwocollision}, it suffices here to derive the subtraction rates, which read
\begin{eqnarray}
\nn	\frac{d\Gamma(p,\omega)}{d\omega}\Big\vert^\mathrm{coll\,subtr.}
_\mathrm{semi-coll}&=&\frac{g^2\crr \, }{2\pi p}(1\pm n(\omega))
(1\pm n(p-\omega))\int\frac{d^2\qp}{(2\pi)^2}
\int\frac{d^2\kp}{(2\pi)^2}\frac{\kp^2\mathcal{C}_F(\kp)}{\cf \qp^4}\\
\nn&&\times\left\{\begin{array}{cc}
\frac{1+(1-x)^2}{x}\big[\cf x^2+\ca(1-x)\big]& q\to qg\\
\frac{d_F}{d_A}(x^2+(1-x)^2)\big[\cf +\ca x(1-x)\big]& g\to q\bar q\\
\frac{1+x^4+(1-x)^4}{x(1-x)}\ca\big[1-x+x^2\big]& g\to gg	
\end{array}\right\},\\
&&
\label{colltosemi}
\end{eqnarray}
where we have relabeled $h^2=p^2\qp^2$.

\section{The diffusion sector at NLO}
\label{sec_long_diff}
We now compute the NLO corrections to the diffusion coefficients
of \Eq{diffexplo},
\begin{equation}
	\qhat_{\sss\mathrm{NLO}} = \mbox{\Eq{qhat_LO}}\, + \delta \qhat \,, \qquad
	\hat{q}_{\sss L,\mathrm{NLO}} = \mbox{\Eq{lofinal}}\, + \delta \ql \,.
\end{equation}
The NLO corrections to $\qhat$ have been
previously calculated,
\cite{CaronHuot:2008ni}:
\beqa
\label{nloqhat}
\delta \qhat=\frac{g^4\crr\ca T^3}{32\pi^2}\frac{\md}{T}\left(3\pi^2+10-4\ln 2\right),
\end{empheq}
so we focus on the corrections to $\ql$.  These corrections will
be the sum of three terms:
\begin{equation}
\delta \ql = \delta \ql \Big|_{\mathrm{loop}}
- \delta \ql \Big|_{\mathrm{subtr.}}^{\mathrm{coll}}
- \delta \ql \Big|_{\mathrm{subtr.}}^{\mathrm{diff}} \,,
\end{equation}
where the
three $\delta \ql$ encode respectively the $\OO(g)$ loop corrections
to longitudinal momentum diffusion, the collinear counterterm obtained in
Eqs.~\eqref{diffexpsoft}-\eqref{collsoftct} and a counterterm for a mistreated region in the LO
calculation of $\ql$.

\begin{figure}[ht]
\begin{center}
\includegraphics[width=14cm]{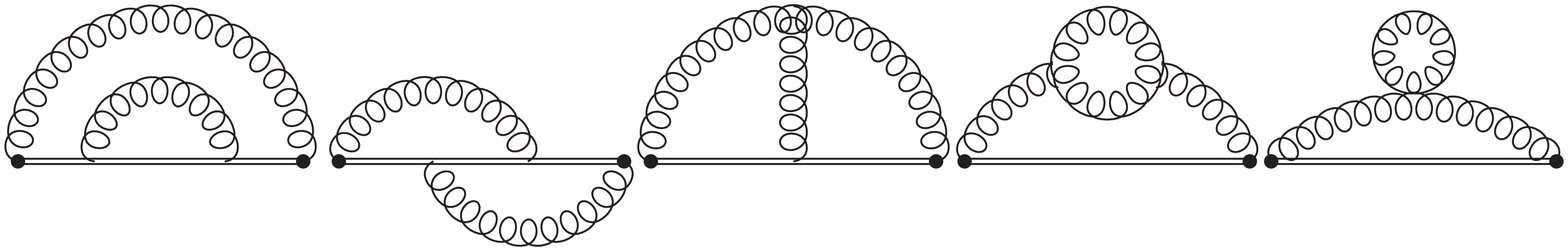}	
\end{center}
\caption{Diagrams contributing to $\delta\ql\Big\vert_\mathrm{loop}$ at NLO.}
\label{fig_nlo_soft_diagrams}	
\end{figure}

We start with $\delta \ql\Big\vert_\mathrm{loop}$, the NLO soft contribution
to \Eq{defqlong} arising from adding one extra soft gluon. A first reduction in the
number of relevant diagrams comes from the fact that, as observed in \cite{simonguy},
we can write $F^{+-}$ as
$F^{+-}=\partial^+A^--[D^-,A^+]$ and use the equation of motion of the
Wilson line, $D_{x^+}^-U(x^+;0)=0$, so that
\begin{equation}
\label{totald}
U(a;x^+)[D^-,A^+(x^+)]U(x^+;b)=\frac{d}{dx^+}\left(U(a;x^+)\,A^+(x^+)\,U(x^+;b)\right),
\end{equation}
\ie\  the commutator acts as a total derivative ($d^-$) and can be discarded
in the $dx^+$ integration,
provided that the boundary term vanishes. This is true in all non-singular gauges,
where the $A^+$ field vanishes at large $x^+$, such as the Coulomb or covariant
gauge (or, trivially, in the singular $A^+=0$ gauge). Using translation invariance and shifting the integration by $-x^+$ the same trick
can be applied to the other field strength insertion, so that in the end
in Coulomb or covariant gauge
we need to worry only about
\begin{equation}
\label{defqlongsimon}
\ql=\frac{g^2 }{d_R}\int_{-\infty}^{+\infty}dx^+\,\mathrm{Tr}\left\langle  U(-\infty;x^+)
\partial^+
A^{-}(x^+)U(x^+;0)\partial^{+}A^{-}(0) U(0;-\infty)\right\rangle,
\end{equation}
where we have suppressed the trivial dependence of gauge fields and Wilson lines
on the constant $x^-$ and $x_\perp$ coordinates.
A second simplification comes from noting that, similarly to leading order,
at NLO operator ordering is not relevant
in the soft sector in this case. Since in a first approximation $G^>\sim G^<\sim G_{rr}\sim G_F\sim
1/g G_R$, all gauge fields must connect
to the Wilson lines as $r$ fields \cite{simonguy},
so that we can replace
the more complicated contour in \Eq{defqlongsimon} with a simpler
adjoint Wilson line, \ie
\begin{equation}
\label{defqlongsimon2}
\ql=\frac{g^2 \crr}{d_A}\int_{-\infty}^{+\infty}dx^+\left\langle  \partial^+
A^{-\,a}(x^+)U^{ab}(x^+;0)\partial^{+}A^{-\,b}(0)\right\rangle.
\end{equation}
Its evaluation
requires the computation of the diagrams shown in
Fig.~\ref{fig_nlo_soft_diagrams}.%
\footnote{%
  3-point and 4-point vertices in these diagrams should be understood
  as including HTL corrections.  However, after we deform the $\qll$
  contour to large (complex) values, the contribution of the HTL
  vertices becomes small and they do not contribute to our final
  calculation.}

$\delta \ql\Big\vert_\mathrm{subtr.}^\mathrm{diff}$ arises from an $\OO(g)$ error we have committed
in the previous determination of $\ql$ to LO. Namely, we have used and
resummed HTL self-energies in the LO calculation, for instance, in
\Eq{lo}, without worrying about the fact that the HTL loop
integration extends down to zero momentum, where the hard
approximations used to simplify the calculation of the HTL break down.
In other words, the last two diagrams in Fig.~\ref{fig_nlo_soft_diagrams}
have already been included in our LO calculation, but using
approximations which are invalid for small loop momentum.  To fix
this, we should subtract off the large-momentum limiting behavior of
these diagrams when we evaluate them in the NLO computation.

We present the details of the calculation of both terms in App.~\ref{app_nlo}. Here we just mention
that the general structure corresponds to what was found for the soft contribution to the photon rate
\cite{Ghiglieri:2013gia}. Schematically, the same sum-rule technology can be applied: the Wilson line
propagators depend only on the minus components of the momenta, so
that we can again deform the contour when integrating the plus
component, which we call $\qll$. This corresponds to expanding those diagrams for large, complex $\qll$.
The leading contribution should be of order $(\qll)^0$ and the subleading one of order $(\qll)^{-1}$.
Higher-order terms are suppressed  and can be neglected. The leading, $\OO((\qll)^0)$ term, once integrated
along the contour, will give rise to a linear divergence. An analogous
linear divergence appears in
$\delta\ql\Big\vert^\coll_\mathrm{subtr.}$, as shown in
\Eq{collsoftct}.  As expected, these linear divergences
cancel.

A term behaving as $\OO(1/\qll)$ at large $\qll$ has an interpretation
of an asymptotic mass, which is why our LO result, \Eq{lofinal}, can be
written in terms of the LO asymptotic mass.  Therefore it is not
surprising that the NLO correction is found by substituting the NLO
form of the asymptotic mass
$\mmg \to \mmg+\delta\mmg$, as defined in \Eq{minftyNLO}, into
\Eq{lofinal}, and then expanding to linear order in $\delta \mmg$:
\begin{equation}
\label{soft_guess}
\frac{\mmg}{\qp^2+\mmg} \; \to \;
\frac{\mmg+\delta\mmg}{\qp^2+\mmg+\delta \mmg}
\simeq \frac{\mmg}{\qp^2+\mmg}+\delta\mmg\frac{\qp^2}{(\qp^2+\mmg)^2}.
\end{equation}
{}From \Eq{soft_guess} we thus obtain
\beqa
\delta\ql\Big\vert_\mathrm{loop}-\delta\ql\Big\vert^\mathrm{coll}_\mathrm{subtr.}-
\delta\ql\Big\vert^\mathrm{diff}_\mathrm{subtr.}
=&g^2\crr  T
\int\frac{d^2\qp}{(2\pi)^2}
\frac{\qp^2\delta\mmg}{(\qp^2+\mmg)^2}\nn\\
=&\frac{g^2\crr  T\delta\mmg}{4\pi}\left[\ln
\left(\frac{\left(\mupp^\mathrm{NLO}\right)^2}{\mmg}\right)-1\right].
\label{finallongdiffnlo}
\end{empheq}
This simpleminded argument indeed reproduces the detailed explicit
calculation of Appendix \ref{app_nlo}.

\Eq{finallongdiffnlo} depends on a regulator $\mupp^\NLO$.
As we will show, the dependence of this term on the regulator and the
dependence of the semi-collinear region will cancel.
This completes the evaluation of the diffusion sector to NLO.

\section{Conversion processes at NLO}
\label{sec_nlo_conv}
According to \Eq{convcorr}, the NLO corrections to the conversion sector take the following form:
\beqe
\label{nloconv}
\delta C_a^\mathrm{conv}[\PP]=\sum_{b\ne a}\bigg[\PP^a(\p)\delta \Gammac_{a\to b}(p)	-
\PP^b(\p)\frac{d_b}{d_a}\delta \Gammac_{b\to a}(p)\bigg],
\end{empheq}
where the sum is understood to give rise to the structure of Eqs.~\eqref{defconvq}-\eqref{defconvgluon}.
The NLO conversion rates are composed of three parts, namely
\begin{equation}
\label{defnloconvrate}
\delta \Gammac_{a\to b}(p)=\delta \Gammac_{a\to b}(p)\Big\vert_\mathrm{loop}-
\delta \Gammac_{a\to b}(p)\Big\vert_\mathrm{subtr.}^\coll-
\delta \Gammac_{a\to b}(p)\Big\vert_\mathrm{subtr.}^\mathrm{conv}.
\end{equation}
The first term on the right-hand side comes from the soft-gluon loop correction to the rates, as
defined by the Wilson-line operators~\eqref{convratefermion}-\eqref{convratesqbar}. The second
is the subtraction term from the collinear region, as obtained in \Eq{jmcollconvfinal},
and the third subtracts the Hard Thermal Loop approximated
leading-order calculation, result~\eqref{convratefermionlo}, in
complete analogy to $\delta \ql\big\vert_\mathrm{subtr.}^\mathrm{diff}$ encountered in
the previous section.

The first and the third term can then be evaluated from Eqs.~\eqref{convratefermion}-\eqref{convratesqbar},
by adding one extra soft gluon to the LO term in \Eq{convratefermionlo}.
A key observation is that the power-counting arguments that lead to the simplified form
for $\ql$ given by \Eq{defqlongsimon2} apply here as well: all soft gluons must connect
to the Wilson line as $r$ fields, so that their ordering is not relevant \cite{simonguy}.
This implies that the fundamental and adjoint
Wilson lines appearing in Eqs.~\eqref{convratefermion}-\eqref{convratesqbar} can be simplified to NLO
to a simpler antifundamental line connecting the soft fermionic fields, \ie
\begin{eqnarray}
\label{convratefermionsimple}
\Gammac_{q\to g}(p)&=&-\frac{ g^2\cf}{8d_Fp}\int_{-\infty}^{+\infty}dx^+\left\langle\mathrm{Tr}
\big[
\bar\psi(x^+,0,0_\perp)\slashed{v} U_F(0,0,0;x^+,0,0_\perp)\psi(0)\big]\right\rangle,\\
\Gammac_{g\to q}(p)&=&\frac{d_F}{d_A}\Gammac_{q\to g}(p).
\label{convrategluonsimple}
\end{eqnarray}
This corresponds to an effective abelianization of these operators, which, we note, are the same as those
appearing in the fermionic sector of the Hard Thermal Loop action \cite{Braaten:1991gm}. Indeed, an altogether similar
abelianization happens for instance when obtaining the effective $qqg$ HTL vertex in QCD.

\begin{figure}[ht]
	\begin{center}
		\includegraphics[width=12cm]{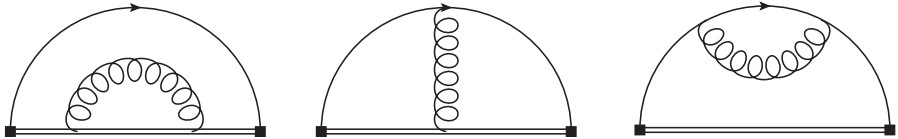}
	\end{center}
	\caption{Diagrams obtained from \Eq{convratefermionsimple}
at NLO. The double line is the fundamental Wilson line and the black squares
are the insertion of the soft fermion fields.}
	\label{fig_nlo_conv}
\end{figure}

The diagrams necessary for the NLO evaluation of
Eqs.~\eqref{convratefermionsimple} and \eqref{convrategluonsimple}
are shown in Fig.~\ref{fig_nlo_conv}. 
However, we note that, in their abelianized forms, Eqs.~\eqref{convratefermionsimple} and
\eqref{convrategluonsimple} correspond, up to the prefactors,
to the soft-sector
contribution to the NLO photon rate in \cite{Ghiglieri:2013gia}. Therefore
we can directly use that result, which
was obtained using the same sum-rule techniques employed in the previous section. Indeed, as we observed
there, the two results are remarkably similar, the only difference being given by the different
asymptotic masses. Here the relevant one is the quark one and we then have
\beqe
\delta\Gammac_{q\to g}(p)=\frac{ g^2\cf}{4p}
\int\frac{d^2\qp}{(2\pi)^2}\frac{\qp^2\,\delta \mmf}{(\qp^2+\mmf)^2}=\frac{ g^2\cf\delta\mmf}{16\pi p}
\left[\ln
\left(\frac{\left(\mupp^\mathrm{NLO}\right)^2}{\mmf}\right)-1\right],
\label{finalnloconv}
\end{empheq}
where $\delta\mmf$ is given by \Eq{minftyNLO} and
the linear divergence in the collinear counterterm
canceled an opposite one coming
from the loop corrections. The logarithmic UV divergence has been
treated with the
same UV regulator $\mupp^\NLO$ used in the previous section for $\delta\ql$.

\section{The semi-collinear region}
\label{sec_semi}
\begin{figure}[ht]
\begin{center}
\includegraphics[width=14cm]{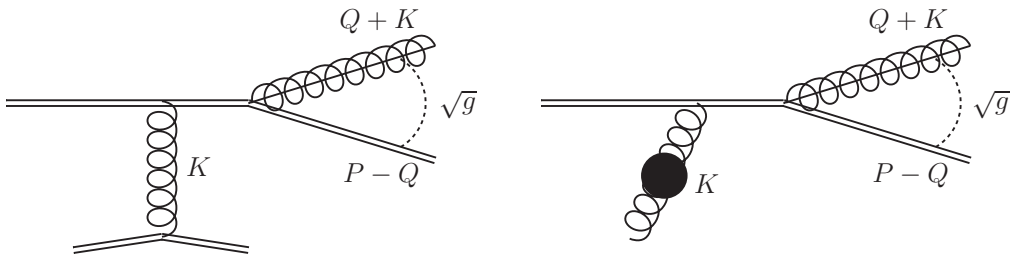}	
\end{center}
\caption{Diagrams for two typical semi-collinear processes. In the first case the soft gluon
is in the space-like Landau cut, whereas in the second case it is on its time-like plasmon pole,
represented by the black blob.}
\label{fig_semicoll}
\end{figure}

As we anticipated in Sec.~\ref{sec_nlo_intro}, semi-collinear processes
can be seen as $\onetwo$ splitting processes where the virtuality and correspondingly
the opening angle are larger. Two examples are drawn in Fig.~\ref{fig_semicoll}.
The scalings of this region are as follows: $K\sim gT$ is soft, whereas the two final-state particles
are quasi-collinear, i.e. with an increased virtuality and opening angle with respect to the collinear sector.
The leading contribution then comes from $\qll\sim T,\,\qm\sim gT,\,\qp^2\sim gT^2$, $Q^2\sim gT^2$
or, in the case of a democratic splitting,
$\qll\sim E,\,\qm\sim gT,\,\qp^2\sim gTE$, $Q^2\sim gTE$. Naive power-counting arguments would suggest
that the semi-collinear
 region should contribute to leading-order, as it is the largest slice of phase space
where a soft gluon can attach to a $\onetwo$ process. However, once all diagrams are summed
and squared, a cancellation, discussed in \cite{Arnold:2001ba} in the context of photon
radiation, introduces an extra $\OO(g)$ suppression%
\footnote{%
  The cancellation occurs because the transverse momentum of the split
  particles $\pp^2$ is larger than the disturbance from the scattering
  $\qp^2$.  In the limit that the disturbance is arbitrarily small, we
  would not expect it to induce a splitting.
  This cancellation can be seen at work
  in our derivation of Eq.~\eqref{colltosemiqh} from
  Eq.~\eqref{implsubst}. Upon enforcing semi-collinear kinematics on
  the latter, i.e. $\bh\gg p\bkp,\omega\bkp,(p-\omega)\bkp$, all terms
  in square brackets vanish at first order in that expansion and only
  the next one gives a nonzero contribution. Without that cancellation
  the semi-collinear rate would indeed be leading-order.
}. 
Furthermore, as we shall show, the contribution from
time-like soft gluons, e.g.\ plasmons, is now allowed.

The contribution $\delta C_a^\mathrm{semi-coll}$ to the collision operator can be written
in the same way as the collinear one, as given by Eqs.~\eqref{onetwocollisionq}-\eqref{onetwocollision},
with the replacement of the collinear rates with semi-collinear ones. For instance, for quarks and
antiquarks it reads
\beqa
\nn-\delta C_{q,\bar q}^\mathrm{semi-coll}[\PP](\p)=&\int_{-\infty}^{+\infty}d\omega \PP_{q,\bar q}
((p+\omega)\hat\p)
\frac{d\Gamma^{q}_{qg}(p+\omega,\omega)}{d\omega}\Big\vert_\mathrm{semi-coll}\\
\label{semicollop}
&\hspace{-2.5cm}-\PP_{q,\bar q}(\p)
\frac{d\Gamma^{q}_{qg}(p,\omega)}{d\omega}\Big\vert_\mathrm{semi-coll}+\PP_{g}((p+\omega)\hat\p)
\frac{\da}{\df}\frac{d\Gamma^{g}_{q\bar q}(p+\omega,\omega)}{d\omega}\Big\vert_\mathrm{semi-coll}.
\end{empheq}
The derivation of the semi-collinear rates then
requires the evaluation of processes of the form of Fig.~\ref{fig_semicoll}, with $p,\qll\gg
\qp\gg\kp,\kl$. Actually we have already evaluated these diagrams using the collinear
expansion, since it is precisely these diagrams which give rise to the
linear-in-collisions expressions we found in
Subsec.~\ref{sub_coll_semi}.  In particular, the subtraction term from the
collinear region,  \Eq{colltosemi}, was derived
by making an expansion in $\qll \gg \qp$, and it still applies, under one condition.
In evaluating the collision sector,
we treated $\qp \sim \kp \sim gT$, leading to $\delta E \sim g^2 T$.
This let us neglect $\delta E$ when working out the kinematics of the
soft gluons, so that $\cc_R(\kp)$ (see for instance \Eq{C_LO}) is defined for $\km=0$ and
hence only space-like gluons contribute to it.
But if $\qp^2 \sim g T^2,\qll\sim T$, or $\qp^2 \sim g TE,\qll\sim E$, then $\delta E \sim gT$ and
can no longer be neglected, opening up the time-like-gluon sector.  In particular, when we put $(P-Q)$ and
$(Q+K)$ on shell, we find, exactly as in the photon case \cite{Ghiglieri:2013gia},
which we refer to for more details, that
\begin{equation}
\delta((Q+K)^2)=
\frac{\delta(k^--\delta E)}{2\vert \qll\vert}+\OO(\sqrt{g})\,, \qquad
\delta E = \frac{p\,\qp^2}{2\qll(p-\qll)}\approx\frac{p\,\qp^2}{2\omega(p-\omega)},
\end{equation}
where the $\OO(\sqrt{g})$ correction comes from $\bqp\cdot\bkp$ and always vanishes in the
angular integrations. We see that $\delta E$ is exactly what we have used in Sec.~\ref{sub_coll_semi}.
Therefore we must re-derive \Eq{colltosemi} with these
somewhat different kinematics.  A straightforward computation\footnote{Interestingly,
the computation can also be performed using standard Soft Collinear Effective Theory (SCET)
\cite{Bauer:2000ew,Bauer:2000yr,Bauer:2001ct,Bauer:2001yt,Bauer:2002nz,Beneke:2002ph}.
Indeed, we have $Q=(\qll,\qm,\qp)\sim
\Lambda(1,\lambda^2,\lambda)$ and $K\sim \Lambda(\lambda^2,\lambda^2,\lambda^2)$,
where $\Lambda$ is the large scale, $E$ or $T$, $\lambda\ll 1$ is the expansion parameter,
either $\lambda\sim\sqrt{gT/E}$ or $\lambda\sim\sqrt{g}$. These are then the standard scalings
of SCET${}_\mathrm{I}$. However, due to the cancellations mentioned in this Section, the $\OO(\lambda)$
soft-collinear couplings \cite{Chay:2002vy,Manohar:2002fd,Beneke:2002ph,Bauer:2002aj} are necessary. }
shows that the findings
in the case of photon radiation \cite{Ghiglieri:2013gia} generalize to the present case.
Namely, the quantity
\begin{equation}
\label{qhat}
\frac{\hat{q}}{g^2 \crr} \equiv \frac{1}{g^2 \crr} \int \frac{d^2
\kp}{(2\pi)^2} \,\kp^2 \,\cc_R(\kp)
= \int \frac{d^4 K}{(2\pi)^3} \delta(k^-) \kp^2 G_{rr}^{--}(K) \,,
\end{equation}
physically interpreted as the transverse momentum diffusion coefficient and present
in \Eq{colltosemi}, should be replaced with its finite $\delta E$
generalization,
\begin{equation}
\label{semiint}
\Isc\equiv\int\frac{d^4K}{(2\pi)^3}\delta(k^--\delta E)\left[	\kp^2 G^{--}_{rr}(K) + 2G_T^{rr}(K)
\left(\delta E^2-\kl\delta E\frac{\kp^2}{k^2}\right)\right],
\end{equation}
which goes into \Eq{qhat} for $\delta E\to 0$  and
corresponds to the leading-order soft term (in Coulomb gauge) in the evaluation of the operator
\begin{equation}
\label{Isc}
\qhat(\delta E)= \frac{g^2\crr}{\da}\int_{-\infty}^\infty dx^+\,e^{ix^+\delta E} \,
 \langle v^\mu {F_\mu}^{\nu\,a}(x^+,0,0_\perp)
U_{ A}^{ab}(x^+,0,0_\perp;0,0,0_\perp)
v^\rho F^b_{\rho\nu}(0)\rangle,
\end{equation}
which was first introduced in the photon case
\cite{Ghiglieri:2013gia}. In principle in the present case
a more complicated ``three-pole'' operator should be needed
\cite{CaronHuot:2008ni}.  However, at leading and next-to-leading order it would reduce
to a set of three two-body exchanges of the form of~\eqref{Isc}, with
the appropriate Casimir factors \cite{CaronHuot:2008ni}.

\Eq{semiint} can be evaluated using Euclidean techniques, yielding \cite{Ghiglieri:2013gia}\footnote{The Euclidean evaluation
combines the time-like (plasmon) and space-like (scattering) contributions. From Eq.~\eqref{semiint} it
follows that, once $k^-$ is integrated over the $\delta$-function, plasmons contribute for $k^+>\kp^2/(2\delta E)$,
while space-like gluons contribute for $k^+<\kp^2/(2\delta E)$. In order to disentangle
the two contributions one would have to proceed numerically.} 
\begin{equation}
\label{semiintexp}
\Isc=T\int\frac{d^2\kp}{(2\pi)^2}\bigg[\frac{\md^2\kp^2}
{(\kp^2+\delta E^2)(\kp^2+\delta E^2+\md^2)}+\frac{2\delta E^2}{\kp^2+\delta E^2}\bigg].
\end{equation}
However, since this momentum region has overlap with both the
collinear and the hard regions, there are two subtractions which must
be conducted, corresponding to the treatments already included in
those leading-order calculations.  Therefore we must compute the
behavior of this momentum region under each of those limiting
kinematics and subtract them.  The collinear case is treated by
subtracting $\qhat$ from $\qhat(\delta E)$.  For the hard region, we
take the soft, bare limit ($\nbe(k^0)\to T/k^0$, $\rho(K)\to\rho^{(0)}(K)
=2\pi\,\mathrm{sgn}(k^0)\delta(K^2)$) of \Eq{semiint}, yielding
\begin{equation}
\nn \hspace{-2ex} \Isc\Big\vert_{\rm hard}\!=\int\frac{d^4K}{(2\pi)^3}
\delta(k^-{-}\delta E)2G_T^{(0)rr}(K) \delta E^2%
= T\int\frac{d^2\kp}{(2\pi)^2}\frac{2\delta E^2}{\kp^2+\delta E^2},
\label{finalssgluonbare}
\end{equation}

The full semi-collinear rate is then obtained by replacing
\begin{equation}
\label{semicollrule}
\frac{\qhat}{g^2\crr}\to\Isc-\frac{\qhat}{g^2\crr}-\Isc\Big\vert_{\rm hard},
\end{equation}
in \Eq{colltosemi}, which yields
\beqa
\nn	\frac{d\Gamma(p,\omega)}{d\omega}\Big\vert_\mathrm{semi-coll}=&\frac{g^4\crr T }{2\pi p}
(1\pm n(\omega))(1\pm n(p-\omega))\int\frac{d^2\qp}{(2\pi)^2}
\int\frac{d^2\kp}{(2\pi)^2}\frac{1}{\qp^4}\\
\nn&\times\left\{\begin{array}{cc}
\frac{1+(1-x)^2}{x}\big[\cf x^2+\ca(1-x)\big]& q\to qg\\
\frac{d_F}{d_A}(x^2+(1-x)^2)\big[\cf +\ca x(1-x)\big]& g\to q\bar q\\
\frac{1+x^4+(1-x)^4}{x(1-x)}\ca\big[1-x+x^2\big]& g\to gg	
\end{array}\right\}\\
&\times
\bigg[\frac{\md^2\kp^2}
{(\kp^2+\delta E^2)(\kp^2+\delta E^2+\md^2)}-\frac{\md^2}{\kp^2+\md^2}\bigg].
\label{jmsemicoll}
\end{empheq}
We stress that the collision operator has the same form as
Eqs.~\eqref{onetwocollisionq}-\eqref{onetwocollision}.

The $\qp$ integration in \Eq{jmsemicoll} is to be understood as IR-regulated by $\mupp^\NLO$.
In App.~\ref{app_semi} we show how the small-$\omega$-and-$\qp$ region gives rise to IR logarithms
that cancel the $\mupp^\NLO$ dependence of the diffusion and conversion sectors. We also give some
details of how the transverse integrations can be carried out analytically. The $\omega$
integration remains to be performed numerically.

\section{Summary and conclusions}
\label{sec_concl}

The main aim of this paper has been to show how the propagation
of highly energetic quarks and gluons through the QGP can be described
at leading- and next-to-leading order by a Boltzmann equation
encoding the interaction between these hard particles and the thermal
and soft constituents of the plasma. Sec.~\ref{sec_lo_amy}
has been devoted to a brief review of the LO kinetic approach
introduced in \cite{Arnold:2002zm} and implemented in MARTINI.
As \Eq{boltzmann} summarizes, the two processes it
incorporates are $\twotwo$ scatterings with the thermal medium
constituents and $\onetwo$ collinear splittings induced by the
soft background.

In Sec.~\ref{sec_reorganize} we have shown how this approach
is not optimal beyond leading order, where the distinction between the two
classes would blur and the resummed matrix-element approach to $\twotwo$
scattering would become cumbersome. With these motivations,
we have reorganized the LO collision operator into four separate processes
which provide a sufficient description at NLO. They are \emph{large-angle scatterings}, \ie\
$\twotwo$ scatterings with $\OO(1)$ angles or equivalently $\OO(T)$ or larger
transferred momentum, \emph{diffusion} processes, caused by soft gluon
exchanges, which preserve the identity of the hard particles while slightly
affecting their momentum, \emph{conversion} processes which instead
turn quarks into gluons and vice versa through soft quark exchange and finally
\emph{collinear} processes, corresponding at LO to $\onetwo$ processes.
In Sec.~\ref{sub_large} we described in detail our description of large-angle
processes, which require regularization to be kept separate from diffusion
and conversion ones, as shown in Eqs.~\eqref{el2reg} and \eqref{el2quarkreg}.
Sec.~\ref{sub_diff} has been dedicated to diffusion processes, which
are described by an effective Fokker-Planck equation,
\Eq{diff}. The three
physical effects of \emph{drag} (energy loss), \emph{longitudinal}
and \emph{transverse momentum broadening} are encoded in three corresponding
coefficients. The requirements that the Fokker-Planck picture be equivalent
to the Boltzmann one and that
it approach equilibrium can be used to write the drag coefficient
in terms of the other two, as per \Eq{equilibrate}. The two
momentum diffusion coefficients can then effectively be described by
field strength correlators along Wilson lines on the light-cone direction
of propagation of the hard particle, as in \Eq{defqlong}.
The calculation of the transverse momentum diffusion
coefficient $\qhat$ is mapped to a Euclidean one
\cite{CaronHuot:2008ni}, whereas for the longitudinal momentum diffusion
coefficient $\ql$ we introduce a sum rule which, through
the analytical properties
of amplitudes at light-like separations, makes it sensitive
only to the gluon dispersion  relation close to the light cone
(see \Eq{lofinal}).
Similarly, conversion processes are shown in Sec.~\ref{sec_lo_conv}
to be described by effective Wilson line operators (Eqs.~\eqref{convratefermion}
and \eqref{convrategluon}), which are also computed through an equivalent
light-cone sum rule mapping them to the quark dispersion relation. The
UV log-divergence of the diffusion and conversion processes cancels with the
opposite IR one in large-angle scatterings.

In Sec.~\ref{sec_nlo_intro} we introduced the NLO extension of this reorganized
approach. All processes, with the exception of large-angle scatterings,
are sensitive to $\OO(g)$ corrections arising from the interactions with the
soft background. Furthermore, some care is necessary in avoiding double
countings in slices of the phase space, which were included at LO,
where some particles become soft, introducing the need for a set of
subtractions. The remaining sections are then devoted to the details of
each process at NLO. In Sec.~\ref{sec_nlo_coll} we discuss the collinear
region, which is sensitive to $\OO(g)$ corrections in the interactions
with the soft background that induce the splitting, as well as in the
dispersion relation of the hard and thermal particles. We further identify
all necessary subtractions.

Sec.~\ref{sec_long_diff} is dedicated to $\OO(g)$ corrections to diffusion.
In treating $\qhat$, we employ the NLO determination of
\cite{CaronHuot:2008ni}, whereas for $\ql$ we perform the calculation
using the light-cone sum rules introduced before. The details are to be
found in App.~\ref{app_nlo}. The result is surprisingly simple: it just amounts
to considering the soft correction to the gluon dispersion relation close
to the light-cone (see \Eq{soft_guess}). Similarly, conversion processes
are dealt with using the fermionic analogue of the same sum rule and
require the inclusion of the soft correction to the quark asymptotic mass, as
in \Eq{finalnloconv}. Both $\ql$ and the conversion rate at NLO
show an UV logarithmic divergence, which is removed once a new process,
which only starts to contribute at NLO, is considered, the \emph{semi-collinear
process}. As illustrated in Sec.~\ref{sec_semi}, this process appears as a
bridge between the diffusion/conversion sector on one side and the collinear
on the other. Indeed, while retaining a collinear kinematics, it shows
relaxed constraints, going beyond strict collinearity and allowing
the interactions with the soft background to be not just space-like (soft
scatterings) but also time-like (plasmon absorption/emission). For its
evaluation a modified form of $\qhat$, $\qhat(\delta E)$, is
introduced in \Eq{Isc}. It accounts for the changes in the small
light-cone component $p^-$, which are no longer negligible.
Euclidean techniques are used for its computation, as per
\Eq{semiintexp}.

We would like to emphasize the importance of
Euclidean techniques,
which map the calculation of $\cc(\kp)$, $\qhat$, $\qhat(\delta E)$, $\delta \mmg$ and $\delta \mmf$ into
simpler calculations in dimensionally-reduced EQCD.  Similarly, light-cone sum rules
reduce the computation of $\ql$ and of the conversion rates to
the determination of the gluon and quark asymptotic masses at leading- and
next-to-leading order. Without these recent theoretical developments,
rooted in the causal properties of amplitudes at light-like separations,
the calculations presented here would have required extensive, cumbersome
numerical integrations over the intricate structures of loops composed
of HTL propagators and vertices. Furthermore, as we have mentioned,
Euclidean techniques also allow lattice determinations. The first measurements
of $\qhat$ and $\cc(x_\perp)$  have recently been reported
\cite{Laine:2013lia,Panero:2013pla}, opening up a new avenue of research. All
other Euclidean operators can be computed on the lattice in the same way,
creating the tantalizing possibility of a factorized approach to kinetics,
where perturbation theory is used at the thermal and hard scales to compute
the large-angle scatterings and the splittings, whereas the 3D lattice
is employed at the soft (and ultrasoft) scale to determine non-perturbatively
the transverse diffusion processes and the scatterings leading to collinear
radiation.

A very important point we have not addressed in this paper, leaving it
to future work, is the impact
of the NLO corrections we have introduced on calculations of jet modification and
their comparison to experimental data. As we mentioned, the Monte Carlo
event generator MARTINI implements a kinetic approach corresponding to the
one described in Sec.~\ref{sec_lo_amy}. This makes it an ideal candidate for
the inclusion of the NLO corrections. Indeed, the reorganization of the LO
collision operator in terms of large-angle, diffusion, conversion and collinear
processes is underway, as well as the implementation of the NLO corrections.
This could also be easily complemented by the inclusion of non-perturbative
input, such as the existing determination of $\qhat$ and, should they become
available, future determinations of $\qhat(\delta E)$ and of the asymptotic
masses. It would also be interesting to study the angular structures of
jets with this numerical implementation and compare it with the
recent order-of-magnitude perturbative estimates from \cite{Kurkela:2014tla}.

We remark that it is difficult for us to gauge a priori
the impact of NLO corrections
relative to LO. The recent NLO calculations of the thermal photon
\cite{Ghiglieri:2013gia} and low-mass dilepton \cite{Ghiglieri:2014kma} rates,
which include many of the features presented here, such
as Euclidean techniques, light-cone sum rules, semi-collinear and collinear
processes, showed how the NLO corrections naturally grouped into two
classes of large, and largely canceling, contributions. The positive
corrections were due to NLO modifications
to the collinear processes, caused by the increased soft scattering rate and
the reduced asymptotic masses, while semi-collinear and conversion processes
decreased the rate by a similar magnitude. The large cancellation between
these contribution is mostly accidental and furthermore depends
significantly on the details of the medium, such as the numbers of colors and
flavors. So, while we anticipate similar cancellations for the present energy
loss case, we are at present unable to quantify their impact in more detail.

Finally, we believe that the approach presented here should go much of the way
towards making possible NLO kinetic theory calculations of
the shear viscosity and other transport coefficients of QCD.  However, we have not
resolved the issue of keeping track of where the energy from a soft
scattering shows up amongst the other (thermal) particles, which so far
prevents us from a true NLO calculation of QCD transport coefficients.
We hope to return to this issue in the future.
We do note, however,
that for cases where the momentum dependence of the off-equilibrium
distributions is isotropic, such as studies of isotropic thermalization
\cite{York:2014wja,Kurkela:2014tea}, an extension to NLO appears within reach.

\section*{Acknowledgments}
We would like to thank Simon Caron-Huot and Aleksi Kurkela
for useful conversations.
This work was supported
in part by the Institute for Particle Physics (Canada), the Natural
Sciences and Engineering Research Council (NSERC) of Canada,
the Swiss National Science Foundation
(SNF) under grant 200020\_155935
and a grant from the U.S. Department of Energy, DE-FG-02-08ER4154.

\appendix
\section{Notation}
\label{appnotate}

We now summarize our notation.
We will use capital letters  for four-vectors, lowercase
italic letters for the modulus of the spatial three-vectors, and the
mostly-plus metric $\eta_{\mu\nu} = \mathrm{Diag}\:[{-}{+}{+}{+}]$,
so that $P^2=p^2-p^2_0$.

For convenience
we will mostly work in the Keldysh, or \ra,
basis of the real-time formalism for the computation
of thermal expectation values. The two elements of this basis are defined as
$\phi_r\equiv(\phi_1+\phi_2)/2$,
$\phi_a\equiv\phi_1-\phi_2$, $\phi$ being a generic field and the subscripts 1 and 2
labeling the time-ordered and anti-time-ordered branches of the Schwinger-Keldysh contour respectively.
The propagator is a $2\times2$ matrix, where one entry is always zero and only one entry depends on the
thermal distribution, \emph{i.e.},
\begin{equation}
\label{raprop}
D=\left(\begin{array}{cc} D_{rr}&D_{ra}\\D_{ar}&D_{aa}\end{array}\right)
=\left(\begin{array}{ccc} \left(\frac12\pm n(p^0)\right)(D_R-D_A)&&D_{R}\\D_{A}&&0\end{array}\right),
\end{equation}
where $D_R$ and $D_A$ are the retarded and advanced propagators, the plus (minus) sign refers
to bosons (fermions). $n(p^0)$ is the corresponding thermal distribution, either
$\nbe(p^0)=(\exp(p^0/T)-1)^{-1}$ for bosons or $\nfd(p^0)=(\exp(p^0/T)+1)^{-1}$ for fermions.
We also define the spectral function as the difference of the retarded and
advanced propagators, $\rho \equiv D_R - D_A$.
We will denote the gluon propagator by $G$ and the quark one $S$.

We will adopt strict Coulomb gauge throughout.  The treatment of soft
momenta in propagators and vertices requires the use of Hard Thermal
Loop (HTL) resummation \cite{Braaten:1989mz}.  For convenience we list
the Coulomb gauge retarded HTL resummed propagators for fermions and
gluons in the next section.

\section{Hard Thermal Loop propagators}
\label{app_props}

In this section we detail our conventions for the HTL
propagators. Fermion propagators are most easily written in terms of
components with positive and negative chirality-to-helicity ratio. The retarded
fermion propagator reads
\begin{equation}
\label{htlfermiondef}
S_{R}(P)=h^+_\bp S^+_{R}(P)+h^-_\bp S^-_{R}(P)\,,
\end{equation}
where
\begin{equation}
\label{htlfermion}
S^{\pm}_R(P)=\frac{i}{p^0\mp (p+\Sigma^\pm(p^0/p))}
= \left.\frac{i}{\displaystyle p^0\mp\left[p+\frac{\mmf}{2p}
\left(1-\frac{p^0\mp p}{2p}\ln\left(\frac{p^0+p}{p^0-p}
\right)\right)\right]}\right\vert_{p^0=p^0+i\epsilon},
\end{equation}
where the upper (lower) sign refers to the positive (negative)
chirality-to-helicity component. The  projectors are
$h^\pm_\bp\equiv(\gamma^0\mp\vec\gamma\cdot\hat p)/2$. Here $\mmf$ is
the fermionic asymptotic mass squared, defined such that the
large-momentum dispersion relation for helicity=chirality fermions is
$p_0^2 = p^2 + \mmf$.  We similarly define the asymptotic gluonic
mass $\mmg$.  At leading order, their values are
\begin{equation}
\label{loasym}
\mmg=\frac{\md^2}{2}=\frac{g^2T^2}{6}\left(\nc+\frac{\nf}{2}\right),
\qquad
\mmf=2 m_q^2 = \cf\frac{g^2T^2}{4},
\end{equation}
where we have also shown the relations to the more commonly used Debye
mass $\md$ and quark ``mass'' $m_q$.

Gluons are described in the strict Coulomb gauge by
\begin{eqnarray}
\label{htllong}
G^{00}_R(Q)&=&\frac{i}{\displaystyle q^2+\md^2\left(1-\frac{q^0}{2q}\ln\frac{q^0+q+i\epsilon}{q^0-q+i\epsilon}\right)},\\
\nn G^{ij}_R(Q)&=&(\delta^{ij}-\hat q^i\hat q^j)G^T_R(Q)=
\left.\frac{i(\delta^{ij}-\hat q^i\hat q^j)}
     {\displaystyle q_0^2-q^2-\mmg \left(\frac{q_0^2}{q^2}
       -\left(\frac{q_0^2}{q^2}-1\right)\frac{q^0}{2q}
       \ln\frac{q^0{+}q}{q^0{-}q}\right)}\right\vert_{q^0=q^0+i\epsilon}.\\
&& 	\label{htltrans}
\end{eqnarray}
The other components of the propagators in the \ra basis can be obtained
through \Eq{raprop}.

\section{Longitudinal momentum diffusion from Wilson lines}
\label{app_wline}
\Eq{defqlong} is based on eikonalization, which naturally happens since $p$ is considered
infinitely larger than all other scales at leading order. As such, it can be easily verified that the
perturbative expansion of \Eq{defqlong} agrees with the rate-based definition~\eqref{longrate}
at leading and next-to-leading order. We believe that, in the presence
of a consistent UV regulator%
\footnote{%
  We use a UV cutoff $\mupp$ in this paper; for a more rigorous
  treatment we could use dimensional regularization, or the
  introduction of a mass with the limit $p\to \infty$ taken holding
  $m/p$ small but finite, which produces a ``dead cone'' which renders
  radiative effects finite.}
\Eq{defqlong} is correct to all
orders in $g$ at the leading order in $1/p$, up to possible Wilson lines along the $x^-$ direction
at $x^+=-\infty$. Indeed, we believe that \Eq{defqlong}
can be rigorously obtained in dimensional regularization
using SCET, analogously
to what has been done in \cite{D'Eramo:2010ak,Benzke:2012sz} for $\qhat$. We sketch  here
a simplistic derivation.
Since we are interested in the differential-in-$q^z$ rate for a fast particle propagating with $p^0=p^z$,
it is natural to expect from the eikonal approximation a correlator of the form
\begin{equation}
(2\pi)\frac{d\Gamma}{dq^z}=(2\pi)\frac{d\Gamma}{d\qll}
=\lim_{L\to\infty}\frac{1}{L}\int dx^- e^{i\qll x^-}\frac{1}{d_R}
\big\langle
\mathrm{Tr}U(-L/2,L/2;x^-)	U(L/2,-L/2;0)\big\rangle,
\label{roughrate}
\end{equation}
where we have used the fact that in the infinite-$p$ limit $q^z=\qll$
and
for simplicity we have introduced
\begin{equation}
U(a^+,b^+;c^-)=P\exp\left(ig \int_{b^+}^{a^+} dl^+ A^{-}(l^+,c^-)\right),
\end{equation}
and for further convenience
\begin{equation}
\tilde{U}(a^-,b^-;c^+)=P\exp\left(ig \int_{b^-}^{a^-} dl^- A^{+}(l^-,c^+)\right).
\end{equation}
The Wilson line at $x^-=0$
is supported on the time-ordered branch of the Schwinger-Keldysh contour and conversely the other one
is supported on the anti-time ordered branch, corresponding to the amplitude and conjugate amplitude
entering the definition of the rate. Indeed, so far the techniques used in
\cite{D'Eramo:2010ak,Benzke:2012sz} are exactly applicable here as well, so that \Eq{roughrate}
is also formally justified within SCET. It is however not gauge-invariant. Following
the steps of \cite{Benzke:2012sz}, we conjecture this form for its gauge-invariant dressing:
\begin{figure}[ht]
\begin{center}
\includegraphics[width=12cm]{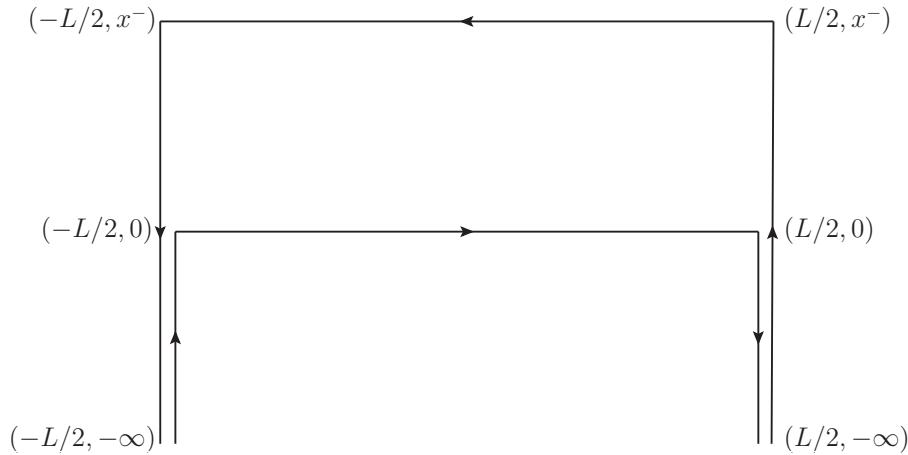}	
\end{center}
\caption{The Wilson loop giving rise to $\ql$. The horizontal axis is the $+$ axis and the vertical
one is the $-$ one. The points are given in $(x^+,x^-)$ coordinates, the constant transverse one
is not shown.}
\label{fig_wloop}
\end{figure}
\begin{eqnarray}
\hspace{-0.5cm}\nn(2\pi)\frac{d\Gamma}{d\qll}&=&
\lim_{L\to\infty}\frac{1}{L}\int dx^- e^{i\qll x^-}\frac{1}{d_R}\big\langle
\mathrm{Tr}\tilde{U}(-\infty,x^-;-L/2)U(-L/2,L/2;x^-)	\\
&&\times
\tilde{U}(x^-,-\infty;L/2)\tilde{U}(-\infty,0;L/2)U(L/2,-L/2;0)
\tilde{U}(0,-\infty;-L/2)\big\rangle.
\label{longwloop}
\end{eqnarray}
The operator defined by \Eq{longwloop} is sketched in Fig.~\ref{fig_wloop}.
This particular ordering corresponds to having the upper three connected Wilson lines
on the anti-time ordered branch of the Schwinger-Keldysh contour and the lower three on
the time-ordered one.
The ``handle'' on the bottom right corner can be trivially annihilated, but the same is
not true for the one at the bottom left, since time-like separated fields appear between
the two vertical Wilson lines there.

Finally, by using the definition of $\ql$ and convoluting \Eq{longwloop} with $(\qll)^2$,
the latter can be replaced by derivatives which,
when acting on the Wilson loop, introduce the $F^{+-}$ electric fields. Once the $\qll$
integration is taken (with infinite cutoff, hence the strict validity in dimensional regularization only),
the Wilson line operator~\eqref{longwloop} is squeezed to the form of
\Eq{defqlong} plus a surviving ``handle'' along $x^-$ at $x^+=-\infty$.
This handle is irrelevant in
non-singular gauges and even in the light-cone gauge $A^-=0$ it can be neglected at LO and
NLO.\footnote{\label{foot_sing_gauge}
In the $A^-=0$ gauge the leading-order term arises from the
$<\partial^-A^+(x^+)\partial^-A^+(0)>$ propagator. At NLO only soft gluon corrections to that propagator
can contribute. Soft gluons connecting the propagator and the handle cannot contribute: as we have
remarked, in the soft limit $G^>\sim G^<\sim G^{rr}\sim1/g\,G_R$, so that these soft gluons have to
connect to the handle as $r$ fields and their contribution cancels between the two branches of the handle.
}
The same would not be true for $d\pll/dt$ (in the $p\to\infty$ limit), where we would encounter a
single $F^{+-}$ insertion (at $x^+$) and the handle would be critical in obtaining a gauge-invariant
leading-order result.

\section{Leading-order matching}
\label{app_lo_matching}
In this section we shall prove how the diffusion+conversion+large-angle scattering is equivalent
to the dressed $\twotwo$ processes of \cite{Arnold:2002zm}.
\subsection{Diffusion matching}
For simplicity, we only consider diffusion matching for
the $q_1q_2\lra q_1q_2 $ contribution previously illustrated in  \Eq{el2}. In the prescription
of \cite{Arnold:2002zm}, that process is treated by using the identity
\begin{equation}
\label{amysubst}
\frac{s^2{+}u^2}{t^2} = \half + \half \, \frac{(s{-}u)^2}{t^2} \,,
\end{equation}
and the replacement
\begin{equation}
\frac {(s{-}u)^2}{t^2} \longrightarrow
\left|
G^R_{\mu\nu}(P{-}P')(P{+}P')^\mu (K{+}K')^\nu
\right|^2 ,
\label{eq:sut}
\end{equation}
where $G^R(Q)$ is the retarded HTL propagator, as given by Eqs.~\eqref{htllong} and \eqref{htltrans}.
Upon plugging this into \Eq{el2reg},
putting the IR regulator to zero, introducing instead an UV regulator $T\gg\mu_{\qpt}\gg gT$ and
consistently expanding for $\omega,\qpt\sim gT$, we get
\begin{eqnarray}
C_{q_1}^\mathrm{large}[\PP] & \supset & \frac{\cf g^4}{32 \pi^3}
\int_{-\infty}^{+\infty} d\omega\int_{0}^{\mu_{\qpt}} d\qpt \frac{\qpt}{q}
\int_{0}^\infty dk \nfd(k) \, [1-\nfd(k)] \times
\\ \nn &&
\left\{\left(2\left\vert G^L_R(Q)\right\vert^2+\frac{\qpt^4}{q^4}\left\vert G^T_R(Q)\right\vert^2\right)
k\left[k+\omega\left(1-\frac
  kp\right)\right]+\order{\omega^2,\qpt^2}\right\}
\times  \\ \nn &&
\left[
\frac{\omega T-\omega^2(1{-}2\nfd(k))}{2T^2} \PP^{q_1}(p) 
+\frac{\omega T-\omega^2(1{-}\nfd(k))}{T}\frac{d\PP^{q_1}(p)}{dp}
-\frac{\omega^2}{2}\frac{d\PP^{q_1}(p)}{dp^2}
\right],
\label{el2softscalar}
\end{eqnarray}
where $G(Q)$ is understood to be $G(\omega,q=\sqrt{\omega^2+\qpt^2})$ and we have omitted
the stimulation factor and its derivatives on the outgoing hard leg, as they are all
exponentially suppressed.  Up to higher-order corrections we
can put the lower integration limit for the $k$ integration to zero%
\footnote{
  \label{foot_bosons}The same is not possible
  when there are bosonic degrees of freedom associated with $k$, due
  to Bose enhancements. There, one needs to consider this region with
  care; this region is part of the semi-collinear processes.}.
The terms within the first set of curly brackets come from the expansion of the $\phi$-averaged
matrix element, whereas those in the second set come from the expansion of the distribution
functions.  We remark that
the square moduli of the propagators on the first line are even functions of $\omega$.
Hence, the terms that would naively be of leading order in this expansion in $g$, \ie\ those
multiplying $\PP$ and its first derivative on the third line, vanish in the integration.
Keeping only the surviving, even-in-$\omega$ pieces and performing the $k$ integration we have
\begin{eqnarray}
\nn C_{q_1}^\mathrm{large}[\PP]	&\supset&-\frac{g^2T^2}{6}\frac{\cf g^2}{32\pi^2}
\int_{-\infty}^{+\infty} d\omega\int_{0}^{\mu_{\qpt}} d\qpt \,\qpt
\frac{\rho^{--}(Q)}{\md^2\omega}\\
&&\times\omega^2\Bigg\{
\PP^{q_1}(p) \, \frac{2}{p}
+\frac{d\PP^{q_1}(p)}{dp}\left[\left(1+\frac {2T}{p}\right)
\right] +T\frac{d\PP^{q_1}(p)}{dp^2}
\Bigg\}.
\label{el2softeven3s}
\end{eqnarray}
where we have also used the following relations, based on the explicit form of the propagator in
Eqs.~\eqref{htllong} and \eqref{htltrans}:
\begin{eqnarray}
\nn	\rho^L(Q)&=&G^L_R(Q)-G^L_A(Q)
=\pi \md^2 \frac{\omega}{q}\left\vert G^L_R \right\vert^2,\\
\rho^T(Q)&=&G^T_R(Q)-G^T_A(Q)=\pi \md^2\frac{\omega}{2q} \left(1-\frac{\omega^2}{q^2}\right)
\left\vert G^T_R \right\vert^2.
\label{rhotor}
\end{eqnarray}
We also remark that the next terms in the soft expansion, \ie\ those $\OO(\omega^2,\qpt^2)$,
which naively would contribute to relative $\OO(g)$, give rise again to a vanishing odd
integration and thus contribute only to $\OO(g^2)$. For this reason
they can be neglected.%
\footnote{%
  In order to obtain the explicit form of these $\OO(\omega^2,\qp^2)$ terms, the prescription illustrated
  in Eqs.~\eqref{amysubst} and \Eq{eq:sut} is no longer sufficient, as Hard Thermal Loops need
  to be included also on less IR-sensitive terms. However, on general grounds, the expansion can only
  give rise to even powers of $\omega$ at that order, as we have
  checked explicitly.
}
Similarly, the next order in the expansion of the distribution function is also odd and vanishes.
Hence, a genuine $\OO(g)$ correction can only arise from adding soft gluons to these diagrams.

An analogous expression can be obtained in the case of a $\p$-dependent $\PP$. It reads
\begin{eqnarray}
\nn C_{q_1}^\mathrm{large}[\PP]	&\supset&-\frac{g^2T^2}{6}\frac{\cf g^2}{32\pi^2}
\int_{-\infty}^{+\infty} d\omega\int_{0}^{\mu_{\qpt}} d\qpt \,\qpt
\frac{\rho^{--}(Q)}{\md^2\omega}\\
&&\times\Bigg\{
\PP^{q_1}(\p) \, \omega^2\frac{2}{p}
+\frac{d\PP^{q_1}(\p)}{dp^z}\left[\omega^2\left(1+\frac {2T}{p}\right)
-\frac{T}{p}\qpt^2\right] \nn\\
&&+T\left(\omega^2\frac{d\PP^{q_1}(\p)}{d(p^z)^2}+\frac{\qpt^2}{2}\nabla^2_{p_\perp}\PP^{q_1}(\p)\right)
\Bigg\}.
\label{el2softeven3}
\end{eqnarray}

Now let us look at the coefficients entering \Eq{diff}, as defined in Eqs.\eqref{dragrate},
\eqref{longrate} and \eqref{transrate}. The differential rates appearing there can be easily inferred
from the loss term of the collision operator~\eqref{el2}. Applying the same steps that led
to \Eq{el2softscalar} we have that the contribution from scattering with a quark $q_2$ to $d\pll/dt$
for a quark reads
\begin{eqnarray}
\nn \frac{d\pll}{dt}\Big\vert_{q_1}	&\supset&-\frac{g^4}{(2\pi)^3}
\frac{\cf}{4}\int_{-\infty}^{+\infty} d\omega\int_{0}^{\mu_{\qpt}} d\qpt \frac{\qpt}{q}
\int_{0}^\infty dk\,q^z\Bigg\{\left(2\left\vert G^L_R(Q)\right\vert^2+\frac{\qpt^4}{q^4}
\left\vert G^T_R(Q)\right\vert^2\right)\\
&&\times k\left[k+\omega\left(1-\frac kp\right)\right]+\order{\omega^2,\qpt^2}
\Bigg\}  \nfd(k) \, [1-\nfd(k+\omega)],
\label{eloss}
\end{eqnarray}
where  we have for clarity left the stimulation
factor $[1-\nfd(k+\omega)]$ unexpanded in $\omega\sim g$. Using \Eq{qptsoft},
\ie\ $q^z=\omega+\qpt^2/(2p)$, one sees again that the naive leading order in $g$
leads to a vanishing $\omega$ integration, which is at the base of the Einstein relation
relating $d\pll/dt$ and $\ql$ in the $p\to\infty$ limit. The leading, even-in-$\omega$ terms
then yield, upon performing the $k$ integration and using again \Eq{rhotor}
\begin{equation}
\frac{d\pll}{dt}\Big\vert_{q_1}	\supset-\frac{g^2T^2}{6} \frac{\cf g^2}{32\pi^2}
\int_{-\infty}^{+\infty} d\omega\int_{0}^{\mu_{\qpt}} d\qpt\,\qpt
\frac{\rho^{--}(Q)}{\md^2\omega}\left[\omega^2 \left(
1-\frac{2T}{p}\right)+\qpt^2\frac{T}{p}\right],
\label{elossevenfinal}
\end{equation}
where we have again not considered the $\OO(g^2)$ correction from the $\OO(\omega^2,\qpt^2)$
terms in the expansion of the matrix elements and from the expansion of the distribution functions.

Returning to $\ql$, it is immediate to see that, at leading order, only  the $\omega^2$ term in
$(q^z)^2$ contributes and other terms are actually suppressed by a factor of $g^2$. Hence
$\ql$ reads at LO
\begin{equation}
\ql\Big\vert_{q_1}\supset\frac{g^2T^3}{6}\frac{\cf g^2}{16\pi^2}
\int_{-\infty}^{+\infty} d\omega\int_{0}^{\mu_{\qpt}} d\qpt\,\qpt
\,\omega^2\frac{\rho^{--}(Q)}{\md^2\omega}.
\label{longdifffinal}
\end{equation}
Similarly, for $\qhat$ one has $\qp^2=\qpt^2$, up to odd corrections
or $\OO(g^2)$ terms, so that one obtains the well known result (see \Eq{qhat_LO})
\begin{equation}
\qhat\Big\vert_{q_1}\supset\frac{g^2T^3}{6}\frac{\cf g^2}{16\pi^2}
\int_{-\infty}^{+\infty} d\omega\int_{0}^{\mu_{\qpt}} d\qpt\,\qpt
\,\qpt^2\frac{\rho^{--}(Q)}{\md^2\omega}.
\label{qhatdifffinal}
\end{equation}
We furthermore remark that in $\ql$ and $\qhat$ corrections in $1/p$ enter only at $\OO(g^2)$, due
again to the $\OO(g)$ term being odd in $\omega$. For this same reason the evaluation in $(\omega,\qpt)$
and $(\qll,\qp)$ coordinates is equivalent. This justifies our evaluation of the ($p$-independent) NLO
corrections to $\ql$ and $\qhat$ in the latter coordinate set.

Let us obtain
the complete leading-order $d\pll/dt$, $\ql$ and $\qhat$. To this end,
one has $2(\nf-1)$ quarks and antiquarks
that are distinguishable from $q_1$, and hence $4(\nf-1)$ contributions in the form of
Eqs.~\eqref{elossevenfinal}, \eqref{longdifffinal} and \eqref{qhatdifffinal}, the extra factor of $2$
coming from the sum over final states (see footnote~\ref{foot_final}). The contribution
from $q_1 q_1$ scattering accounts for  two times those equations, as the $u$-channel contribution
is identical, and the $q_1 \bar q_1$ accounts for another two due again to
final state symmetries.
Hence the contribution from all quark scatterings account for a factor of $4\nf$. Using altogether
similar steps one can show that the contribution from $q_1g$ scatterings amounts to a factor of $8\nc$,
so that the complete leading-order expressions are
\begin{eqnarray}
\label{elosssummed}
\frac{d\pll}{dt}\Big\vert_{q}	&=&- \frac{\cf g^2}{8\pi^2}
\int_{-\infty}^{+\infty} d\omega\int_{0}^{\mu_{\qpt}} d\qpt\,\qpt
\frac{\rho^{--}(Q)}{\omega}\left[\omega^2 \left(
1-\frac{2T}{p}\right)+\qpt^2\frac{T}{p}\right],\\
\label{longdiffsummed}
\ql\Big\vert_{q}&=&\frac{\cf g^2}{4\pi^2}
\int_{-\infty}^{+\infty} d\omega\int_{0}^{\mu_{\qpt}} d\qpt\,\qpt
\,\omega^2\frac{T}{\omega}\rho^{--}(Q),\\
\label{qhatdifsummed}
\qhat\Big\vert_{q}&=&\frac{\cf g^2}{4\pi^2}
\int_{-\infty}^{+\infty} d\omega\int_{0}^{\mu_{\qpt}} d\qpt\,\qpt
\,\qpt^2\frac{T}{\omega}\rho^{--}(Q),
\end{eqnarray}
which agree with Eqs.~\eqref{lofinal} and \eqref{qhat_LO}.
In the case where the hard particle is a gluon, one
obtains the same expressions with $\cf$ replaced by $\ca$.

We can now see explicitly, by comparing Eqs.~\eqref{elosssummed}, \eqref{longdiffsummed} and
\eqref{qhatdifsummed}, that the equilibration condition
\Eq{equilibrate} is obeyed at leading order. Finally,
let us take \Eq{diff}, and substitute the equilibration condition.
This yields, for a $p$-dependent $\PP$,
\begin{equation}
\label{diffexpscalar}
C_a^\mathrm{diff}[\PP]=-\frac{\ql}{Tp}\PP(p)-\frac{d\PP(p)}{dp}\ql\left(\frac{1}{p}
+\frac{1}{2T}\right)-\half\ql\frac{d^2\PP(p)}{dp^2},
\end{equation}
which matches with the structure of \Eq{el2softeven3s}. In the $\p$-dependent case
we recover instead Eq.~\eqref{diffexplo},
which also matches with \Eq{el2softeven3}. We have thus explicitly shown
how the effective diffusion picture of \Eq{diff} matches exactly at leading order
with the standard treatment of dressed matrix elements.

\subsection{Conversion matching}
\label{app_lo_conv}
Let us consider more in detail the $t$-channel quark exchange to  the $q_1\bar q_1\lra gg$ process,
as introduced in \Eq{el2quarkreg}. The resummation of HTLs in the $t$ propagator, as per the
prescription of \cite{Arnold:2002zm,Arnold:2003zc}, gives
\begin{equation}
\int_0^{2\pi}\frac{d\phi}{2\pi}\frac ut\to
-\frac{pk}{q^2}
\left[\left(\omega{-}q\right)^2S^+_R(Q)S^+_A(Q)+\left(\omega+ q\right)^2S^-_R(Q)S^-_A(Q)\right]
\left(1+\order{\frac{\omega}{T},\frac{\omega}{p}}\right),
\label{softuovertresum}
\end{equation}
where we have used our parameterization~\eqref{htlfermiondef} of the quark propagator.
Although not immediately obvious in a naive expansion, all $\OO(g)$ corrections do take
the form of an odd function of $\omega$, whereas the leading order, \ie\ the terms in square brackets,
are even. By using the explicit form of the propagator in \Eq{htlfermion} the expression above
simplifies to
\begin{equation}
\int_0^{2\pi}\frac{d\phi}{2\pi}\frac ut\to
\frac{2pk}{\mmf}
\left[\left(q-\omega\right)\rho_+(Q)+\left(q+\omega\right)\rho_-(Q)\right]
\left(1+\order{\frac{\omega}{T},\frac{\omega}{p}}\right),
\label{softuovertresum2}
\end{equation}
where $\rho_\pm(Q)=S^\pm_R(Q) -S^\pm_A(Q)$.
Similarly the expansion of the statistical factors, as in \Eq{statexpand}, leads to
odd terms in $\omega$ as the only possible $\OO(g)$ corrections.

Hence, summing all contributions\footnote{This amounts to the $u/t$ and $t/u$ terms for
$q_1\bar q_1\lra gg$, as well as the $s/u$ one for $q_1g\lra q_1g$. The non-underlined $u/s$ there
can be easily shown not to contribute at leading and next-to-leading order in $g$, whereas the $t/u$ and
$s/u$ terms become identical to \Eq{softuovertresum2}.}, the conversion part of the collision operator
for a quark $i$ reads
\begin{eqnarray}
\nn C_{q_i}^\mathrm{conv}[\PP]	&=&\frac{g^4\cf^2}{8\pi^4\mmf p}
\Bigl\{\PP^{q_1}(p)-\PP^{g}(p) \Bigr\}
\int_{0}^\infty \!\!dk\, k \, \nfd(k)  \, [1+\nbe(k)] \int_{-\infty}^{\infty} d\omega\int_0^{\mu_{\qpt}}  d\qpt \,\qpt \\
&&\times
\left[\rho_+(Q)\left(1-\frac\omega q\right)+\rho_-(Q)\left(1+\frac\omega q\right)\right]
\left(1+\order{g^2}\right),
\label{el2quarkregconv}
\end{eqnarray}
and corrections are naturally suppressed by $g^2$ because of the even $\omega$ integration.
Carrying out the $k$ integration leads to
\begin{eqnarray}
\nn C_{q_i}^\mathrm{conv}[\PP]	&=&\frac{g^2}{16\pi^2}
\frac{\cf}{p}\Bigl\{
\PP^{q_1}(p)-\PP^{g}(p)
\Bigr\}\int_{-\infty}^{\infty} d\omega\int_0^{\mu_{\qpt}} d\qpt \,\qpt
\\
&&\times
\left[\rho_+(Q)\left(1-\frac\omega q\right)+\rho_-(Q)\left(1+\frac\omega q\right)\right]
\left(1+\order{g^2}\right),
\label{el2quarkregconv2}
\end{eqnarray}
Finally, the $\omega$ integration can be performed using the sum rule in
\cite{Ghiglieri:2013gia,Besak:2012qm}, leading to
\begin{eqnarray}
C_{q_i}^\mathrm{conv}[\PP]	&=&\frac{g^2}{8\pi}
\frac{\cf}{p}\Bigl\{
\PP^{q_1}(p)-\PP^{g}(p)
\Bigr\}\int_0^{\mu_{\qpt}} d\qpt \,\qpt \frac{\mmf}{\qpt^2+\mmf},
\label{el2quarkregconv3}
\end{eqnarray}
which matches with Eqs.~\eqref{defconvq} and \eqref{convratefermionlo}.

The conversion operator for gluons can be easily checked using the same approach.
The case of a $\p$-dependent distribution function is also a straightforward generalization. It
too matches with the results of Sec.~\ref{sec_lo_conv}.


\section{Solving the integral equation in position space
at LO and NLO}
\label{sub_nlo_coll}
The most convenient way to solve \Eq{defimplfull}\footnote{The $g\lra q\bar q$ case is
again not dealt with explicitly.}
is by Fourier transforming $\bh$ and $\bqp$ into impact-parameter
variables, as first proposed in \cite{Aurenche:2002pd}.  In this way
the convolution over the collision kernel $\cc(\kp)$
diagonalizes, turning an integral equation into a differential equation.
Furthermore,  the source on the left-hand side becomes a boundary condition at $\b=0$
and the desired final integral, \Eq{jmcoll}, becomes a boundary
value of the ODE solution.  Specifically, defining
\begin{equation}
\bff(\b) = \int \frac{d^2h}{(2\pi)^2} e^{i\b \cdot \bh}\bff(\bh)\,,
\end{equation}
we have
\begin{equation}
\label{b_want}
{\rm Re}\:\int \frac{d^2h}{(2\pi)^2} 2 \bh \cdot \bff(\bh)
= {\rm Im}( 2 \nabla_{b} \cdot \bff(0))\,,
\end{equation}
and \Eq{defimplfull} becomes
\begin{eqnarray}
\nn-2i\bfnabla \delta^2(\b)&=&\frac{i}{2p\omega(p-\omega)}\left(p(p-\omega)m_{\infty\,\omega}^2+
p\,\omega\,m_{\infty\,p-\omega}^2-\omega(p-\omega)m_{\infty\,p}^2-\nabla_\b^2\right)\bff(\b)\\
&&+
\left(\cc'_R(\vert \omega\vert\,b)-\frac{\cc'_A(\vert \omega\vert\,b)}{2}
+\frac{\cc'_A(\vert p\vert\,b)}{2}+\frac{\cc'_A(\vert p-\omega\vert b)}{2}\right)\bff(\b),
\label{bspace}
\end{eqnarray}
with
\begin{equation}
\label{eq:C(b)}
\cc_R'(\vert \omega \vert\,b) \equiv \int \frac{d^2\kp}{(2\pi)^2}
\Big( 1 - e^{i \omega \b \cdot \bkp} \Big) \cc_R(\kp)\,.
\end{equation}

As we mentioned in Sec.~\ref{sub_intro_nlo_coll},
for generic kinematics $\OO(g)$ corrections enter then in two places:
the effective thermal masses squared $m^2_{\infty\,p}$ and the collision kernel
$\cc(\kp)$ get $\OO(g)$ corrections which modify \Eq{bspace},
\begin{eqnarray}
m^2_{\infty\,p,\rm LO+NLO}& =& m^2_{\infty\,p} +\delta m^2_{\infty\,p}, \\
\cc'_{R\,{\rm LO+NLO}}(b) &=& \cc_R'(b) + \delta \cc_R'(b).
\end{eqnarray}
The NLO thermal masses have been given in \Eq{minftyNLO}.
The NLO collision kernel is computed in
\cite{CaronHuot:2008ni} in momentum space;
the Fourier transformation into impact parameter space has been performed in
\cite{Ghiglieri:2013gia}.  The expressions are sufficiently cumbersome
that we have decided not to repeat them here.
In \cite{CaronHuot:2008ni} it was also explicitly
shown that ``three-pole'' contributions are absent at NLO, so that the sum of
two-body  (dipole) interactions on the second line of \Eq{bspace} still holds.

\Eq{bspace} is then solved perturbatively, by treating $\bff(\b)$
formally as an expansion in powers of $\delta m_\infty, \delta \cc$;
$\bff(\b) = \bff_0(\b)+\bff_1(\b)+\ldots$, and expanding to first order.
The zero-order expression is just \Eq{bspace}, while at the linear order
the expression reads
\begin{eqnarray}
\nn 0 &=& \left( \frac{i }{2p\omega(p-\omega)} \Big(  (p(p-\omega)m_{\infty\,\omega}^2+
p\,\omega\,m_{\infty\,p-\omega}^2-\omega(p-\omega)m_{\infty\,p}^2-\nabla_\b^2 \Big)\right.\\
&&\nn\left.+ \cc'_R(\vert \omega\vert\,b)-\frac{\cc'_A(\vert \omega\vert\,b)}{2}
+\frac{\cc'_A(\vert p\vert\,b)}{2}+\frac{\cc'_A(\vert p-\omega\vert b)}{2}\right) \bff_1(\b)\\
&&\nn+\left( \frac{i }{2p\omega(p-\omega)} \Big(  (p(p-\omega)\delta m_{\infty\,\omega}^2+
p\,\omega\,\delta m_{\infty\,p-\omega}^2-\omega(p-\omega)\delta m_{\infty\,p}^2 \Big)\right.\\
&&\left.+ \delta\cc'_R(\vert \omega\vert\,b)-\frac{\delta\cc'_A(\vert \omega\vert\,b)}{2}
+\frac{\delta\cc'_A(\vert p\vert\,b)}{2}+\frac{\delta\cc'_A(\vert p-\omega\vert b)}{2}\right)
\bff_0(\b) \,,
\label{bspace3}
\end{eqnarray}
where the leading order solution $\bff_0(\b)$ acts as a source term in the
differential equation for $\bff_1(\b)$. We refer to
\cite{Ghiglieri:2013gia,Ghisoiu:2014mha,Ghiglieri:2014kma}
for details on the boundary conditions and the numerical evaluation of these
equations.

\section{Longitudinal momentum diffusion at NLO}
\label{app_nlo}
In this appendix we present the details of the calculation of $\ql$ to NLO. We will not
explicitly consider diagrams with HTL vertices: as in the photon case, their contribution
can be shown to be suppressed once the contour is deformed away from the real axis. Furthermore,
when performing such deformations, we will not explicitly keep track
of contributions from certain causality-violating poles
at $\qll=\qm/2\pm i\qp$ ($q^2=0$), which are artifacts of our gauge
choice and cancel in the final sum over diagrams, as they must
\cite{Ghiglieri:2013gia}.
\subsection{The rainbow diagram}

Let us first go through the diagrams shown in
Fig.~\ref{fig_nlo_soft_diagrams},
contributing to $\delta\ql\Big\vert_\mathrm{loop}$ (we will drop the
${}_\mathrm{loop}$ label to avoid clutter).
\begin{figure}[ht]
\begin{center}
\includegraphics[width=5cm]{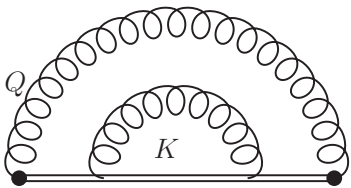}
\end{center}
\caption{The rainbow diagram}
\label{fig_rainbow}
\end{figure}
We will label first ``rainbow'' diagram, shown in
Fig.~\ref{fig_rainbow}, $r$. Its contribution reads
\begin{eqnarray}
\nn\delta\ql\Big\vert_{r}&=&-g^4\crr\int_{-\infty}^{+\infty}dx^+
\int_0^{x^+}dx^{+\prime}\int_0^{x^{+\prime}}dx^{+\prime\prime}
\int\frac{d^4Q}{(2\pi)^4}\int\frac{d^4K}{(2\pi)^4}\\
\label{defhardself}
&&\times e^{-iq^-x^+}e^{-i\km(x^{+\prime}-x^{+\prime\prime})}(\qll)^2G^{--}_{rr}(Q)G^{--}_{rr}(K),
\end{eqnarray}
where, as we remarked in Sec.~\ref{sec_long_diff}, the specific ordering
of the two propagators is not relevant to NLO, as long as they receive a Bose enhancement.
The Wilson line integrations yield
\begin{equation}
\label{hardselfmom2}
\delta\ql\Big\vert_{r}=g^4\crr \ca
\int\frac{d^4Q}{(2\pi)^4}\int\frac{d^4K}{(2\pi)^4}
\left(\frac{i(\qll)^2}{(q^-+i\epsilon)^2(q^-+\km+i\epsilon)}-\mathrm{adv}\right)
G^{--}_{rr}(Q)G^{--}_{rr}(K),
\end{equation}
where ``adv'' stands for the advanced $-i\epsilon \to +i\epsilon$
counterpart of the first term in round brackets.
As in Sec.~\ref{sub_diff} and Fig.~\ref{fig_contour},
we set out to perform the $\qll$ integration in the complex plane.
The integral is very sensitive to large $\qll$ due to the
$(\qll)^2$ in the numerator; but, contrary to the
leading-order case, $\qm$ is not fixed to be zero; also,
$G_{rr}^{--}(Q)$ contains the statistical function
$\nbe(q^0) \simeq T/q^0 = T/(\qll + \qm/2)$.  Applying some numerator
algebra to these terms, we obtain
\begin{equation}
\frac{T(\qll)^2}{\qll+q^-/2}=T\qll-\frac{Tq^-}{2}+\frac{T(q^-)^2}{4(\qll+q^-/2)}.
\end{equation}
The first term yields the contour deformation, the second will vanish as we shall show
(no poles and no contour
contributions) and the third can be dealt with using Euclidean technology.

We start with the contribution from the first term, with additional label
$(a)$ for arc.
Upon deforming $\qll$ away from the real
axis, the retarded propagator turns into \Eq{arcexpand},
so that
\begin{eqnarray}
	\nn	\delta\ql\Big\vert^{(a)}_{r}&=&g^4\crr \ca T
\int_{\calr}\frac{d^4Q}{(2\pi)^4}\int\frac{d^4K}{(2\pi)^4}
\left(\frac{i}{(q^-+i\epsilon)^2(q^-+\km+i\epsilon)}-\mathrm{adv}\right)\\
&&\times G^{--}_{rr}(K)\frac{i}{\qll}\left(1+\frac{q^-}{\qll}\right)\frac{2\qll q^--\mmg}{2\qll q^--\qp^2-\mmg}\Big\vert_{\calr}+\cala,
\label{starthardselfa}
\end{eqnarray}
where $\calr$ and $\cala$ are the retarded and advanced deformed contours, as defined in
Sec.~\ref{sub_diff}. The contribution from the latter is not shown explicitly.
Let us define $\dep$ and, for later convenience, $\depq$ and $\depqm$ as
\begin{equation}
\label{defdep}
\dep\equiv\frac{\qp^2+\mmg}{2\qll},\qquad
\depq\equiv\frac{(\bqp+\bkp)^2+\mmg}{2\qll},\qquad\depqm\equiv\frac{(\bqp-\bkp)^2+\mmg}{2\qll}.
\end{equation}
When deforming above the $\qll$ axis we can then close the $\qm$ contour in the lower half-plane,
picking the pinched $\qm=\dep$ retarded pole (and conversely
for the $\cala$ contribution).
This  yields
\begin{eqnarray}
	\nn	\delta\ql\Big\vert^{(a)}_{r}&=&-ig^4\crr \ca T
\int_{\calr}\frac{d\qll d^2\qp}{(2\pi)^3}\int\frac{d^4K}{(2\pi)^4}
\left(-\frac{i}{\dep^2(\dep+\km-i\epsilon)}\right)\\
&&\times G^{--}_{rr}(K)\frac{i}{\qll}\left(1+\frac{\dep}{\qll}\right)\frac{\qp^2}{2\qll}+\cala.
\end{eqnarray}
The final expression, up to order $1/\qll$ terms, reads
\begin{equation}
	\delta\ql\Big\vert^{(a)}_{r}=g^4\crr \ca T
\int_{\calr}\frac{d\qll}{2\pi}\int\frac{d^2\qp}{(2\pi)^2}\int\frac{d^4K}{(2\pi)^4}
\frac{\qp^2G^{--}_{rr}(K)}{2(\qll)^2\dep^2}\left(\pi\delta(\km)+\frac{i\dep}{(\km-i\epsilon)^2}\right)+\cala\,,
\end{equation}
where we have used the symmetries of the integrand to express the leading-order
term as a $\delta$-function of $\km$.

We now inspect the second term, labeled $(s)$
\begin{equation}
	\delta\ql\Big\vert^{(s)}_{r}=-\frac{g^4\crr \ca}{2}
\int\! \frac{d^4Q}{(2\pi)^4}\int \! \frac{d^4K}{(2\pi)^4}
\left(\frac{i}{(q^-{+}i\epsilon)(q^-{+}\km+{i}\epsilon)}-\mathrm{adv}\right)
\rho^{--}_{rr}(Q)G^{--}_{rr}(K).
\label{laddersecond}
\end{equation}
When deforming on $\calr$ and $\cala$ we have
\begin{eqnarray}
	\nn\delta\ql\Big\vert^{(s)}_{r}&=&\frac{g^4\crr \ca}{2}
\int_{\calr}\frac{d\qll}{(2\pi)}\int\frac{dq^-d^2\qp}{(2\pi)^3}\int\frac{d^4K}{(2\pi)^4}
\left(\frac{i}{(q^--i\epsilon)(q^-+\km-i\epsilon)}\right)\\
&&\times	G^{--}_{rr}(K)\frac{i}{(\qll)^2}\left(1+\frac{q^-}{\qll}\right)\frac{2\qll q^--\mmg}
{2\qll(q^--\dep+i\epsilon)}+\cala.
\end{eqnarray}
The $\qm$ integration can be performed as before, yielding
\begin{eqnarray}
	\nn\delta\ql\Big\vert^{(s)}_{r}&=&-i\frac{g^4\crr \ca}{2}
\int_{\calr}\frac{d\qll}{(2\pi)}\int\frac{d^2\qp}{(2\pi)^2}\int\frac{d^4K}{(2\pi)^4}
\left(\frac{i}{(\dep)(\dep+\km-i\epsilon)}\right)\\
&&\times	G^{--}_{rr}(K)\frac{i}{(\qll)^2}\left(1+\frac{q^-}{\qll}\right)\frac{\qp^2}
{2\qll}+\cala,
\end{eqnarray}
which goes like $1/(\qll)^2$ and hence is irrelevant. This can be easily understood by noting
that the pinched poles in $\qm$ force $\qm\sim1/\qll$, so that the factor of $\qm/\qll$ of this term
with respect to \Eq{starthardselfa} behaves like $1/(\qll)^2$.

Finally, we look at the Euclidean term, labeled $(e)$, which reads
\begin{equation}
	\delta\ql\Big\vert^{(e)}_{r}=g^4\crr \ca
\int\frac{d^4Q}{(2\pi)^4}\int\frac{d^4K}{(2\pi)^4}
2\pi\delta(q^-+\km)
\frac{1}{4}G^{--}_{rr}(Q)G^{--}_{rr}(K).
\label{hardselfe}
\end{equation}
This term will be canceled by an opposite term in another diagram.

\subsection{The crossed rainbow diagram}
\begin{figure}[ht]
\begin{center}
\includegraphics[width=5cm]{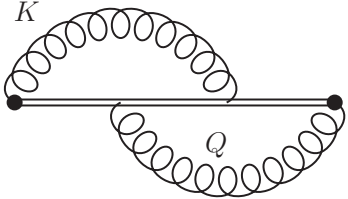}
\end{center}
\caption{The crossed rainbow diagram}
\label{fig_cross}
\end{figure}
The amplitude of this diagram, shown in Fig.~\ref{fig_cross} and labeled $+$ for cross, reads
\begin{eqnarray}
\nn\delta\ql\Big\vert_{+}&=&+g^4\crr \ca\int_{-\infty}^{+\infty}dx^+
\int_0^{x^+}dx^{+\prime}\int_0^{x^{+\prime}}dx^{+\prime\prime}
\int\frac{d^4Q}{(2\pi)^4}\int\frac{d^4K}{(2\pi)^4}\\
\label{defcross}
&&\times e^{-iq^-(x^+-x^{+\prime\prime})}e^{-i\km x^{+\prime}}\qll\kl G^{--}_{rr}(Q)G^{--}_{rr}(K).
\end{eqnarray}
The sign is opposite to \Eq{defhardself} because of the different ordering of the
color matrices. Doing the $x^+$ integrations we obtain
\begin{eqnarray}
\nonumber		\delta\ql\Big\vert_{+}&=&-g^4\crr \ca
\int\frac{d^4Q}{(2\pi)^4}\int\frac{d^4K}{(2\pi)^4}
\left(\frac{i}{(\qm+i\epsilon)(\km+i\epsilon)(\qm+\km+i\epsilon)}
-\mathrm{adv}\right)\\
&&\hspace{4.5cm}\times \qll\kl G^{--}_{rr}(Q)G^{--}_{rr}(K).
\label{crossdelta2}
\end{eqnarray}
Again $G^{--}_{rr}(Q) G^{--}_{rr}(K)$ contain statistical functions
$\frac{T^2}{q^0 k^0} = \frac{4T^2}{(2\qm+\qll)(2\km+\kl)}$.  We handle this by
performing the following algebra:
\begin{equation}
\frac{4T^2\qll\kl}{(2\qll{+}q^-)(2\kl{+}\km)}=T^2-\frac{T^2q^-}{2\qll{+}q^-}
-\frac{T^2\km}{2\kl{+}\km}+\frac{T^2q^-\km}{(2\qll{+}q^-)(2\kl{+}\km)}.
\end{equation}
The first term will not contribute: deforming the $\qll$ integral
\begin{eqnarray}
\nn	\delta\ql\Big\vert_{+}^{(1)}&=&g^4\crr \ca T^2
\int_{\calr}\frac{d^4Q}{(2\pi)^4}\int\frac{d^4K}{(2\pi)^4}
\left(\frac{i}{(\qm-i\epsilon)(\km-i\epsilon)(\qm+\km-i\epsilon)}
\right)\rho^{--}_{rr}(K)\\
&&\times\frac{i}{(\qll)^2}\left(1+\frac{q^-}{\qll}\right)\frac{2\qll q^--\mmg}
{2\qll(\qm-\dep+i\epsilon)}+\cala\,,
\end{eqnarray}
the $\qm$ integration can be closed below, yielding
\begin{eqnarray}
\nn	\delta\ql\Big\vert_{+}^{(1)}&=&-ig^4\crr \ca T^2
\int_{\calr}\frac{d\qll d^2\qp}{(2\pi)^3}\int\frac{d^4K}{(2\pi)^4}
\left(\frac{i}{\dep(\km-i\epsilon)(\km+\dep-i\epsilon)}
\right)\rho^{--}_{rr}(K)\\
&&\times\frac{i}{(\qll)^2}\left(1+\frac{q^-}{\qll}\right)\frac{\qp^2}
{2\qll}+\cala\,.
\end{eqnarray}
The $\km$ integration can be closed in the upper half-plane, giving
\begin{eqnarray}
\nn\delta\ql\Big\vert_{+}^{(1)}&=&g^4\crr \ca T^2
\!\int_{\calr} \!\frac{d\qll d^2\qp}{(2\pi)^3}\int\frac{d\kl d^2\kp}{(2\pi)^3}
\frac{i}{\dep^2}\left[ G_R^{--}(\km{=}0)-G_R^{--}(\km{=}-\dep)
\right]\\
&&\times\frac{i}{(\qll)^2}\left(1+\frac{q^-}{\qll}\right)\frac{\qp^2}
{2\qll}+\cala\,.
\end{eqnarray}
This vanishes on $\calr$, because the square bracket is at least linear in $\dep$.

The second and third term are identical to \Eq{laddersecond} and thus vanish.
Only the last term contributes, yielding
\begin{equation}
\delta\ql\Big\vert_{+}=-g^4\crr \ca
\int\frac{d^4Q}{(2\pi)^4}\int\frac{d^4K}{(2\pi)^4}
\frac{2\pi\delta(q^-+\km)}{4}G^{--}_{rr}(Q)G^{--}_{rr}(K),
\label{crossfinal}
\end{equation}
which cancels \Eq{hardselfe}.

\subsection{The cat eye diagram}

\begin{figure}[ht]
\begin{center}
\includegraphics[width=5cm]{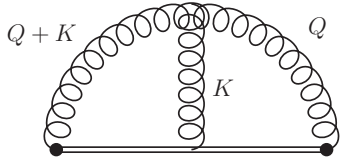}
\end{center}
\caption{The cat-eye diagram}
\label{fig_cateye}
\end{figure}
The diagram is shown in Fig.~\ref{fig_cateye}, and will be labeled $c$
for cat-eye.  Using only the bare vertex, the graph yields
\begin{eqnarray}
  \nn\delta\ql\Big\vert_{c}\!\! &=& g^4\crr\ca \!
  \int_{-\infty}^{+\infty} \!\! dx^+ \! \int_0^{x^+} \!\!\! dx^{+\prime}
  \int\! \frac{d^4Q}{(2\pi)^4}\int\! \frac{d^4K}{(2\pi)^4}
  e^{-i(q^-x^++\km x^{+\prime})} \Gamma_{\mu\nu\rho}(-Q,-K,Q{+}K)\\
\nn	 &&\times
\qll(\qll+\kl)\bigg[G_A^{-\rho}(Q{+}K)G_{rr}^{-\nu}(K)G_{rr}^{-\mu}(Q)
  + G_{rr}^{-\rho}(Q{+}K)G_{R}^{-\nu}(K) G_{rr}^{-\mu}(Q)\\
&& + G_{rr}^{-\rho}(Q{+}K)G_{rr}^{-\nu}(K)G_{R}^{-\mu}(Q)\bigg],
\end{eqnarray}
where we have defined the three-gluon vertex as
\begin{equation}
\label{threegluon}
gf^{abc}\Gamma^{\mu\nu\rho}(P,Q,K)\equiv -gf^{abc}\left[g^{\mu\nu}(P-Q)^\rho
+g^{\nu\rho}(Q-K)^\mu+g^{\rho\mu}(K-P)^\nu\right].
\end{equation}
$P,Q,K$ are all inflowing in the vertex, $P$ is associated with $a$ and $\mu$ and
similarly for the others.
The $x^+$ and $x^{+\prime}$ integrals yield
\begin{eqnarray}
  \nn\delta\ql\Big\vert_{c} \!\! & = & g^4\crr\ca \!
  \int \! \frac{d^4Q}{(2\pi)^4} \! \int \! \frac{d^4K}{(2\pi)^4} \!
\left(\frac{1}{(\qm{+}i\epsilon)(\qm{+}\km{+}i\epsilon)}-\mathrm{adv}\right)
\Gamma_{\mu\nu\rho}(-Q,-K,Q{+}K)\\
\nn	 &&\times
\qll(\qll+\kl)\bigg[G_A^{-\rho}(Q{+}K)G_{rr}^{-\nu}(K)G_{rr}^{-\mu}(Q) + G_{rr}^{-\rho}(Q{+}K)G_{R}^{-\nu}(K)
G_{rr}^{-\mu}(Q)\\
&& + G_{rr}^{-\rho}(Q{+}K)G_{rr}^{-\nu}(K)G_{R}^{-\mu}(Q)\bigg].
\end{eqnarray}
Let us look at the $r/a$ structure of the propagators. Suppressing
Lorentz indices and using $G_{rr}(K) = \nbe(K)(G_R(K)-G_A(K))$,
$\nbe(K) \simeq T/k^0$, the last two lines can be rewritten as
\begin{eqnarray}
\nn&& T \qll G_{rr}(K)\big[G_R(Q)G_R(Q{+}K)-G_A(Q)G_A(Q{+}K)\big]\\
\nn&&+T^2 \rho(Q)\big[G_R(K)G_R(Q{+}K)-G_A(K)G_A(Q{+}K)\big]\\
\nn&&-T\frac{\qm}{2}G_{rr}(Q)\big[\rho(K)G_A(Q{+}K)
+\rho(Q{+}K)G_R(K)\big]\\
\nn&&-T\frac{\qm+\km}{2}G_{rr}(K)\big[\rho(Q)G_A(Q{+}K)+\rho(Q{+}K)G_R(Q)\big]\\
\nn&&-T\frac{\qm+\km}{2}G_{rr}(Q{+}K)\big[\rho(Q)G_R(K)-\rho(K)G_R(Q)\big]\\
\nn&&+\frac{\qm(\qm+\km)}{4}\bigg\{G_{rr}(Q)\big[G_{rr}(K)G_A(Q{+}K)+G_R(K)G_{rr}(Q{+}K)\big]\\
&&+G_{rr}(K)G_{rr}(Q{+}K)G_R(Q)\bigg\}.
\label{propdecomp}
\end{eqnarray}
We start by dealing with the first line, which has the highest power of
$\qll$ in the numerator. We label its contribution $(1)$. It reads
\begin{eqnarray}
\nn\delta\ql\Big\vert_{c}^{(1)}&=&	g^4\crr\ca		\int\frac{d^4Q}{(2\pi)^4}\int\frac{d^4K}{(2\pi)^4}
\left(\frac{1}{(\qm+i\epsilon)(\qm+\km+i\epsilon)}-\mathrm{adv}\right)T\qll G_{rr}^{-\nu}(K)
\\
&&\times\Gamma_{\mu\nu\rho}(-Q,-K,Q{+}K)
\bigg[G_{R}^{-\mu}(Q)G_R^{-\rho}(Q{+}K)
-G_{A}^{-\mu}(Q)G_A^{-\rho}(Q{+}K)\bigg].\nn\\
&&\label{startcateyeone}
\end{eqnarray}
Having obtained a fully retarded (advanced) function of $\qll$ we can now expand
on $\calr$ ($\cala$). Similar comments about pinching poles apply here as well:
we expect $G_R^T(Q)$ and $G_R^T(Q{+}K)$ to introduce poles for $q^-=\dep$ and
$\qm+\km=\depq$ respectively. Indeed we obtain %
\begin{eqnarray}
\nn\delta\ql\Big\vert_{c}^{(1)}&=&	g^4\crr\ca T	
\int_{\calr}\frac{d\qll}{2\pi}\int\frac{dq^-d^2\qp}{(2\pi)^3}\int\frac{d^4K}{(2\pi)^4}
\left(\frac{1}{(\qm+i\epsilon)(\qm+\km+i\epsilon)}-\mathrm{adv}\right)\\
\nn	&&\times\bigg[ 2(\qp^2+\bqp\cdot\bkp)G^T_R(Q)G^T_R(Q{+}K)G_{rr}^{--}(K)
+\order{\frac{1}{\qll}}\bigg]
+\cala.\\
&&\label{catexpanded}
\end{eqnarray}
Expanding the transverse propagators to order $1/(\qll)^2$ we have
\begin{align}
\nn G^T_R(Q)\to& \frac{i}{2\qll(\qm-\dep+i\epsilon)},\\
\qquad G^T_R(Q{+}K)\to&
\frac{i}{2\qll(\qm{+}\km{-}\depq{+}i\epsilon)}\left(1-\frac{\qm+\km}{\qll(\qm{+}\km{-}\depq{+}i\epsilon)}\right),
\label{transprops}
\end{align}
and considering only the $(1/\qll)^0$ terms in \Eq{catexpanded} we have
\begin{eqnarray}
\nn\delta\ql\Big\vert_{c}^{(1)}&=&	-g^4\crr\ca T	
\int_{\calr}\frac{d\qll}{2\pi}\int\frac{dq^-d^2\qp}{(2\pi)^3}\int\frac{d^4K}{(2\pi)^4}
\left(\frac{1}{(\qm+i\epsilon)(\qm+\km+i\epsilon)}-\mathrm{adv}\right)\\
\nn	&&\hspace{-1cm}\times\bigg[\frac{ (\qp^2+\bqp\cdot\bkp)G_{rr}^{--}(K) }
{2(\qll)^2(q^-{-}\dep+i\epsilon)(q^-{+}\km{-}\depq+i\epsilon)}
\left(1-\frac{\qm+\km}{\qll(\qm{+}\km{-}\depq{+}i\epsilon)}\right)\bigg]\\
&&+\cala.\label{catexpanded2}
\end{eqnarray}
Rewriting the terms in round brackets as $\delta$-functions gives
\begin{eqnarray}
\nn\delta\ql\Big\vert^{(1)}_{ca}&=&	-ig^4\crr\ca T	
\int_{\calr}\frac{d\qll}{2\pi}\int\frac{dq^-d^2\qp}{(2\pi)^3}\int\frac{d^4K}{(2\pi)^4}
2\pi P\frac{1}{\km}\left(\delta(\qm+\km)-\delta(\qm)\right)\\
\nn	&&\hspace{-1cm}\times\bigg[\frac{ (\qp^2+\bqp\cdot\bkp)G_{rr}^{--}(K) }
{2(\qll)^2(q^-{-}\dep+i\epsilon)(q^-{+}\km{-}\depq+i\epsilon)}
\left(1-\frac{\qm+\km}{\qll(\qm{+}\km{-}\depq{+}i\epsilon)}\right)\bigg]\\
&&+\cala.
\end{eqnarray}
where $P$ denotes the principal value. This yields
\begin{eqnarray}
\nn \delta\ql\Big\vert^{(1)}_{ca}&=&	-ig^4\crr\ca T	
\int_{\calr}\frac{d\qll}{2\pi}\int\frac{d^2\qp}{(2\pi)^2}\int\frac{d^4K}{(2\pi)^4}
\frac{ (\qp^2+\bqp\cdot\bkp)G_{rr}^{--}(K) }
{2(\qll)^2}\\
\nn &&\times P\frac{1}{\km}\bigg[\frac{ 1}
{\dep(\km-\depq+i\epsilon)}\left(1-\frac{\km}{\qll(\km{-}\depq{+}i\epsilon)}\right)\\
&&+\frac{ 1}
{(\km+\dep-i\epsilon)\depq}\bigg]
+\cala.
\end{eqnarray}
With some algebra and making pinches explicit we have
\begin{eqnarray}
\nn\delta\ql\Big\vert^{(1)}_{ca}&=&	-ig^4\crr\ca T	
\int_{\calr}\frac{d\qll}{2\pi}\int\frac{d^2\qp}{(2\pi)^2}\int\frac{d^4K}{(2\pi)^4}
\frac{ (\qp^2+\bqp\cdot\bkp)G_{rr}^{--}(K) }
{4(\qll)^2}\\
\nn	&&\times\bigg[\frac{ 1}
{\dep(\km-\depq+i\epsilon)}\left(1-\frac{\km}{\qll(\km{-}\depq{+}i\epsilon)}\right)
\left(\frac{2}{\km+i\epsilon}+2\pi i\delta(\km)\right)\\
&&	+\frac{ 1}
{(\km+\dep-i\epsilon)\depq}\left(\frac{2}{\km-i\epsilon}-2\pi i\delta(\km)\right)\bigg]
+\cala,
\end{eqnarray}
which gives, upon expanding the non-pinched denominators
\begin{eqnarray}
\nn\delta\ql\Big\vert^{(1)}_{ca}&=&	-ig^4\crr\ca T	
\int_{\calr}\frac{d\qll}{2\pi}\int\frac{d^2\qp}{(2\pi)^2}\int\frac{d^4K}{(2\pi)^4}\frac{ (\qp^2+\bqp\cdot\bkp)G_{rr}^{--}(K)}
{2(\qll)^2\dep\depq}
\\
&&\times\bigg[-2\pi i\delta(\km)	+\frac{ \depq}
{(\km+i\epsilon)^2}
+\frac{ \dep}
{(\km-i\epsilon)^2}
\bigg]
+\cala.\label{finalcat}
\end{eqnarray}
We have dropped terms that are $\OO(1/(\qll)^2)$.  The terms suppressed by one further power
of $1/\qll$ in \Eq{catexpanded} turn out to be either completely independent
of $\qm$ on $\calr$, and hence vanishing when its integration is done, or proportional to
the terms in \Eq{catexpanded2} times $\km/\qll$, $\qm/\qll$ or $\kl/\qll$. From the
previous calculation it should be clear that the only way they could contribute at order $1/\qll$
on $\calr$ would be if both pinches ($\qm$ and $\qm+\km$) were taken. In the first two cases
that is not possible, because the factors of $\qm$ or $\km$ at the numerator eliminate
either of the two pinched poles and in the last case the resulting $\kl$ integration is
odd once $\km$ is set to zero.

We now consider the second line in \Eq{propdecomp}, which we label $(2)$:
\begin{eqnarray}
\hspace{-1.5em}
\delta\ql\Big\vert_{c}^{(2)}\!\!&=&\! g^4\crr\ca
\! \int \! \frac{d^4Q}{(2\pi)^4} \! \int \! \frac{d^4K}{(2\pi)^4}
\left(\frac{1}{(\qm{+}i\epsilon)(\qm{+}\km{+}i\epsilon)}-\mathrm{adv}\right)T^2 \rho^{-\mu}(Q)
\\
\nn &&\times\Gamma_{\mu\nu\rho}(-Q,-K,Q{+}K)
\bigg[G_{R}^{-\nu}(K)G_R^{-\rho}(Q{+}K)
  -G_{A}^{-\nu}(K)G_A^{-\rho}(Q{+}K)\bigg].
\end{eqnarray}
We can now deform the $\kl$ integration, obtaining
\begin{eqnarray}
\nn\delta\ql\Big\vert_{c}^{(2)}&=&-	g^4\crr\ca	T^2
\int_{\calr}\frac{d\kl d\km d^2\kp}{(2\pi)^4}\int\frac{d^4Q}{(2\pi)^4}
\left(\frac{1}{(\qm+i\epsilon)(\qm+\km+i\epsilon)}-\mathrm{adv}\right)
\\
&&\times
\frac{2(\kp^2+\bkp\cdot\bqp)G^T_R(K)G^T_R(Q{+}K)\rho^{--}(Q)}{\kl},
\end{eqnarray}
which is very similar to what we had before, due to the symmetries
of the vertex. Higher-order terms in the expansion will not be relevant,
as this contribution is one power smaller on the arc. Hence, replacing
the transverse propagators with their leading-order expressions~\eqref{transprops}
and rewriting the terms in round brackets as $\delta$-functions we have
\begin{eqnarray}
\nn\delta\ql\Big\vert_{c}^{(2)}&=&+	ig^4\crr\ca	T^2
\int_{\calr}\frac{d\kl d\km d^2\kp}{(2\pi)^4}\int\frac{d^4Q}{(2\pi)^4}
2\pi P\frac{1}{\km}(\delta(\qm+\km)-\delta(\qm))
\\
&&\times
\frac{(\kp^2+\bkp\cdot\bqp)\rho^{--}(Q)}{2(\kl)^3
(\km-\delta E_\bq+i\epsilon)(\km+\qm-\depq+i\epsilon)},
\end{eqnarray}
which yields
\begin{eqnarray}
\hspace{-2em}
  \nn\delta\ql\Big\vert_{c}^{(2)}&=&-	ig^4\crr\ca	T^2
\int_{\calr}\frac{d\kl d\km d^2\kp}{(2\pi)^4}\int\frac{d\qll d^2\qp}{(2\pi)^3}
P\frac{1}{\km}\frac{(\kp^2+\bkp\cdot\bqp)}{2(\kl)^3}
\\
&&\times\bigg[
\frac{\rho^{--}(-\km,\qll,\qp)}{
(\km-\delta E_\bq+i\epsilon)\depq}+\frac{\rho^{--}(0,\qll,\qp)}{
(\km-\delta E_\bq+i\epsilon)(\km-\depq+i\epsilon)}\bigg].
\end{eqnarray}
The second term on the bottom line vanishes under the $\qll$ integration,
as it is odd. Similarly, the first term yields
\begin{eqnarray}
\nn\delta\ql\Big\vert_{c}^{(2)}&=&-g^4\crr\ca	T^2
\int_{\calr}\frac{d\kl d^2\kp}{(2\pi)^3}\int\frac{d\qll d^2\qp}{(2\pi)^3}
\frac{(\kp^2+\bkp\cdot\bqp)}{2(\kl)^3\depq\delta E_\bq}
\\
&&\times\bigg[G^{--}_R(-\delta E_\bq,\qll,\qp)-G_R^{--}(0,\qll,\qp)  \bigg],
\end{eqnarray}
which vanishes, as the $\qll$ integration can only pick up the residue of the Coulomb gauge
poles, which is $\OO(\delta E_\bq)$ and thus makes the $\kl$ integration vanish.

Finally, terms with $\qm$ or $\qm+\km$ at the numerator in \Eq{propdecomp} vanish
again for the loss of $\qll$ at the
numerator and of a pinched pole at the denominator. The last term trivially vanishes.
The entire result is hence given by \Eq{finalcat}.

\subsection{Self-energy diagrams}
\begin{figure}[ht]
\begin{center}
\includegraphics[width=5cm]{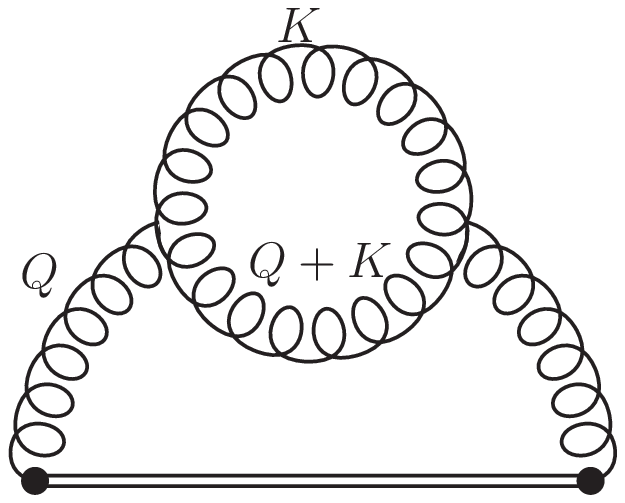}
\includegraphics[width=5cm]{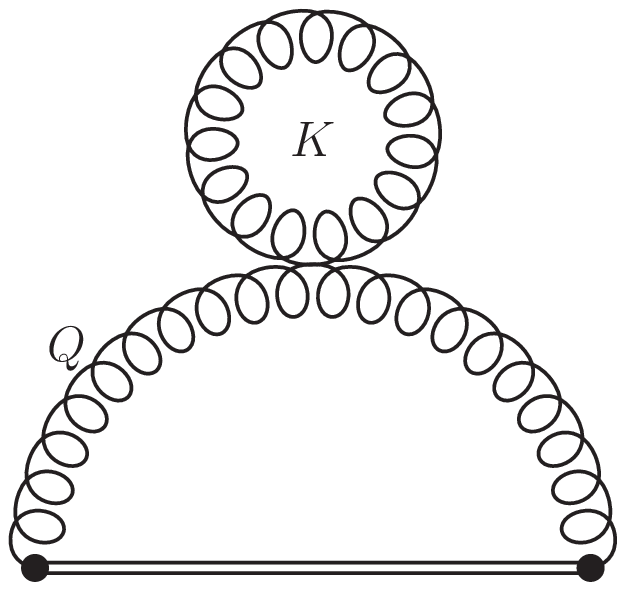}
\end{center}
\caption{The \emph{loop diagram} on the left and the \emph{tadpole diagram}
on the right.}
\label{fig_loop}
\end{figure}

We analyze separately the two diagrams show in Fig.~\ref{fig_loop}, the
\emph{loop diagram} on the left and the \emph{tadpole diagram} on the right.
\subsubsection{The loop diagram}
The amplitude is labeled by $l$ and reads
\begin{eqnarray}
\nn\delta\ql\Big\vert_{\mathrm{l}}&=&\frac{g^4\crr \ca }{2}
\int\frac{d\qll d^2\qp}{(2\pi)^3}\int\frac{d^4K}{(2\pi)^4}T\qll
\Gamma_{\mu\nu\rho}(Q,-Q-K,K)\Gamma_{\mu'\nu'\rho'}(-Q,Q{+}K,-K)\\
\nn&&\times\bigg \{G^{-\mu}_R(Q)G^{-\mu'}_R(Q)	
\big[G_{rr}^{\nu\nu'}(Q{+}K)G_A^{\rho\rho'}(K)+G_{R}^{\nu\nu'}(Q{+}K)
G_{rr}^{\rho\rho'}(K)\big]-\mathrm{adv}\bigg\}_{q^-=0},\\
&&\label{defself}	
\end{eqnarray}
where $1/2$ is a symmetry factor. We now perform a shift $K\to K-Q$ in the first term
in square brackets on the second line. In principle one should be careful in performing such operations, as
the integrals here are not finite. Indeed, as we anticipated, we will need to subtract the HTL counterterm,
which, however, is obtained by performing the same shift, as we will show in
Sec.~\ref{app_htl_ct}. We then have
\begin{eqnarray}
\label{selfshifts}	
\delta\ql\Big\vert_{\mathrm{l}}&=&\frac{g^4\crr \ca }{2}
\int\frac{d\qll d^2\qp}{(2\pi)^3}\int\frac{d^4K}{(2\pi)^4}T\qll
\bigg \{G^{-\mu}_R(Q)G^{-\mu'}_R(Q)\\
\nn&&\hspace{-1.2cm}\times\bigg[\Gamma_{\mu\nu\rho}(Q,-Q-K,K)\Gamma_{\mu'\nu'\rho'}(-Q,Q{+}K,-K)
  G_{R}^{\nu\nu'}(Q{+}K)G_{rr}^{\rho\rho'}(K)  \\ \nn
&&\hspace{-1cm}	
+\Gamma_{\mu\nu\rho}(Q,-K,K-Q)\Gamma_{\mu'\nu'\rho'}(-Q,K,-K+Q)G_{rr}^{\nu\nu'}(K)G_A^{\rho\rho'}(K-Q)\bigg]
-\mathrm{adv}\bigg\}_{q^-=0}.
\end{eqnarray}
We are now free to deform the contour, since in this case there are no statistical factor poles, as
$\qm$ is set to zero. We then have
\begin{eqnarray}
\nn\delta\ql\Big\vert_{\mathrm{l}}&=&g^4\crr \ca
\int_{\calr}\frac{d\qll}{2\pi}\int\frac{d^2\qp}{(2\pi)^2}\int\frac{d^4K}{(2\pi)^4}\frac{2T\qll\qp^2}
{(\qp^2+\mmg)^2}\bigg[\left(G_R^T(K+Q)+G^T_A(K-Q)\right)\\
\nn &&\times G_{++}^{rr}(K)	-\frac{(\km+2\kl)}{2\qll}\left(G^T_A(K-Q)-G_R^T(K+Q)\right)G_{L}^{rr}(K)\\
&&+i\frac{G_T^{rr}(K)}{(\qll)^2}\left(\frac12-\frac{(\bqp\cdot\bkp)^2}{2\qp^2 k^2}\right)+\order{1/(\qll)^2}\bigg]_{q^-=0}
+\cala.
\label{selfexpandend}	
\end{eqnarray}
Terms that had a linear term in the azimuthal angle at the numerator have been neglected.
Since there are no pinching poles in $\km$ we can safely expand the $K\pm Q$ propagators,
yielding %
for their sum, up to order $1/(\qll)^2$
\begin{equation}
G_R^T(K+Q)+G^T_A(K-Q)=\frac{\pi\delta(\km)}{\qll}+\frac{i}{2\qll}\left[
\frac{\depq}{(\km+i\epsilon)^2}+\frac{\depqm}{(\km-i\epsilon)^2}-\frac{2\kl}{\qll}P\frac{1}{\km}
\right],
\end{equation}
whereas the difference is
\begin{equation}
G_A^T(K-Q)-G^T_R(K+Q)=-\frac{i}{\qll}P\frac{1}{\km}+\order{\frac{1}{(\qll)^2}}.
\end{equation}
Plugging this back in \Eq{selfexpandend} we obtain
\begin{eqnarray}
\nn\delta\ql\Big\vert_{\mathrm{l}}&=&2g^4\crr \ca T
\int_{\calr}\frac{d\qll}{2\pi}\int\frac{d^2\qp}{(2\pi)^2}\int\frac{d^4K}{(2\pi)^4}\frac{\qp^2}
{(\qp^2+\mmg)^2}\bigg\{G_{++}^{rr}(K)\bigg[\pi\delta(\km)\\
\nn&&\left.+\frac{i}{2}\left(
\frac{\depq}{(\km+i\epsilon)^2}+\frac{\depqm}{(\km-i\epsilon)^2}-\frac{2\kl}{\qll}P\frac{1}{\km}
\right)\right]+\frac{i(\km+2\kl)}{2\qll}G_{L}^{rr}(K)P\frac{1}{\km}\\
&&+i\frac{G_T^{rr}(K)}{\qll}\left(\frac12-\frac{(\bqp\cdot\bkp)^2}{2\qp^2 k^2}\right)
+\order{1/(\qll)^2}\bigg\}
+\cala.
\label{selfexpandend2}	
\end{eqnarray}
\subsubsection{The tadpole diagram}
The amplitude, labeled by $t$, reads
\begin{eqnarray}
\nn	\delta\ql\Big\vert_{\mathrm{t}}&=&\frac{-ig^4\crr \ca}{2}
\int\frac{d\qll d^2\qp}{(2\pi)^3}\int\frac{d^4K}{(2\pi)^4}T\qll
\bigg [G^{-\mu}_R(Q)G^{-\nu}_R(Q)G_{rr}^{\rho\sigma}(K)\\
&&\times\left(2g_{\mu\nu}g_{\rho\sigma}-g_{\mu\rho}g_{\nu\sigma}
-g_{\mu\sigma}g_{\nu\rho}\right) -\mathrm{adv}\bigg]_{q^-=0}.
\label{deftad}	
\end{eqnarray}
Expanding on  $\calr$ we have
\begin{eqnarray}
\nn\delta\ql\Big\vert_{\mathrm{t}}&=&\frac{-ig^4\crr \ca T}{2}
\int_{\calr}\frac{d\qll}{2\pi}\int\frac{d^2\qp}{(2\pi)^2}\int\frac{d^4K}{(2\pi)^4}\frac{2}{\qll(\qp^2+\mmg)^2}
\bigg[\qp^2G_L^{rr}(K)\\
&&\left.-\left(\qp^2+\frac{(\bqp\cdot\bkp)^2}{k^2}\right)
G_T^{rr}(K)+\order{\frac{1}{(\qll)^2}}\right]+\cala.
\label{tadexpand}	
\end{eqnarray}
\subsubsection{Summary}
Summing Eqs.~\eqref{selfexpandend2}, \eqref{tadexpand}
we obtain
\begin{eqnarray}
\nn	\delta\ql\Big\vert_{\mathrm{t+l-ct}}&=&g^4\crr \ca T
\int_{\calr}\frac{d\qll}{2\pi}\int\frac{d^2\qp}{(2\pi)^2}\int\frac{d^4K}{(2\pi)^4}\frac{\qp^2}
{4(\qll)^2\dep^2}\bigg\{G_{++}^{rr}(K)2\pi \delta(\km)\\
&&+
\frac{2i\dep G_{++}^{rr}(K)}{(\km-i\epsilon)^2}
+\frac{i}{\qll}\bigg[\frac{\kp^2 G_{rr}^{--}(K)}{(\km-i\epsilon)^2}
+ G_T^{rr}(K)\bigg(2-\frac{2\kl\kp^2}{k^2(\km-i\epsilon)}\bigg)\bigg]
\bigg\}\nn\\
&&+\cala,\label{finalself}
\end{eqnarray}
where we have used the fact that $\delta E_{\bq\pm\bk}=\dep+\kp^2/(2\qll)$, up to vanishing terms
in the azimuthal integration and that the square bracket on the second line, which can be identified
with the NLO contribution to $Z_g$ in Coulomb gauge
\cite{CaronHuot:2008uw,Ghiglieri:2013gia}, is purely real,
so that the prescription used for
the $\km$ poles at the denominator there is irrelevant.
\subsection{The subtraction term}
\label{app_htl_ct}
We now turn to the computation of the subtraction counterterm $\delta
\ql\Big\vert^\mathrm{diff}_\mathrm{subtr.}$, \ie\ the soft part
of the HTL self-energy. To this end, we need only the gluon loop, as soft fermions are not Bose enhanced
and do not contribute to relative $\OO(g)$. The contribution
from the loop diagram is, after the previously-discussed shift
\begin{eqnarray}
\nn\delta\ql\Big\vert_{\mathrm{l\,subtr.}}^\mathrm{diff}&=&\frac{g^4\crr \ca }{2}
\int\frac{d^4Q}{(2\pi)^4}\int\frac{d^4K}{(2\pi)^4}T\qll
\Gamma_{\mu\nu\rho}(0,-K,K)\Gamma_{\mu'\nu'\rho'}(0,K,-K)2\pi\delta(\qm)\\
&&\hspace{-2cm}\times\bigg\{G^{-\mu}_R(Q)G^{-\mu'}_R(Q)	\bigg[
G_{R}^{(0)\,\nu\nu'}(Q{+}K)G_{rr}^{(0)\,\rho\rho'}(K)
+G_{rr}^{(0)\,\nu\nu'}(K)
G_A^{(0)\,\rho\rho'}(K{-}Q)\bigg]-\mathrm{adv}\bigg\},\nn\\
&&\label{defloopct}
\end{eqnarray}
where  the vertices are treated in the HTL approximation, \ie\ $Q\ll K$,
and the $G^{(0)}$ propagators
on the second line are bare and in the soft approximation, $\nbe(k^0)\to T/k^0$. This yields
\begin{equation}
\delta\ql\Big\vert_{\mathrm{l\,subtr.}}^\mathrm{diff}\!\!=ig^4\crr \ca T
\!\int_{\calr}\!\! \frac{d\qll d^2\qp}{(2\pi)^3\qll}
\!\int \!\! \frac{d^4K}{(2\pi)^4}
\frac{G_{rr}^{(0)\,T}(K)}{(\qp^2+\mmg)^2}
\bigg[\qp^2 -\frac{(\bqp\cdot\bkp)^2}{k^2}\bigg]+\cala,
\label{loopct}
\end{equation}
where we have used the fact that in Coulomb gauge the bare longitudinal spectral
density vanishes and the transverse one puts $K$ on shell. Furthermore, consistently
with the HTL approximation,
in $G^{(0)\rho\rho'}(K\pm Q)$ one has only to keep the leading terms in $K\gg Q$,
\ie\ $(Q\pm K)^2\to\pm 2Q\cdot K$, $(\bq\pm\bk)^2\to k^2$. Other terms in the propagators and
vertices do not contribute, as discussed in \cite{Ghiglieri:2013gia}.

The tadpole contribution is trivially obtained from \Eq{tadexpand}
by replacing the resummed $K$ propagator with its bare counterpart, \ie
\begin{equation}
  \delta\ql\Big\vert_{\mathrm{t\,subtr.}}^\mathrm{diff}
  \!\! = ig^4\crr \ca T
  \! \int_{\calr} \! \frac{d\qll d^2\qp}{(2\pi)^3\qll}
  \! \int \! \frac{d^4K}{(2\pi)^4}
\frac{G^{(0)\,T}_{rr}(K)}{(\qp^2{+}\mmg)^2}
\bigg[\qp^2+\frac{(\bqp\cdot\bkp)^2}{k^2}\bigg]
+\cala,
\label{tadct}	
\end{equation}
so that the sum is
\begin{equation}
\delta\ql\Big\vert_{\mathrm{subtr.}}^\mathrm{diff}=ig^4\crr \ca T
\int_{\calr}\frac{d\qll d^2\qp}{(2\pi)^3}\int\frac{d^4K}{(2\pi)^4}
\frac{2\qp^2G_{rr}^{(0)\,T}(K)}{\qll(\qp^2+\mmg)^2}
+\cala.
\label{HTLct}
\end{equation}

\subsection{Summary}
Summing the contributions from all diagrams and subtracting the counterterm~\eqref{HTLct}
we obtain
\begin{eqnarray}
\label{almostfinalsoft}
  \delta\ql\Big\vert_{\mathrm{loop}} \!\!\!
  -\delta\ql\Big\vert_{\mathrm{subtr.}}^\mathrm{diff} \!\! &=&
g^4\crr \ca T
\! \int_{\calr} \! \frac{d\qll}{2\pi}
\! \int \! \frac{d^2\qp}{(2\pi)^2}
\! \int \! \frac{d^4K}{(2\pi)^4}
\\ && \times  \nn
\bigg\{
\frac{G^{--}_{rr}(K)\pi\delta(\km)}{(\qll)^2\dep}
\left(\frac{\qp^2}{\dep} -\frac{(\qp^2+\bqp\cdot\bkp)}{\depq}\right)
\\
\nn&& + \frac{i\qp^2}
{4(\qll)^3\dep^2}\bigg[\frac{\kp^2\delta G_{rr}^{--}(K)}{(\km-i\epsilon)^2}
+2\delta G_T^{rr}(K)\bigg(1-\frac{\kl\kp^2}{k^2(\km{-}i\epsilon)}\bigg)\bigg]
\bigg\}\! +\cala,
\end{eqnarray}
where $\delta G\equiv G-G^{(0)}$ is the difference
between  resummed and bare propagators; in Coulomb gauge \Eq{HTLct} is equivalent
to the bare part of \Eq{finalself}.
After subtracting the collinear  counterterm\footnote{A shift in the integration variable is
necessary, see footnotes 9-11 in \cite{Ghiglieri:2013gia}.} given in \Eq{collsoftct}
and performing the $K$ and $\qll$ integrations as in \cite{CaronHuot:2008ni,Ghiglieri:2013gia}
we obtain \Eq{finallongdiffnlo}.

\section{Semi-collinear integrations}
\label{app_semi}

Let us consider \Eq{jmsemicoll}. As mentioned, we put an IR cutoff $\mupp^\mathrm{NLO}$
on $\qp$, which is the same cutoff
used for the diffusion and conversion processes .
We find it is simpler to use $\delta E$ as an integration variable, so that,
after performing the $d^2\kp$ integration and the $\delta E$ integration
with cutoff $\delta E_\mu\equiv(\mupp^\mathrm{NLO})^2\vert p
\vert/(2\vert \omega(p-\omega)\vert)$  we have
\begin{eqnarray}
\nn &&\int\frac{d^2\qp}{(2\pi)^2}\frac{1}{\qp^4}
\int\frac{d^2\kp}{(2\pi)^2}
\bigg[\frac{\md^2\kp^2}
{(\kp^2+\delta E^2)(\kp^2+\delta E^2+\md^2)}-\frac{\md^2}{\kp^2+\md^2}\bigg]\\
\nn&=&\frac{\md p}{32\pi^2\vert  \omega(p-\omega)\vert}\left[-2\pi + \frac{ \delta E_\mu}{\md}
\ln\frac{\delta E_\mu^2+\md^2}{\delta E_\mu^2}+\frac{\md}{\delta E_\mu}\ln\frac{\md^2}{\delta E_\mu^2+\md^2}
+4 \arctan\frac{\delta E_\mu}{\md}\right]\\
&\equiv&\frac{\md p}{32\pi^2\vert  \omega(p-\omega)\vert}
I_\perp\left(\frac{\delta E_\mu}{\md}\right),\label{transintmu}
\end{eqnarray}
so that \Eq{jmsemicoll} turns into
\begin{eqnarray}
\nn	\frac{d\Gamma(p,\omega)}{d\omega}\Big\vert_\mathrm{semi-coll}&=&\frac{g^4\crr T\md}{64\pi^3
\vert  \omega (p-\omega)\vert}
(1\pm n(\omega))(1\pm n(p-\omega)) I_\perp\left(\frac{\delta E_\mu}{\md}\right)\\
&&\times\left\{\begin{array}{cc}
\frac{1+(1-x)^2}{x}\big[\cf x^2+\ca(1-x)\big]& q\to qg\\
\frac{d_F}{d_A}(x^2+(1-x)^2)\big[\cf +\ca x(1-x)\big]& g\to q\bar q\\
\frac{1+x^4+(1-x)^4}{x(1-x)}\ca\big[1-x+x^2\big]& g\to gg	
\end{array}\right\}.
\label{jmsemicolltrans}
\end{eqnarray}

Let us now explicitly obtain the logarithmic sensitivity to the diffusion region. Upon expanding
\Eq{semicollop} and its gluonic equivalent around $\omega\sim 0$ we find
\begin{equation}
\label{diffexpsemicoll}
\delta C_{a\,\mathrm{soft\,gluon}}^\mathrm{semi-coll}[\PP]
=-
\left[\frac{1}{Tp}\PP(\p)+\left(\frac{1}{p}
+\frac{1}{2T}\right)\frac{d\PP(\p)}{dp^z}+\half\frac{d^2\PP(\p)}{d(p^z)^2}\right]
\delta\ql\Big\vert_\mathrm{semi-coll},
\end{equation}
where
\begin{equation}
\label{qlsemisoft}
\delta\ql\Big\vert_\mathrm{semi-coll}=\int_{\vert\omega\vert\siml T}d\omega \omega^2
\frac{g^4\crr\ca T^2\md}{32\pi^3
\vert  \omega^3 \vert}I_\perp\left(\frac{\left(\mupp^\NLO\right)^2}{2\vert\omega\vert\md}\right),
\end{equation}
\ie\ the expected diffusion structure (see for instance Eq.~\eqref{diffexplo} or
App.~\ref{app_lo_matching}) has appeared.
Regulating the $d\omega$ integral with an $\OO(T)$ UV regulator\footnote{Its exact value
is not relevant, as we are only interested in the constant in front of the logarithm.}
and expanding for small
$\mupp^\NLO$ one obtains
\begin{equation}
\label{qlsemisoftfinal}
\delta\ql\Big\vert_\mathrm{semi-coll}=
\frac{g^4\crr\ca T^2\md}{8\pi^2}\left[\ln\frac{(\mupp^\NLO)^2}{2\md T}
+\order{(\mupp^\NLO)^2}\right],
\end{equation}
which indeed cancels the $\mupp^\NLO$ dependence of \Eq{finallongdiffnlo}.

Similarly, as in \Eq{collsubtrconv}, we can take the small-$\omega$ (or small $p-\omega$)
limit for the final-state quarks in \Eq{semicollop}
and examine the overlap with the conversion sector. We take
as example the $q\to g$ rate, the opposite being the same times $d_F/d_A$. We obtain
\begin{equation}
\delta\Gammac_{q\to g}(p)\Big\vert_\mathrm{semi-coll}=
\int_{\vert p-\omega\vert\siml T}d\omega\frac{g^4\cf^2 T\md}{128\pi^3 p
\vert p- \omega \vert}
I_\perp\left(\frac{\left(\mupp^\NLO\right)^2}{2\vert p-\omega\vert\md}\right).
\label{semicollconv}
\end{equation}
Performing the same integration and expansion as before we then have
\begin{equation}
\delta\Gammac_{q\to g}(p)\Big\vert_\mathrm{semi-coll}=
\frac{g^4\cf^2 T^2\md}{32\pi^2p}\left[\ln\frac{(\mupp^\NLO)^2}{2\md T}
+\order{(\mupp^\NLO)^2}\right],
\label{semicollconvfinal}
\end{equation}
which removes the  $\mupp^\NLO$ dependence of \Eq{finalnloconv}.

\bibliographystyle{JHEP}
\bibliography{eloss.bib}

\providecommand{\href}[2]{#2}\begingroup\raggedright\begin{thebibliography}{10}

\bibitem{Roland:2014jsa}
G.~Roland, K.~Safarik, and P.~Steinberg, {\it {Heavy-ion collisions at the
  LHC}},  {\em Prog.Part.Nucl.Phys.} {\bf 77} (2014) 70--127.

\bibitem{d'Enterria:2009am}
D.~d'Enterria, {\it {Jet quenching}},  {\em Springer Verlag,
  Landholt-Boernstein} {\bf Vol. 1-23A} (2009)
  [\href{http://arxiv.org/abs/0902.2011}{{\tt arXiv:0902.2011}}].

\bibitem{Majumder:2010qh}
A.~Majumder and M.~Van~Leeuwen, {\it {The Theory and Phenomenology of
  Perturbative QCD Based Jet Quenching}},  {\em Prog.Part.Nucl.Phys.} {\bf A66}
  (2011) 41--92, [\href{http://arxiv.org/abs/1002.2206}{{\tt
  arXiv:1002.2206}}].

\bibitem{Mehtar-Tani:2013pia}
Y.~Mehtar-Tani, J.~G. Milhano, and K.~Tywoniuk, {\it {Jet physics in heavy-ion
  collisions}},  {\em Int.J.Mod.Phys.} {\bf A28} (2013) 1340013,
  [\href{http://arxiv.org/abs/1302.2579}{{\tt arXiv:1302.2579}}].

\bibitem{Aad:2010bu}
{\bf ATLAS Collaboration} Collaboration, G.~Aad et~al., {\it {Observation of a
  Centrality-Dependent Dijet Asymmetry in Lead-Lead Collisions at
  $\sqrt{s_{NN}}=2.77$ TeV with the ATLAS Detector at the LHC}},  {\em
  Phys.Rev.Lett.} {\bf 105} (2010) 252303,
  [\href{http://arxiv.org/abs/1011.6182}{{\tt arXiv:1011.6182}}].

\bibitem{Chatrchyan:2011sx}
{\bf CMS Collaboration} Collaboration, S.~Chatrchyan et~al., {\it {Observation
  and studies of jet quenching in PbPb collisions at nucleon-nucleon
  center-of-mass energy = 2.76 TeV}},  {\em Phys.Rev.} {\bf C84} (2011) 024906,
  [\href{http://arxiv.org/abs/1102.1957}{{\tt arXiv:1102.1957}}].

\bibitem{Burke:2013yra}
{\bf JET} Collaboration, K.~M. Burke et~al., {\it {Extracting the jet transport
  coefficient from jet quenching in high-energy heavy-ion collisions}},  {\em
  Phys.Rev.} {\bf C90} (2014), no.~1 014909,
  [\href{http://arxiv.org/abs/1312.5003}{{\tt arXiv:1312.5003}}].

\bibitem{Majumder:2013re}
A.~Majumder, {\it {Incorporating Space-Time Within Medium-Modified Jet Event
  Generators}},  {\em Phys.Rev.} {\bf C88} (2013) 014909,
  [\href{http://arxiv.org/abs/1301.5323}{{\tt arXiv:1301.5323}}].

\bibitem{Zapp:2011ya}
K.~C. Zapp, J.~Stachel, and U.~A. Wiedemann, {\it {A local Monte Carlo
  framework for coherent QCD parton energy loss}},  {\em JHEP} {\bf 1107}
  (2011) 118, [\href{http://arxiv.org/abs/1103.6252}{{\tt arXiv:1103.6252}}].

\bibitem{Schenke:2009gb}
B.~Schenke, C.~Gale, and S.~Jeon, {\it {MARTINI: An Event generator for
  relativistic heavy-ion collisions}},  {\em Phys.Rev.} {\bf C80} (2009)
  054913, [\href{http://arxiv.org/abs/0909.2037}{{\tt arXiv:0909.2037}}].

\bibitem{Renk:2010zx}
T.~Renk, {\it {YaJEM: a Monte Carlo code for in-medium shower evolution}},
  {\em Int.J.Mod.Phys.} {\bf E20} (2011) 1594--1599,
  [\href{http://arxiv.org/abs/1009.3740}{{\tt arXiv:1009.3740}}].

\bibitem{Armesto:2009fj}
N.~Armesto, L.~Cunqueiro, and C.~A. Salgado, {\it {Q-PYTHIA: A Medium-modified
  implementation of final state radiation}},  {\em Eur.Phys.J.} {\bf C63}
  (2009) 679--690, [\href{http://arxiv.org/abs/0907.1014}{{\tt
  arXiv:0907.1014}}].

\bibitem{Vitev:2009rd}
I.~Vitev and B.-W. Zhang, {\it {Jet tomography of high-energy nucleus-nucleus
  collisions at next-to-leading order}},  {\em Phys.Rev.Lett.} {\bf 104} (2010)
  132001, [\href{http://arxiv.org/abs/0910.1090}{{\tt arXiv:0910.1090}}].

\bibitem{He:2011pd}
Y.~He, I.~Vitev, and B.-W. Zhang, {\it {${\cal O}(\alpha_s^3)$ Analysis of
  Inclusive Jet and di-Jet Production in Heavy Ion Reactions at the Large
  Hadron Collider}},  {\em Phys.Lett.} {\bf B713} (2012) 224--232,
  [\href{http://arxiv.org/abs/1105.2566}{{\tt arXiv:1105.2566}}].

\bibitem{Xing:2014kpa}
H.~Xing, Z.-B. Kang, E.~Wang, and X.-N. Wang, {\it {Transverse momentum
  broadening at NLO and QCD evolution of $\hat q$}},  {\em Nucl.Phys.} {\bf
  A931} (2014) 493--498, [\href{http://arxiv.org/abs/1407.8506}{{\tt
  arXiv:1407.8506}}].

\bibitem{Kang:2014ela}
Z.-B. Kang, E.~Wang, X.-N. Wang, and H.~Xing, {\it {Transverse momentum
  broadening in semi-inclusive deep inelastic scattering at next-to-leading
  order}},  \href{http://arxiv.org/abs/1409.1315}{{\tt arXiv:1409.1315}}.

\bibitem{Liou:2013qya}
T.~Liou, A.~Mueller, and B.~Wu, {\it {Radiative $p_\bot$-broadening of
  high-energy quarks and gluons in QCD matter}},  {\em Nucl.Phys.} {\bf A916}
  (2013) 102--125, [\href{http://arxiv.org/abs/1304.7677}{{\tt
  arXiv:1304.7677}}].

\bibitem{Blaizot:2013vha}
J.-P. Blaizot, F.~Dominguez, E.~Iancu, and Y.~Mehtar-Tani, {\it {Probabilistic
  picture for medium-induced jet evolution}},  {\em JHEP} {\bf 1406} (2014)
  075, [\href{http://arxiv.org/abs/1311.5823}{{\tt arXiv:1311.5823}}].

\bibitem{Blaizot:2015lma}
J.-P. Blaizot and Y.~Mehtar-Tani, {\it {Jet Structure in Heavy Ion
  Collisions}},  {\em Int. J. Mod. Phys.} {\bf E24} (2015), no.~11 1530012,
  [\href{http://arxiv.org/abs/1503.05958}{{\tt arXiv:1503.05958}}].

\bibitem{Arnold:2002ja}
P.~B. Arnold, G.~D. Moore, and L.~G. Yaffe, {\it {Photon and gluon emission in
  relativistic plasmas}},  {\em JHEP} {\bf 0206} (2002) 030,
  [\href{http://arxiv.org/abs/hep-ph/0204343}{{\tt hep-ph/0204343}}].

\bibitem{Arnold:2002zm}
P.~B. Arnold, G.~D. Moore, and L.~G. Yaffe, {\it {Effective kinetic theory for
  high temperature gauge theories}},  {\em JHEP} {\bf 0301} (2003) 030,
  [\href{http://arxiv.org/abs/hep-ph/0209353}{{\tt hep-ph/0209353}}].

\bibitem{Jeon:2003gi}
S.~Jeon and G.~D. Moore, {\it {Energy loss of leading partons in a thermal QCD
  medium}},  {\em Phys.Rev.} {\bf C71} (2005) 034901,
  [\href{http://arxiv.org/abs/hep-ph/0309332}{{\tt hep-ph/0309332}}].

\bibitem{Qin:2007rn}
G.-Y. Qin, J.~Ruppert, C.~Gale, S.~Jeon, G.~D. Moore, et~al., {\it {Radiative
  and collisional jet energy loss in the quark-gluon plasma at RHIC}},  {\em
  Phys.Rev.Lett.} {\bf 100} (2008) 072301,
  [\href{http://arxiv.org/abs/0710.0605}{{\tt arXiv:0710.0605}}].

\bibitem{Schenke:2009ik}
B.~Schenke, C.~Gale, and G.-Y. Qin, {\it {The Evolving distribution of hard
  partons traversing a hot strongly interacting plasma}},  {\em Phys.Rev.} {\bf
  C79} (2009) 054908, [\href{http://arxiv.org/abs/0901.3498}{{\tt
  arXiv:0901.3498}}].

\bibitem{CaronHuot:2010bp}
S.~Caron-Huot and C.~Gale, {\it {Finite-size effects on the radiative energy
  loss of a fast parton in hot and dense strongly interacting matter}},  {\em
  Phys.Rev.} {\bf C82} (2010) 064902,
  [\href{http://arxiv.org/abs/1006.2379}{{\tt arXiv:1006.2379}}].

\bibitem{Iancu:2015uja}
E.~Iancu and B.~Wu, {\it {Thermalization of mini-jets in a quark-gluon
  plasma}},  {\em JHEP} {\bf 10} (2015) 155,
  [\href{http://arxiv.org/abs/1506.07871}{{\tt arXiv:1506.07871}}].

\bibitem{Ghiglieri:2013gia}
J.~Ghiglieri, J.~Hong, A.~Kurkela, E.~Lu, G.~D. Moore, and D.~Teaney, {\it
  {Next-to-leading order thermal photon production in a weakly coupled
  quark-gluon plasma}},  {\em JHEP} {\bf 1305} (2013) 010,
  [\href{http://arxiv.org/abs/1302.5970}{{\tt arXiv:1302.5970}}].

\bibitem{Ghiglieri:2014kma}
J.~Ghiglieri and G.~D. Moore, {\it {Low Mass Thermal Dilepton Production at NLO
  in a Weakly Coupled Quark-Gluon Plasma}},  {\em JHEP} {\bf 1412} (2014) 29,
  [\href{http://arxiv.org/abs/1410.4203}{{\tt arXiv:1410.4203}}].

\bibitem{simonguy}
S.~Caron-Huot and G.~D. Moore, {\it {Heavy quark diffusion in QCD and N=4 SYM
  at next-to-leading order}},  {\em JHEP} {\bf 0802} (2008) 081,
  [\href{http://arxiv.org/abs/0801.2173}{{\tt arXiv:0801.2173}}].

\bibitem{Braaten:1989mz}
E.~Braaten and R.~D. Pisarski, {\it {Soft Amplitudes in Hot Gauge Theories: A
  General Analysis}},  {\em Nucl.Phys.} {\bf B337} (1990) 569.

\bibitem{Braaten:1991gm}
E.~Braaten and R.~D. Pisarski, {\it {Simple effective Lagrangian for hard
  thermal loops}},  {\em Phys. Rev.} {\bf D45} (1992) 1827--1830.

\bibitem{Blaizot:2001nr}
J.-P. Blaizot and E.~Iancu, {\it {The Quark gluon plasma: Collective dynamics
  and hard thermal loops}},  {\em Phys.Rept.} {\bf 359} (2002) 355--528,
  [\href{http://arxiv.org/abs/hep-ph/0101103}{{\tt hep-ph/0101103}}].

\bibitem{CaronHuot:2008ni}
S.~Caron-Huot, {\it {O(g) plasma effects in jet quenching}},  {\em Phys.Rev.}
  {\bf D79} (2009) 065039, [\href{http://arxiv.org/abs/0811.1603}{{\tt
  arXiv:0811.1603}}].

\bibitem{Ghiglieri:2015zma}
J.~Ghiglieri and D.~Teaney, {\it {Parton energy loss and momentum broadening at
  NLO in high temperature QCD plasmas}},  {\em Int. J. Mod. Phys.} {\bf E24}
  (2015), no.~11 1530013, [\href{http://arxiv.org/abs/1502.03730}{{\tt
  arXiv:1502.03730}}]. To appear in QGP5, ed. X-N.~Wang.

\bibitem{Baier:1994bd}
R.~Baier, Y.~L. Dokshitzer, S.~Peigne, and D.~Schiff, {\it {Induced gluon
  radiation in a QCD medium}},  {\em Phys.Lett.} {\bf B345} (1995) 277--286,
  [\href{http://arxiv.org/abs/hep-ph/9411409}{{\tt hep-ph/9411409}}].

\bibitem{Baier:1996kr}
R.~Baier, Y.~L. Dokshitzer, A.~H. Mueller, S.~Peigne, and D.~Schiff, {\it
  {Radiative energy loss of high-energy quarks and gluons in a finite volume
  quark - gluon plasma}},  {\em Nucl.Phys.} {\bf B483} (1997) 291--320,
  [\href{http://arxiv.org/abs/hep-ph/9607355}{{\tt hep-ph/9607355}}].

\bibitem{Zakharov:1996fv}
B.~Zakharov, {\it {Fully quantum treatment of the Landau-Pomeranchuk-Migdal
  effect in QED and QCD}},  {\em JETP Lett.} {\bf 63} (1996) 952--957,
  [\href{http://arxiv.org/abs/hep-ph/9607440}{{\tt hep-ph/9607440}}].

\bibitem{Zakharov:1997uu}
B.~Zakharov, {\it {Radiative energy loss of high-energy quarks in finite size
  nuclear matter and quark - gluon plasma}},  {\em JETP Lett.} {\bf 65} (1997)
  615--620, [\href{http://arxiv.org/abs/hep-ph/9704255}{{\tt hep-ph/9704255}}].

\bibitem{Arnold:2003zc}
P.~B. Arnold, G.~D. Moore, and L.~G. Yaffe, {\it {Transport coefficients in
  high temperature gauge theories. 2. Beyond leading log}},  {\em JHEP} {\bf
  0305} (2003) 051, [\href{http://arxiv.org/abs/hep-ph/0302165}{{\tt
  hep-ph/0302165}}].

\bibitem{Kurkela:2015qoa}
A.~Kurkela and Y.~Zhu, {\it {Isotropization and hydrodynamization in weakly
  coupled heavy-ion collisions}},  {\em Phys. Rev. Lett.} {\bf 115} (2015),
  no.~18 182301, [\href{http://arxiv.org/abs/1506.06647}{{\tt
  arXiv:1506.06647}}].

\bibitem{Svetitsky:1987gq}
B.~Svetitsky, {\it {Diffusion of charmed quarks in the quark-gluon plasma}},
  {\em Phys.Rev.} {\bf D37} (1988) 2484--2491.

\bibitem{Moore:2004tg}
G.~D. Moore and D.~Teaney, {\it {How much do heavy quarks thermalize in a heavy
  ion collision?}},  {\em Phys.Rev.} {\bf C71} (2005) 064904,
  [\href{http://arxiv.org/abs/hep-ph/0412346}{{\tt hep-ph/0412346}}].

\bibitem{Braaten:1991jj}
E.~Braaten and M.~H. Thoma, {\it {Energy loss of a heavy fermion in a hot
  plasma}},  {\em Phys.Rev.} {\bf D44} (1991) 1298--1310.

\bibitem{Benzke:2012sz}
M.~Benzke, N.~Brambilla, M.~A. Escobedo, and A.~Vairo, {\it {Gauge invariant
  definition of the jet quenching parameter}},  {\em JHEP} {\bf 1302} (2013)
  129, [\href{http://arxiv.org/abs/1208.4253}{{\tt arXiv:1208.4253}}].

\bibitem{Aurenche:2002pd}
P.~Aurenche, F.~Gelis, and H.~Zaraket, {\it {A Simple sum rule for the thermal
  gluon spectral function and applications}},  {\em JHEP} {\bf 0205} (2002)
  043, [\href{http://arxiv.org/abs/hep-ph/0204146}{{\tt hep-ph/0204146}}].

\bibitem{Braaten:1994na}
E.~Braaten, {\it {Solution to the perturbative infrared catastrophe of hot
  gauge theories}},  {\em Phys.Rev.Lett.} {\bf 74} (1995) 2164--2167,
  [\href{http://arxiv.org/abs/hep-ph/9409434}{{\tt hep-ph/9409434}}].

\bibitem{Braaten:1995cm}
E.~Braaten and A.~Nieto, {\it {Effective field theory approach to high
  temperature thermodynamics}},  {\em Phys.Rev.} {\bf D51} (1995) 6990--7006,
  [\href{http://arxiv.org/abs/hep-ph/9501375}{{\tt hep-ph/9501375}}].

\bibitem{Braaten:1995jr}
E.~Braaten and A.~Nieto, {\it {Free energy of QCD at high temperature}},  {\em
  Phys.Rev.} {\bf D53} (1996) 3421--3437,
  [\href{http://arxiv.org/abs/hep-ph/9510408}{{\tt hep-ph/9510408}}].

\bibitem{Kajantie:1995dw}
K.~Kajantie, M.~Laine, K.~Rummukainen, and M.~E. Shaposhnikov, {\it {Generic
  rules for high temperature dimensional reduction and their application to the
  standard model}},  {\em Nucl.Phys.} {\bf B458} (1996) 90--136,
  [\href{http://arxiv.org/abs/hep-ph/9508379}{{\tt hep-ph/9508379}}].

\bibitem{Kajantie:1997tt}
K.~Kajantie, M.~Laine, K.~Rummukainen, and M.~E. Shaposhnikov, {\it {3-D SU(N)
  + adjoint Higgs theory and finite temperature QCD}},  {\em Nucl.Phys.} {\bf
  B503} (1997) 357--384, [\href{http://arxiv.org/abs/hep-ph/9704416}{{\tt
  hep-ph/9704416}}].

\bibitem{Laine:2013lia}
M.~Laine and A.~Rothkopf, {\it {Light-cone Wilson loop in classical lattice
  gauge theory}},  {\em JHEP} {\bf 1307} (2013) 082,
  [\href{http://arxiv.org/abs/1304.4443}{{\tt arXiv:1304.4443}}].

\bibitem{Panero:2013pla}
M.~Panero, K.~Rummukainen, and A.~{Sch\"afer}, {\it {A lattice study of the jet
  quenching parameter}},  {\em Phys.Rev.Lett.} {\bf 112} (2014) 162001,
  [\href{http://arxiv.org/abs/1307.5850}{{\tt arXiv:1307.5850}}].

\bibitem{CasalderreySolana:2006rq}
J.~Casalderrey-Solana and D.~Teaney, {\it {Heavy quark diffusion in strongly
  coupled N=4 Yang-Mills}},  {\em Phys. Rev.} {\bf D74} (2006) 085012,
  [\href{http://arxiv.org/abs/hep-ph/0605199}{{\tt hep-ph/0605199}}].

\bibitem{Gubser:2006nz}
S.~S. Gubser, {\it {Momentum fluctuations of heavy quarks in the gauge-string
  duality}},  {\em Nucl. Phys.} {\bf B790} (2008) 175--199,
  [\href{http://arxiv.org/abs/hep-th/0612143}{{\tt hep-th/0612143}}].

\bibitem{Peigne:2007sd}
S.~Peigne and A.~Peshier, {\it {Collisional Energy Loss of a Fast Muon in a Hot
  QED Plasma}},  {\em Phys.Rev.} {\bf D77} (2008) 014015,
  [\href{http://arxiv.org/abs/0710.1266}{{\tt arXiv:0710.1266}}].

\bibitem{Arnold:1999uza}
P.~B. Arnold, {\it {Langevin equations with multiplicative noise: Resolution of
  time discretization ambiguities for equilibrium systems}},  {\em Phys.Rev.}
  {\bf E61} (2000) 6091--6098, [\href{http://arxiv.org/abs/hep-ph/9912208}{{\tt
  hep-ph/9912208}}].

\bibitem{Arnold:1999va}
P.~B. Arnold, {\it {Symmetric path integrals for stochastic equations with
  multiplicative noise}},  {\em Phys.Rev.} {\bf E61} (2000) 6099--6102,
  [\href{http://arxiv.org/abs/hep-ph/9912209}{{\tt hep-ph/9912209}}].

\bibitem{Besak:2012qm}
D.~Besak and D.~B{\"{o}}deker, {\it {Thermal production of ultrarelativistic
  right-handed neutrinos: Complete leading-order results}},  {\em JCAP} {\bf
  1203} (2012) 029, [\href{http://arxiv.org/abs/1202.1288}{{\tt
  arXiv:1202.1288}}].

\bibitem{Arnold:2008vd}
P.~B. Arnold and W.~Xiao, {\it {High-energy jet quenching in weakly-coupled
  quark-gluon plasmas}},  {\em Phys.Rev.} {\bf D78} (2008) 125008,
  [\href{http://arxiv.org/abs/0810.1026}{{\tt arXiv:0810.1026}}].

\bibitem{CaronHuot:2008uw}
S.~Caron-Huot, {\it {On supersymmetry at finite temperature}},  {\em Phys.Rev.}
  {\bf D79} (2009) 125002, [\href{http://arxiv.org/abs/0808.0155}{{\tt
  arXiv:0808.0155}}].

\bibitem{Arnold:2001ms}
P.~B. Arnold, G.~D. Moore, and L.~G. Yaffe, {\it {Photon emission from quark
  gluon plasma: Complete leading order results}},  {\em JHEP} {\bf 0112} (2001)
  009, [\href{http://arxiv.org/abs/hep-ph/0111107}{{\tt hep-ph/0111107}}].

\bibitem{Arnold:2001ba}
P.~B. Arnold, G.~D. Moore, and L.~G. Yaffe, {\it {Photon emission from
  ultrarelativistic plasmas}},  {\em JHEP} {\bf 0111} (2001) 057,
  [\href{http://arxiv.org/abs/hep-ph/0109064}{{\tt hep-ph/0109064}}].

\bibitem{Bauer:2000ew}
C.~W. Bauer, S.~Fleming, and M.~E. Luke, {\it {Summing Sudakov logarithms in B
  $\to$ X(s gamma) in effective field theory}},  {\em Phys.Rev.} {\bf D63}
  (2000) 014006, [\href{http://arxiv.org/abs/hep-ph/0005275}{{\tt
  hep-ph/0005275}}].

\bibitem{Bauer:2000yr}
C.~W. Bauer, S.~Fleming, D.~Pirjol, and I.~W. Stewart, {\it {An Effective field
  theory for collinear and soft gluons: Heavy to light decays}},  {\em
  Phys.Rev.} {\bf D63} (2001) 114020,
  [\href{http://arxiv.org/abs/hep-ph/0011336}{{\tt hep-ph/0011336}}].

\bibitem{Bauer:2001ct}
C.~W. Bauer and I.~W. Stewart, {\it {Invariant operators in collinear effective
  theory}},  {\em Phys.Lett.} {\bf B516} (2001) 134--142,
  [\href{http://arxiv.org/abs/hep-ph/0107001}{{\tt hep-ph/0107001}}].

\bibitem{Bauer:2001yt}
C.~W. Bauer, D.~Pirjol, and I.~W. Stewart, {\it {Soft collinear factorization
  in effective field theory}},  {\em Phys.Rev.} {\bf D65} (2002) 054022,
  [\href{http://arxiv.org/abs/hep-ph/0109045}{{\tt hep-ph/0109045}}].

\bibitem{Bauer:2002nz}
C.~W. Bauer, S.~Fleming, D.~Pirjol, I.~Z. Rothstein, and I.~W. Stewart, {\it
  {Hard scattering factorization from effective field theory}},  {\em
  Phys.Rev.} {\bf D66} (2002) 014017,
  [\href{http://arxiv.org/abs/hep-ph/0202088}{{\tt hep-ph/0202088}}].

\bibitem{Beneke:2002ph}
M.~Beneke, A.~Chapovsky, M.~Diehl, and T.~Feldmann, {\it {Soft collinear
  effective theory and heavy to light currents beyond leading power}},  {\em
  Nucl.Phys.} {\bf B643} (2002) 431--476,
  [\href{http://arxiv.org/abs/hep-ph/0206152}{{\tt hep-ph/0206152}}].

\bibitem{Chay:2002vy}
J.~Chay and C.~Kim, {\it {Collinear effective theory at subleading order and
  its application to heavy - light currents}},  {\em Phys.Rev.} {\bf D65}
  (2002) 114016, [\href{http://arxiv.org/abs/hep-ph/0201197}{{\tt
  hep-ph/0201197}}].

\bibitem{Manohar:2002fd}
A.~V. Manohar, T.~Mehen, D.~Pirjol, and I.~W. Stewart, {\it {Reparameterization
  invariance for collinear operators}},  {\em Phys.Lett.} {\bf B539} (2002)
  59--66, [\href{http://arxiv.org/abs/hep-ph/0204229}{{\tt hep-ph/0204229}}].

\bibitem{Bauer:2002aj}
C.~W. Bauer, D.~Pirjol, and I.~W. Stewart, {\it {Factorization and endpoint
  singularities in heavy to light decays}},  {\em Phys.Rev.} {\bf D67} (2003)
  071502, [\href{http://arxiv.org/abs/hep-ph/0211069}{{\tt hep-ph/0211069}}].

\bibitem{Kurkela:2014tla}
A.~Kurkela and U.~A. Wiedemann, {\it {Picturing perturbative parton cascades in
  QCD matter}},  {\em Phys.Lett.} {\bf B740} (2014) 172--178,
  [\href{http://arxiv.org/abs/1407.0293}{{\tt arXiv:1407.0293}}].

\bibitem{York:2014wja}
M.~C.~A. York, A.~Kurkela, E.~Lu, and G.~D. Moore, {\it {UV Cascade in
  Classical Yang-Mills via Kinetic Theory}},  {\em Phys.Rev.} {\bf D89} (2014),
  no.~7 074036, [\href{http://arxiv.org/abs/1401.3751}{{\tt arXiv:1401.3751}}].

\bibitem{Kurkela:2014tea}
A.~Kurkela and E.~Lu, {\it {Approach to Equilibrium in Weakly Coupled
  Non-Abelian Plasmas}},  {\em Phys.Rev.Lett.} {\bf 113} (2014), no.~18 182301,
  [\href{http://arxiv.org/abs/1405.6318}{{\tt arXiv:1405.6318}}].

\bibitem{D'Eramo:2010ak}
F.~D'Eramo, H.~Liu, and K.~Rajagopal, {\it {Transverse Momentum Broadening and
  the Jet Quenching Parameter, Redux}},  {\em Phys.Rev.} {\bf D84} (2011)
  065015, [\href{http://arxiv.org/abs/1006.1367}{{\tt arXiv:1006.1367}}].

\bibitem{Ghisoiu:2014mha}
I.~Ghisoiu and M.~Laine, {\it {Interpolation of hard and soft dilepton rates}},
   {\em JHEP} {\bf 1410} (2014) 83, [\href{http://arxiv.org/abs/1407.7955}{{\tt
  arXiv:1407.7955}}].

\end{thebibliography}\endgroup

\end{document}